\documentclass[useAMS,usenatbib]{aa}  

\usepackage{txfonts}
\usepackage{graphicx}
\usepackage{amsmath}
\usepackage{tabularx}
\usepackage{inputenc}
\usepackage{supertabular}
\usepackage{ltxtable}
\usepackage{subfig}
\usepackage{float}

\usepackage{color}

\begin{document} 

   \title{The RoPES project with HARPS and HARPS-N.}

   \subtitle{I. A system of super-Earths orbiting the moderately active K-dwarf HD 176986 \thanks{Based on: observations made with the Italian Telescopio Nazionale Galileo (TNG), operated on the island of La Palma by the INAF - Fundación Galileo Galilei at the Roche de Los Muchachos Observatory of the Instituto  de  Astrofísica  de  Canarias (IAC);  observations  made  with the  HARPS instrument  on  the ESO 3.6-m telescope at La Silla Observatory (Chile)} \thanks{The data used in this paper (Table~\ref{tab_a1}) is available in electronic form at the CDS via anonymous ftp to cdsarc.u-strasbg.fr (130.79.128.5) or via http://cdsweb.u-strasbg.fr/cgi-bin/qcat?J/A+A/}}

      \author{A. Su\'{a}rez Mascare\~{n}o\inst{1,2} \and
         J.~I. Gonz\'{a}lez Hern\'{a}ndez   \inst{1,3} \and
         R. Rebolo  \inst{1,3,4} \and
         S. Velasco\inst{1,3} \and 
         B. Toledo-Padr\'{o}n\inst{1,3} \and
         S. Udry\inst{2} \and
         F. Motalebi\inst{2} \and
         D. S\'{e}grasan\inst{2} \and
         A. Wyttenbach\inst{2} \and
         M. Mayor \inst{2} \and
         F. Pepe \inst{2} \and
         C. Lovis \inst{2} \and
         N. C. Santos \inst{5,6} \and
         P. Figueira \inst{5} \and
         M. Esposito \inst{7}
}         
    \institute{Instituto de Astrof\'{i}sica de Canarias, E-38205 La Laguna, Tenerife, Spain\\
              \email{asm\_ext@iac.es, jonay@iac.es}            \and                      
               Observatoire Astronomique de l'Université de Genève, 1290 Versoix, Switzerland \\
                             \email{Alejandro.SuarezMascareno@unige.ch}          \and
               Universidad de La Laguna, Dpto. Astrof\'{i}sica, E-38206 La Laguna, Tenerife, Spain \and
               Consejo Superior de Investigaciones Cient{\'\i}ficas, Spain \and
                Instituto de Astrofísica e Ciências do Espa\c{c}o, Universidade do Porto,CAUP, Rua das Estrelas, 4150-762 Porto, Portugal   \and    
             Departamento de F\'{i}sica e Astronomia, Faculdade de Ci\^{e}ncias,Universidade do Porto,Rua Campo Alegre, 4169-007  Porto, Portugal             
                \and 
            INAF - Osservatorio Astronomico di Capodimonte, Via Moiariello, 16, 80131 - NAPOLI}

   \date{Accepted 11/2017}

% \abstract{}{}{}{}{} 
% 5 {} token are mandatory
 
  \abstract
   {We report the discovery of a system of two super-Earths orbiting the moderately active K-dwarf HD 176986. This work is part of the RoPES RV program of G- and K-type stars, which combines radial velocities (RVs) from the HARPS and HARPS-N spectrographs to search for short-period terrestrial planets. HD 176986 b and c are super-Earth planets with masses of 5.74 and 9.18 M$_{\oplus}$, orbital periods of 6.49 and 16.82 days, and distances of 0.063 and 0.119 AU in orbits that are consistent with circular.  The host star is a K2.5 dwarf, and despite its modest level of chromospheric activity (log$_{10}$ (R$_{HK}^{'})$ = -- 4.90 $\pm$ 0.04), it shows a complex activity pattern. Along with the discovery of the planets, we study the magnetic cycle and rotation of the star. HD 176986 proves to be suitable for testing the available RV analysis technique and further our understanding of stellar activity.}

   {}

   \keywords{      
    Planetary Systems --- Techniques: radial velocity --- Stars: activity --- Stars: rotation --- Stars: magnetic cycle --- Stars: individual (HD 176986)
 }

   \maketitle
%
%________________________________________________________________

\section{Introduction}

Unveiling the population of small-mass planets in the range of super-Earths and smaller is a key step in order to understand the formation of planetary systems \citep{Mordasini2012, Benz2014}. The HARPS survey \citep{Mayor2003,Mayor2011, Mayor2014}, the Keck survey \citep{Howard2010} and the Kepler survey \citep{Borucki2011, Fressin2013} have made very remarkable contributions to our knowledge of the population of low-mass and small-radius planets around late-type stars. 

Radial velocity (RV) measurements have been especially effective in detecting low-mass planets around nearby stars  \citep{Mayor2009, Lovis2011, Pepe2011, Santos2014, Affer2016, AngladaEscude2016, Masca2017, Gillon2017, AstudilloDefru2017, Masca2017c, Perger2017}. Low-mass planets are particularly difficult targets when they
are found in systems or in the presence of activity signals at similar timescales, however. Detecting small amplitude signals in these cases demands very many observations over long timescales, even with modern spectrographs like HARPS and HARPS-N, which can provide sub metre-per-second precision in RV measurements \citep{Pepe2011} or the upcoming ESPRESSO mission, which is expected to reach a precision of a few centimetres per second \citep{Pepe2013}. 

At these levels,  signals induced by stellar activity in the RV curves  become an important  limiting factor even in the case of quiet stars. This has to be properly addressed, especially when we aim to detect Earth-like planets \citep{Dumusque2012, Santos2014, Robertson2014, Masca2017b}. Stellar activity can induce apparent Doppler shifts of the stellar spectrum \citep{Saar1997}, which causes periodic signals that range from less than one to dozens of metres per second \citep{Huelamo2008, Pont2011,Dumusque2011, Hatzes2013, Masca2017b}. Long-term magnetic cycles can have a similar effect \citep{Lovis2011}. They can create RV variations that could easily be mistaken for Neptune- or Jupiter-mass planets at large orbital separations. The correct detection and characterization of the different star-induced signals and their effect in the RVs is one of the most important steps in order to detect low-mass exoplanets, and its importance will greatly increase with increased precision, as even in the case of the quietest stars, these signals will surface. Complex activity patterns that combine RV signals induced by magnetic cycles, rotation, and differential rotation might induce detectable signals at these periods, and their harmonics are to be expected in almost every star. 

The star HD 176986 is a bright, K2.5-type star \citep{Gray2006}. It is visible from both the southern and northern hemispheres. It shows a mean level of stellar activity similar to that of the Sun (log$_{10}$ (R$_{HK}^{'})$ = -- 4.90 $\pm$ 0.04), but the observed activity pattern seems to be more complex than what would be expected for a quiet star.  We report the discovery of a multiple system composed of two low-mass planets.  We outline the step-by-step analysis of all the available time-series in order to identify the origin of each of the detected signals, and finally, we perform the global analysis of the RV data. 

\section{The RoPES program}

The RoPES program \textup{(Rocky Planets in Equatorial Stars)} is an effort led by the Instituto de Astrof\'{i}sica de Canarias (IAC) to study a group of $\sim$20 nearby equatorial solar-type and late K-type dwarfs combining HARPS and HARPS-N data. The goal is to combine daily-cadence observations with HARPS-N with the long baseline of the HARPS survey, and the increased precision expected with the arrival of the Laser Frequency Comb in order to search for small amplitude Keplerian signals in a very broad space of planetary parameters. The selected stars have spectral types from G0 to K3, and most of them are around G8 to K1, and magnitudes from m$_{V}$ $\sim$ 3 to $\sim$9, with a mean magnitude m$_{V}$ $\sim$7.5. Most of the selected stars show low chromospheric activity levels around log$_{10}$ (R$_{HK}^{'})$ $\sim$ -- 4.9. Of the $\sim$20 selected stars, three are orbited by previously reported planets, which are in all cases massive planets at large orbital separations. 

We present the first discovery of the RoPES program, a system of two low-mass planets orbiting the moderately active K-type dwarf HD 176986. Table~\ref{tab:parameters} shows the stellar parameters of HD 176986.

\begin{table}
\begin{center}
\caption{Stellar properties of HD176986 \label{tab:parameters}}
\begin{tabular}[center]{l l l}
\hline
Parameter & HD 176986 & Ref. \\ \hline
RA (J2000) & 19:03:05.87 & 1 \\
DEC (J2000) & -11:02:38.13 & 1\\
$\mu \alpha$ ($mas$ yr$^{-1}$)& -126.19 & 1 \\
$\mu \delta$ ($mas$ yr$^{-1}$)& -236.86 & 1 \\
Parallax ($mas$) &      35.90 $\pm$ 0.23 & 2\\
Distance (pc) & 26.4 $\pm$ 0.7 & 2 \\
$m_{B}$  (mag) & 9.39 $\pm$ 0.03 & 3 \\
$m_{V}$  (mag) & 8.45 $\pm$ 0.01 & 3 \\
Spectral Type  & K2.5V & 4\\
L$_{*}$/L$_{\odot}$ & 0.331 $\pm$ 0.027 & 5 \\
T$_{eff}$ (K) & 4931 $\pm$ 77 & 6 \\
$[Fe/H]$ (dex) & 0.03 $\pm$ 0.05 & 6 \\
M$_{*}$ (M$_{\odot}$) & 0.789  $\pm$ 0.019 & 6 \\
R$_{*}$ (R$_{\odot}$) & 0.782  $\pm$ 0.035 & 7 \\
P$_{rot}$ (days) & 35.9 $\pm$ 0.2 & 0 \\
log g (cgs) & 4.44 $\pm$ 0.17 & 6\\
Age (Gyr) & 4.3 $\pm$ 4.0 & 6\\
log$_{10}$ (R$_{HK}^{'})$ & -- 4.90 $\pm$ 0.04 & 0 \\
\hline
\end{tabular}
\end{center}
\textbf{References:} 0 - This work, 1 - \citet{vanLeeuwen2007}, 2 - \citet{Gaia2016}, 3 - \citet{Hog2000}, 4 -\citet{Gray2006}, 5 - \citet{Sousa2008}, 6 - \citet{Tsantaki2013}, 7 - estimated following  \citet{Boyajian2012}.
\end{table}

\subsection{HARPS and HARPS-N Spectroscopy}

The star HD 176986 has been extensively monitored since mid-2004 with HARPS and HARPS-N. It was first followed in the {\footnotesize HARPS} planet-search  programme on Guaranteed Time Observations (GTO, PI: M. Mayor) for six years between 2003 and 2009. The high-precision part of this {\footnotesize HARPS} GTO survey aimed at the detection of very low-mass planets in a sample of quiet solar-type stars that had before been screened for giant planets at a lower precision with the {\footnotesize CORALIE} echelle spectrograph mounted on the 1.2 m Swiss telescope on the same site  \citep{Udry2000}.  The observations were then continued within the ESO Large Programs 183.C-0972+183.C-1005 (PI: S. Udry). Then it was observed with HARPS-N from 2014 onwards as part of the RoPES project, using a daily-cadence observation strategy aimed at the detection of very low-mass planets in close orbits of quiet G and K-type stars. HARPS \citep{Mayor2003} and HARPS-N \citep{Cosentino2012} are two fibre-fed high-resolution echelle spectrographs installed at the 3.6 m ESO telescope in La Silla Observatory (Chile) and at the Telescopio Nazionale Galileo in the Roque de los Muchachos Observatory (Spain), respectively. Both instruments have a resolving power of $R\sim 115\,000$ over a spectral range from $\sim$380 to $\sim$690 nm and have been designed to attain very high long-term RV precision. Both are contained in temperature- and pressure-controlled vacuum vessels to avoid spectral drifts due to temperature and air pressure variations, thus  ensuring their stability. HARPS and HARPS-N are equipped with their own pipeline that provides extracted and wavelength-calibrated spectra, as well as RV measurements and other data products such as cross-correlation functions and their bisector profiles. All observations have been carried out with simultaneous calibration, using the thorium argon (ThAr) lamp or the Fabry Perot (FP), depending on the observation dates. During the HARPS campaigns, our star was typically observed once per night using an exposure time of 900 s, with only a few exceptions. In the HARPS-N campaign, the star was always observed using 3 x 300 s exposures per visit. There was one visit per night during the first years and two visits separated by a few hours during the 2016 and 2017 campaigns. The data were then re-sampled and averaged into one-hour bins. Each of the one-hour bins is considered an individual observation. The combination of these two observational programmes provided 156 HARPS observations and 103 HARPS-N newly acquired observations, obtained in 234 individual nights during 13.2 years of observations.

\subsection{ASAS Photometry}

We also used the photometric data on HD 176986 provided by the All Sky Automated Survey (ASAS) public database. ASAS \citep{Pojmanski1997} is an all sky survey in the $V$ and $I$ bands running since 1998 at Las Campanas Observatory, Chile.  Best photometric results are achieved for stars with V $\sim$8-14, but this range can be extended  by implementing  some quality control on the data. ASAS has produced light curves for about $10^{7}$ stars at $\delta < 28^{\circ}$. The ASAS catalogue supplies ready-to-use light curves with flags indicating the quality of the data. For this analysis we relied  only on  good quality data (grades "A" and "B" in the internal flags). Even after this quality control,  there are still some high-dispersion measurements  that cannot  be explained by a regular stellar behaviour. We rejected these measurements by de-trending the series and eliminating points that deviated more than three times the standard deviation from the median seasonal value. We were left with 411 photometric observations taken over 8.7 yr that show an RMS of 13.4 mmag and a typical uncertainty of 10.4 mmag per exposure. 

\subsection{FastCam lucky imaging observations}

On June 6, 2016, we collected 30,000 individual frames of HD 176986 in the I band using the lucky imaging FastCam instrument
\citep{Oscoz2008} at the 1.5m Carlos S\'anchez Telescope (TCS) at the Teide Observatory in Tenerife, with 30 ms exposure time for each frame. FastCam is an optical imager with a low-noise EMCCD camera that allows obtaining speckle-featuring unsaturated images at a fast frame rate \citep{Labadie2011}.

\begin{figure}
\centering
        \includegraphics[width=9.0cm]{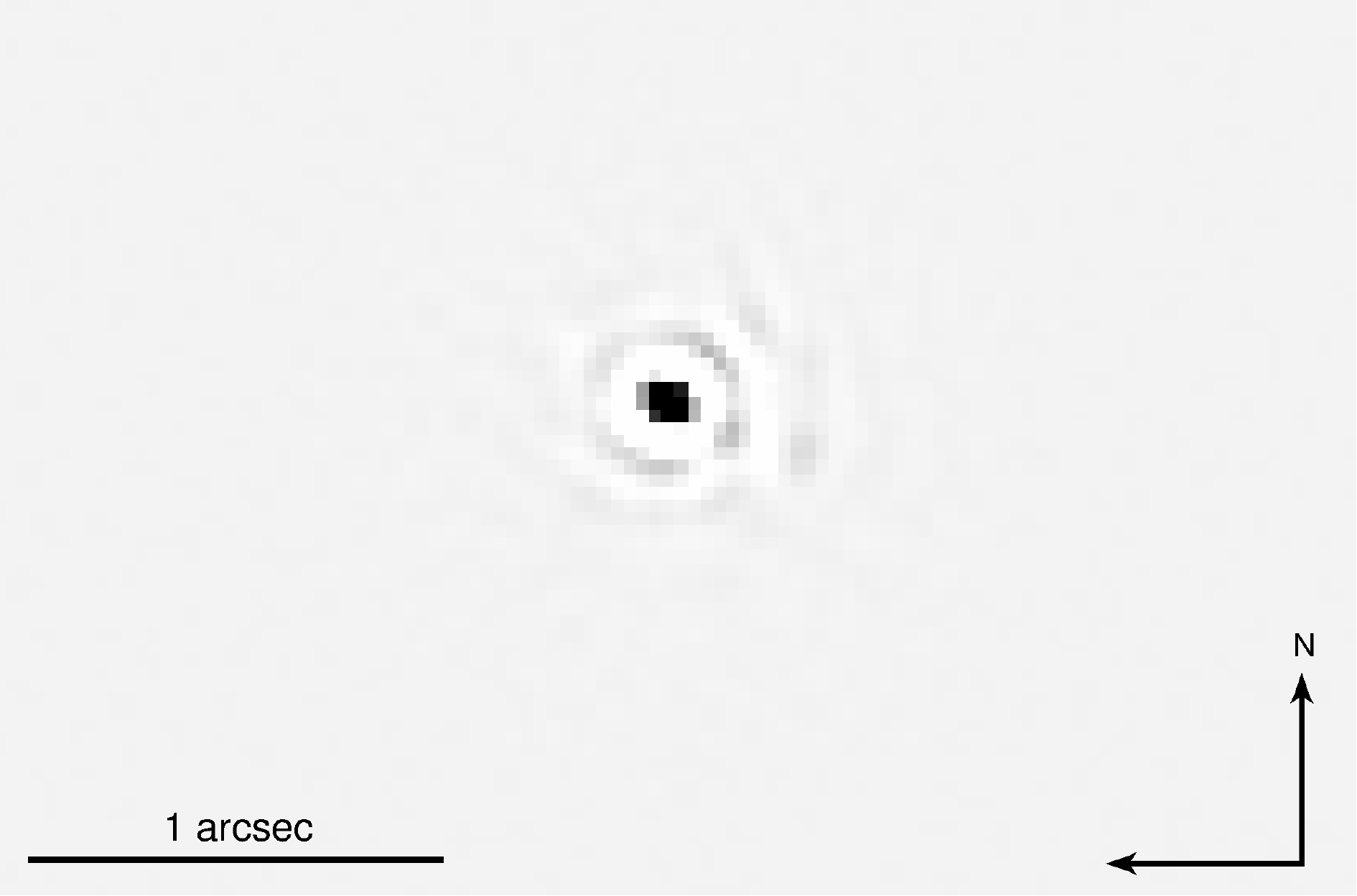}
        \caption{Diffraction-limited image of HD 176986 after lucky imaging processing with a selection of the 10\% best individual TCS/FastCam frames.}
        \label{lucky}
\end{figure}

In order to construct a high-resolution diffraction-limited long-exposure image, the individual frames were bias subtracted, aligned, and co-added using our own lucky imaging  algorithm \citep{Velasco2016}. Figure \ref{lucky} presents the high-resolution image constructed by co-addition of the  best percentage of the images using lucky imaging and shift-and-add algorithms. Owing to the atmospheric conditions of the night, selecting 10\% of the individual frames was found to be the best solution to  produce a deep and diffraction-limited image of the target, resulting in a total integration time of 90 seconds.  The combined image achieved $\Delta m_I=4.5 - 5.0$ at $1 \farcs$ We find no bright contaminant star down to this magnitude limit in the diffraction-limited image.

\section{Determination of Stellar Activity Indicators and Radial Velocities}

\subsection*{S$_{MW}$ Index}

\begin{figure}
\centering
        \includegraphics[width=9.0cm]{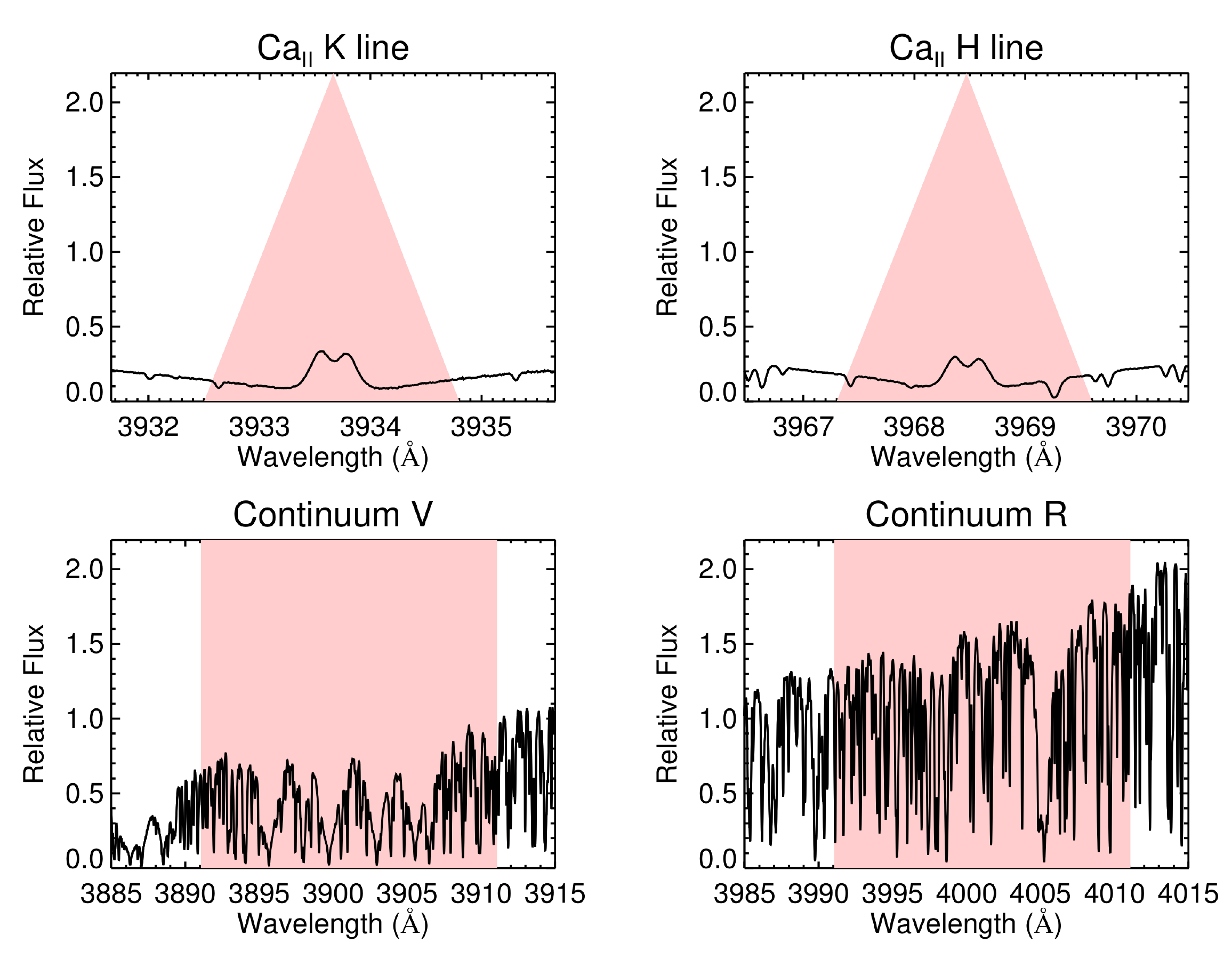}
        \caption{Ca II H\&K filter of the spectrum of the star HD 176986 with the same shape as the Mount Wilson Ca II H\&K passband.}
        \label{Sindex}
\end{figure}

We calculated the Mount Wilson $S$ index (S$_{MW}$) and the $\log_{10}(R'_{HK})$ using the original \citet{Noyes1984} procedure, following \citet{Lovis2011} and \citet{Masca2015}. We defined two triangular passbands with  a full-width at half-maximum (FWHM) of 1.09~{\AA} centred at 3968.470~{\AA} and 3933.664~{\AA} for the Ca II H\&K line cores, and for the continuum, we used two 20~{\AA} wide bands centred at 3901.070~{\AA} (V) and 4001.070~{\AA}(R), as shown in Fig.~\ref{Sindex}.

Then the S$_{MW}$ index is defined as 

\begin{equation}
   S_{MW}=\alpha {{\tilde{N}_{H}+\tilde{N}_{K}}\over{\tilde{N}_{R}+\tilde{N}_{V}}} + \beta,
\end{equation}
\noindent where $\tilde{N}_{H},\tilde{N}_{K},\tilde{N}_{R}$, and $\tilde{N}_{V}$ are the mean fluxes per wavelength interval in each passband,  while $\alpha$ and $\beta$ are calibration constants fixed as $\alpha = 1.111$ and $\beta = 0.0153$ . The S index serves as a measurement of the Ca II H\&K core flux normalized to the neighbour continuum. As a normalized index to compare it to other stars, we computed the $\log_{10}(R'_\textrm{HK})$ following \citet{Masca2015}.

\subsection*{H$_{\alpha}$ Index}

\begin{figure}
        \includegraphics[width=9.0cm]{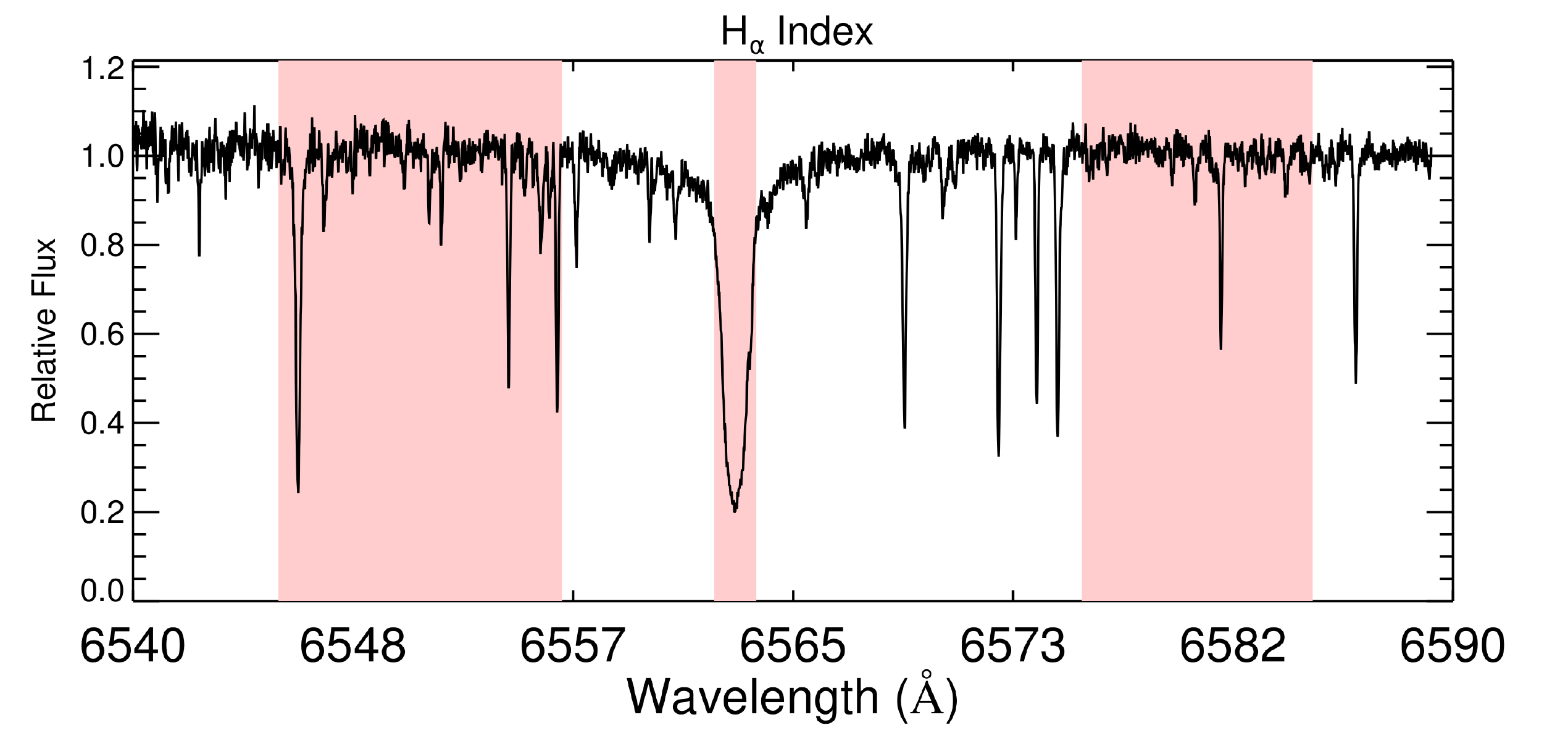}
        \caption{Spectrum of the K-type star HD 176986 showing the H$\alpha$ filter passband and continuum bands.}
        \label{halpha}
\end{figure}

We also used the H$\alpha$ index, with a simpler passband following \cite{GomesdaSilva2011}. It consists of a rectangular bandpass with a width of 1.6~{\AA} and centred at 6562.808~{\AA} (core), and two continuum bands of 10.75~{\AA} and  8.75~{\AA} in width, centred at 6550.87~{\AA} (L) and 6580.31~{\AA} (R), respectively, as shown in Figure~\ref{halpha}. 

Thus, the H$\alpha$ index is defined as

\begin{equation}
   H\alpha_{\rm Index}={{\tilde{H}\alpha_{\rm core} }\over{\tilde{H}\alpha_{L} +\tilde{H}\alpha_{R}}},\end{equation}
 
\noindent where $\tilde{H}\alpha_{\rm core},\tilde{H}\alpha_{L}$ and $\tilde{H}\alpha_{R}$ are the mean fluxes per wavelength interval in each passband.
 
\subsection{Radial Velocities}
 
The RV measurements in the HARPS and HARPS-N standard pipelines are determined by a Gaussian fit of the cross-correlation function (CCF) of the spectrum with a binary stellar template \citep{Baranne1996, PepeMayor2000}.

\section{Analysis}

Using all the available datasets, we searched for periodic signals that are compatible with  planetary companions, stellar rotation, and long-term magnetic cycles. We computed the power spectrum using a least-squares periodogram \citep{Cumming2004} by fitting a sinusoidal model over the whole range of test frequencies using the MPFIT routine \citep{Markwardt2009}, with the power spectral density (PSD) defined following \citet{Zechmeister2009}. The significance of the periodogram peaks was evaluated by calculating the spectral density thresholds for the desired false-alarm probability (FAP) levels using bootstrap randomization \citep{Endl2001} of the data. When we detected a significant signal, we re-computed the periodogram using the superposition of two signals, one with a period fixed at the detected peak of the periodogram, and a new sinusoidal model again evaluated over the whole range of test frequencies. The amplitudes and phases of both signals were set free to ensure the best fit at each test period. We recursively repeated the procedure by adding new signals until no more significant signals were found in the data \citep{Arriagada2013}, with the FAP levels re-calculated at each step. 

During this stage, all the signals detected in the recursive periodograms (in all datasets: RV, FWHM, bisector span, S$_{MW}$ index, H$\alpha$ index, ASAS photometry) were modelled using a double-harmonic sinusoidal model \citep{BerdyuginaJarvinen2005} with the IDL MPFIT routine \citep{Markwardt2009}. This model easily allows accounting for the asymmetry that is expected for activity signals and also maintains a sinusoidal shape for sinusoidal signals. When the origin of the signals was established, RV planetary signals were modelled using Keplerian models, whose  parameters were calculated by performing a Bayesian analysis, using the Markov chain Monte Carlo (MCMC) code built in The Data \& Analysis Center for Exoplanets (DACE)\footnote{https://dace.unige.ch/}.

\subsection{Stitching HARPS and HARPS-N data together}

HARPS and HARPS-N are almost identical spectrographs, but this does not mean that their data can be combined as if they were the same instrument. The HARPS set of fibres was upgraded during May 2015, which also means that HARPS post-upgrade data need to be analysed as data from a different instrument. The offets between the data coming from both instruments are not large, but still need to be estimated. The offsets between the different datasets were estimated as free parameters in the fits, using the HARPS pre-upgrade dataset as reference.  Table ~\ref{tab:offsets} shows the offsets applied to the HARPS post-upgrade and HARPS-N data. The offsets estimated between the different datasets are consistent with previous measurements for similar stars \citep{Desidera2013, LoCurto2015}.

\begin{table}
\begin{center}
\caption{Offsets applied \label{tab:offsets}}
\begin{tabular}[center]{l c c}
\hline
Serie & Offset HARPS post-upgrade & Offset HARPS-N \\ \hline
RV & -- 11.8 m s$^{-1}$ & -- 0.5 m s$^{-1}$\\
FWHM & --31.9 m s$^{-1}$  & 159.5 m s$^{-1}$ \\
Bis. Span & -- 16.8  m s$^{-1}$ & -- 2.5  m s$^{-1}$ \\
S$_{MW}$ Index & 0.0465 &  0.0090 \\
H${\alpha}$ Index & 0.00049 & -- 0.00109  \\
\hline
\end{tabular}
\end{center}
\end{table}

\subsection{Analysis of the radial velocity variability} \label{sect_rv}

\begin{figure}
        \includegraphics[width=9.0cm]{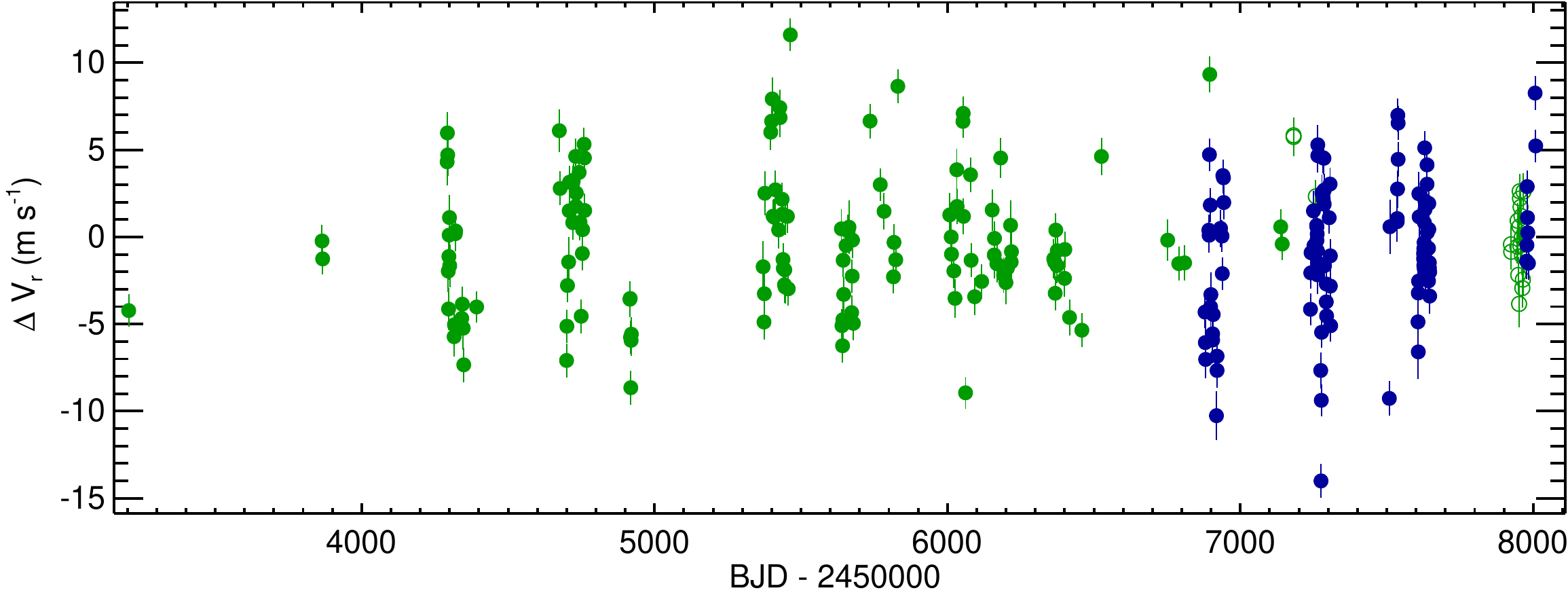}
        \caption{Radial velocity time-series for the star HD 176986 obtained by combining HARPS and HARPS-N measurements. Green filled dots show HARPS pre-upgrade data, green empty dots show HARPS post-upgrade data, and blue filled dots show HARPS-N data.}
        \label{RV_timeseries}
\end{figure}

Our RV dataset consists of 259 RV measurements distributed along 13.2 years with a good coverage of almost the whole baseline. We measured a mean RV of the star of 37.74 km s$^{-1}$, an RMS of the measurements of 3.8 m s$^{-1}$ , and a typical uncertainty of 1.0 m s$^{-1}$ per individual measurement, including photon noise, pointing errors, and drift measurement errors. Figure ~\ref{RV_timeseries} shows the available RV dataset. 

\begin{figure*}
        \includegraphics[width=18cm]{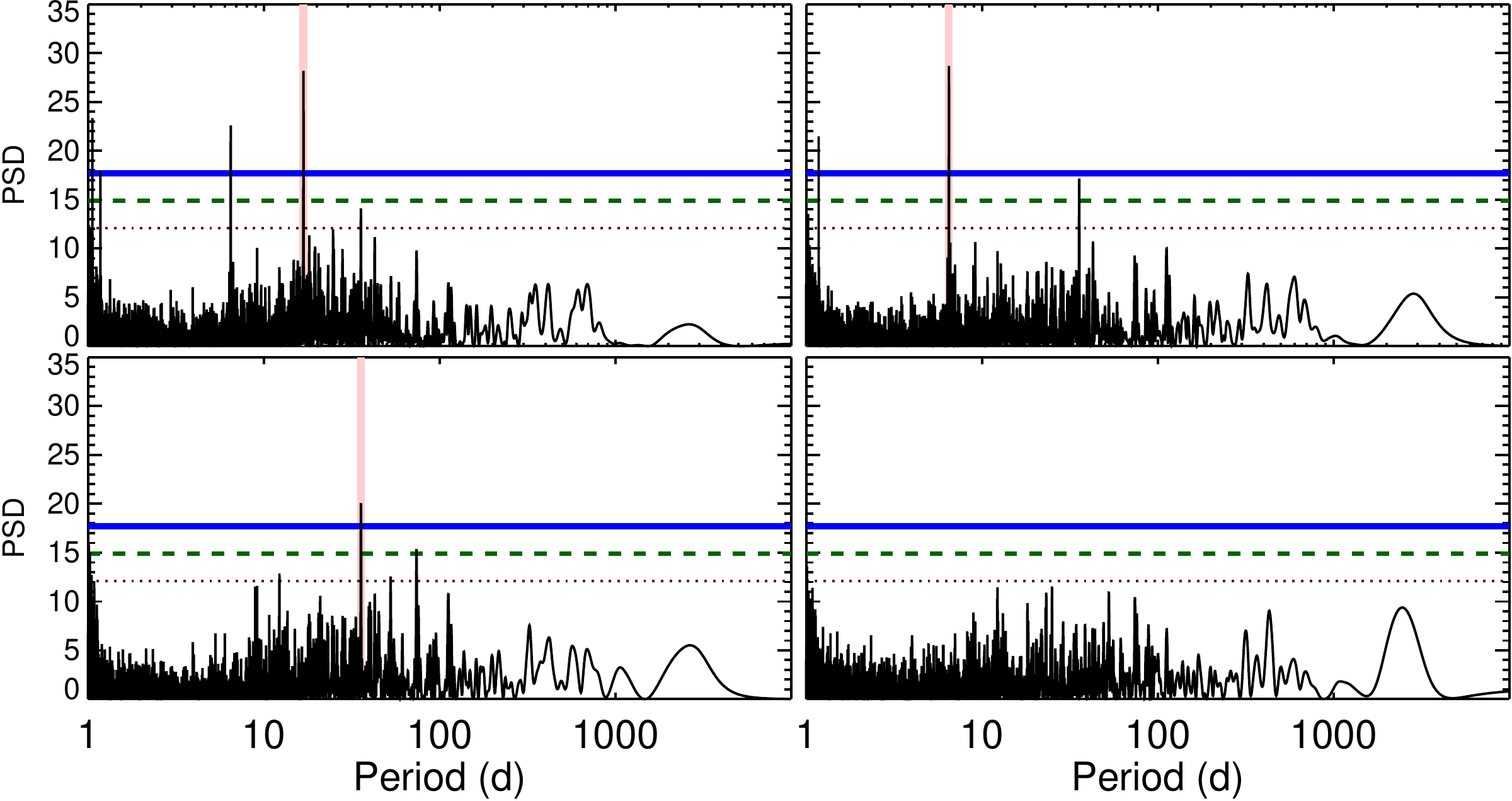}
        \caption{Periodograms of the RV time-series. The shaded region highlights the detected periodicities of 16.83 d (top left), 6.49 d (top right), and 35.74 d (bottom left). The red dotted, green dashed, and blue solid horizontal lines show the 10\%, 1\%, and 0.1\% FAP levels. }
        \label{rv_gls}
\end{figure*}

\begin{figure*}
\begin{minipage}{0.33\textwidth}
        \centering
        \includegraphics[width=6.cm]{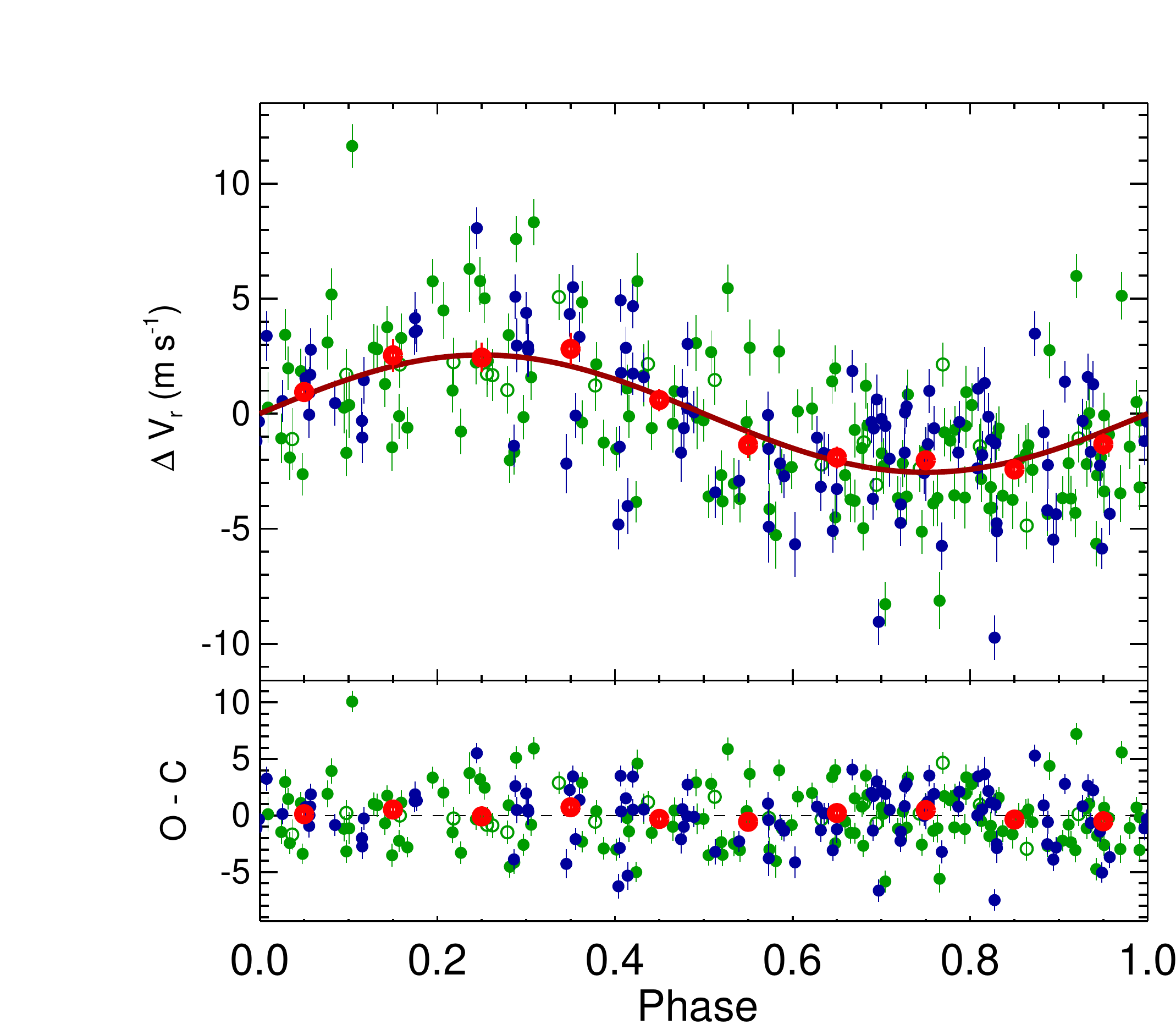}
\end{minipage}%
\begin{minipage}{0.33\textwidth}
        \centering
        \includegraphics[width=6.cm]{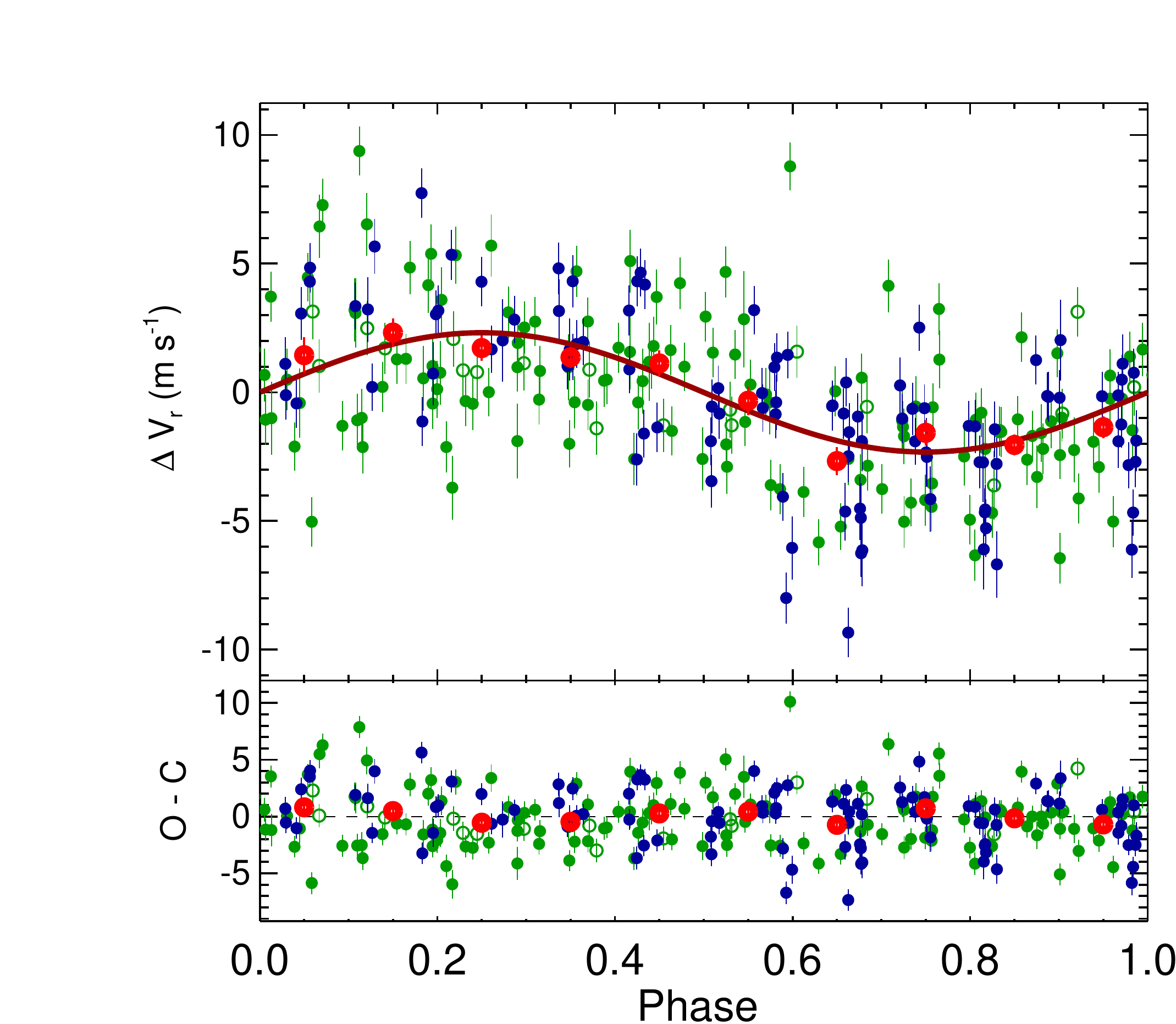}
\end{minipage}%
\begin{minipage}{0.33\textwidth}
        \centering
        \includegraphics[width=6.cm]{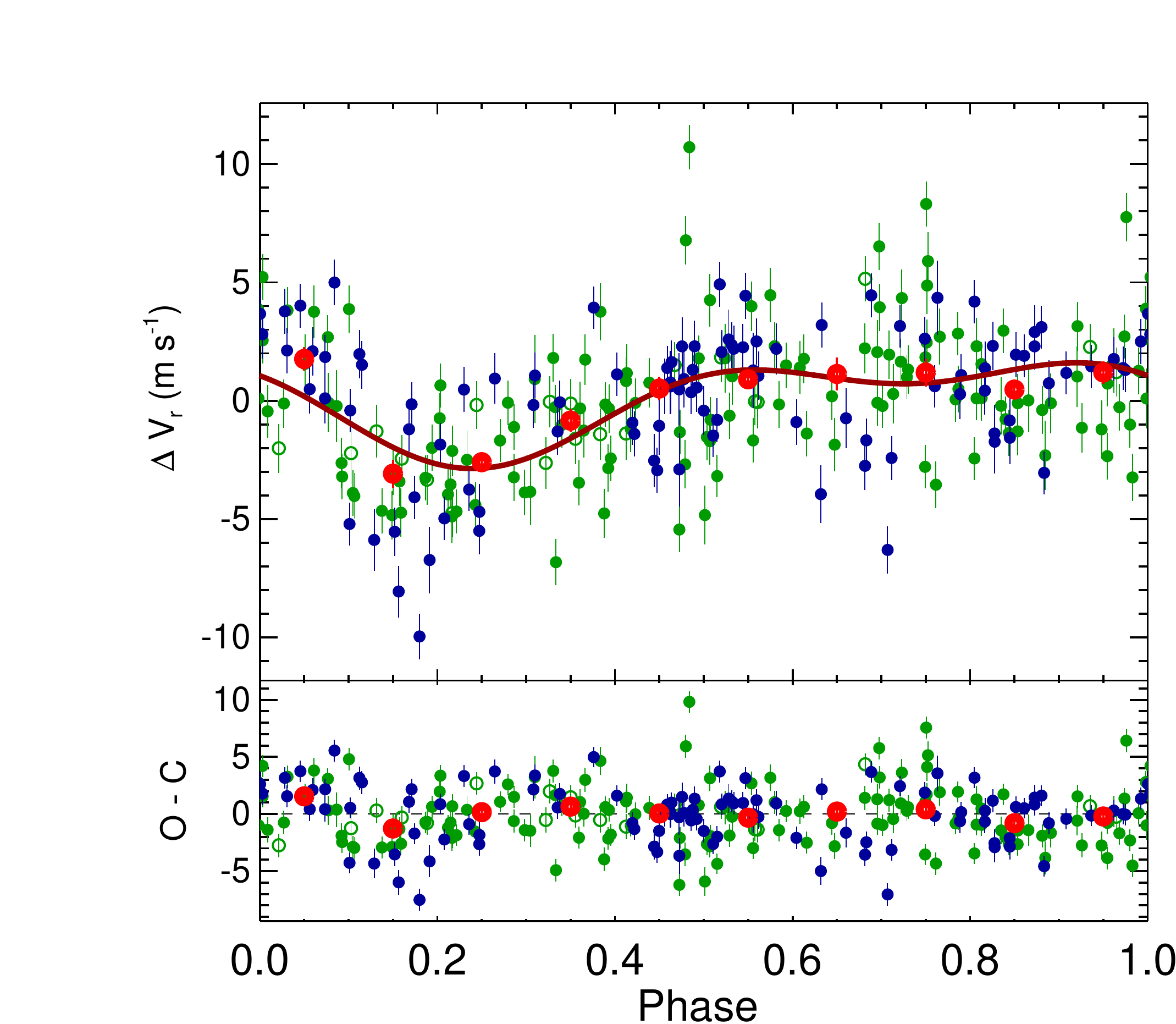}
\end{minipage}%

\caption{Phase-folded curve of the RV time-series using the 16.82 d signal (left panel), the 6.49 d signal (centre panel), and the 35.74 d signal (right panel). Green filled dots show HARPS pre-upgrade data, green empty dots show HARPS post-upgrade data, and blue filled dots show HARPS-N data. Red empty dots are the same points binned in phase with a bin size of 0.1. The error bar of a given bin is estimated using the weighted standard deviation of the binned measurements divided by the square root of the number of measurements included in this bin. The dark
red line shows the best fit to the data using double sinusoidals at P and P/2 for all the signals. In each case, the contribution of the other two detected signals has been subtracted. }
\label{rv_phase}
\end{figure*}

By calculating the recursive periodograms and identifying the most significant signals at every iteration, we detected three dominant signals in the RV data.  The three identified peaks in the periodograms correspond to a 16.82 $\pm$ 0.01 d signal with a semi-amplitude of 2.59 $\pm$ 0.36 m s$^{-1}$, a 6.49 $\pm$ 0.01 d signal with a semi-amplitude of 2.37  $\pm$ 0.35 m s$^{-1}$  , and a 35.74 $\pm$ 0.02 d signal with a semi-amplitude of 1.20 $\pm$ 0.35 m s$^{-1}$. The three signals are detected with FAPs several orders of magnitude smaller than 0.1\%. The signals detected at 16.82 and 6.49 d are consistent with sinusoidals, while the signal at 35.74 d is highly non-sinusoidal. The amplitude ratios between the sinusoidal signals at P/2 and the sinusoidal signals at P are smaller than the 0.1 for the signals at 16.82 and 6.49 d, while for the case of the 35.74 d, the ratio is 0.6. The 35.7 d signal is consistent with previous measurements of the rotation period of the star \citep{Masca2015, Masca2017b}, while the signal at 16.82 days is one day apart from the first harmonic of that measurement of stellar rotation. Figure~\ref{rv_gls} shows the recursive periodograms of the RV time-series; the three signals are highlighted. Figure ~\ref{rv_phase} shows the phase-folded fits to the data after removing the contributions of all the other signals. 

We did not find more significant signals in the RV data. The RMS of the residuals after fitting the three signals is approximately 2.5 m s$^{-1}$. Given the small uncertainty of the individual measurements, there is still room for more signals that we do not detect. These hypothetical signals could be of very low amplitude, especially if the periods are long. Another possibility is that the scatter is caused by undetected quasi-periodic signals or activity signals that may not be stable in time.

\subsection{Analysis of the FWHM variability}

\begin{figure}
        \includegraphics[width=9.0cm]{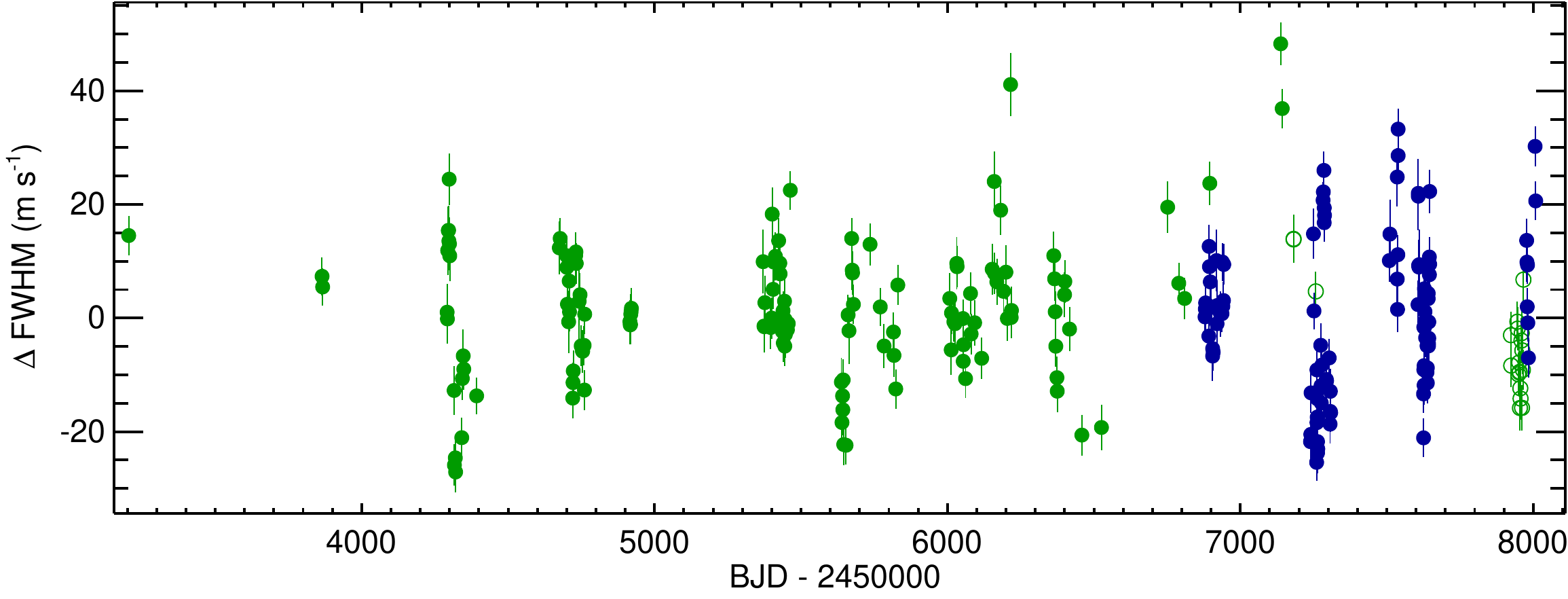}
        \caption{Time-series for FWHM variations of the star HD 176986, obtained by combining HARPS and HARPS-N data. See Fig.~\ref{RV_timeseries} for reference.}
        \label{FWHM_timeseries}
\end{figure}

The FWHM measurements show a mean measurement of 6.13 km s$^{-1}$ with an RMS of the measurements of 12.5 m s$^{-1}$ and a typical uncertainty of the measurements of 3.7 m s$^{-1}$ (see Fig.~\ref{FWHM_timeseries}).

\begin{figure}
        \includegraphics[width=9.0cm]{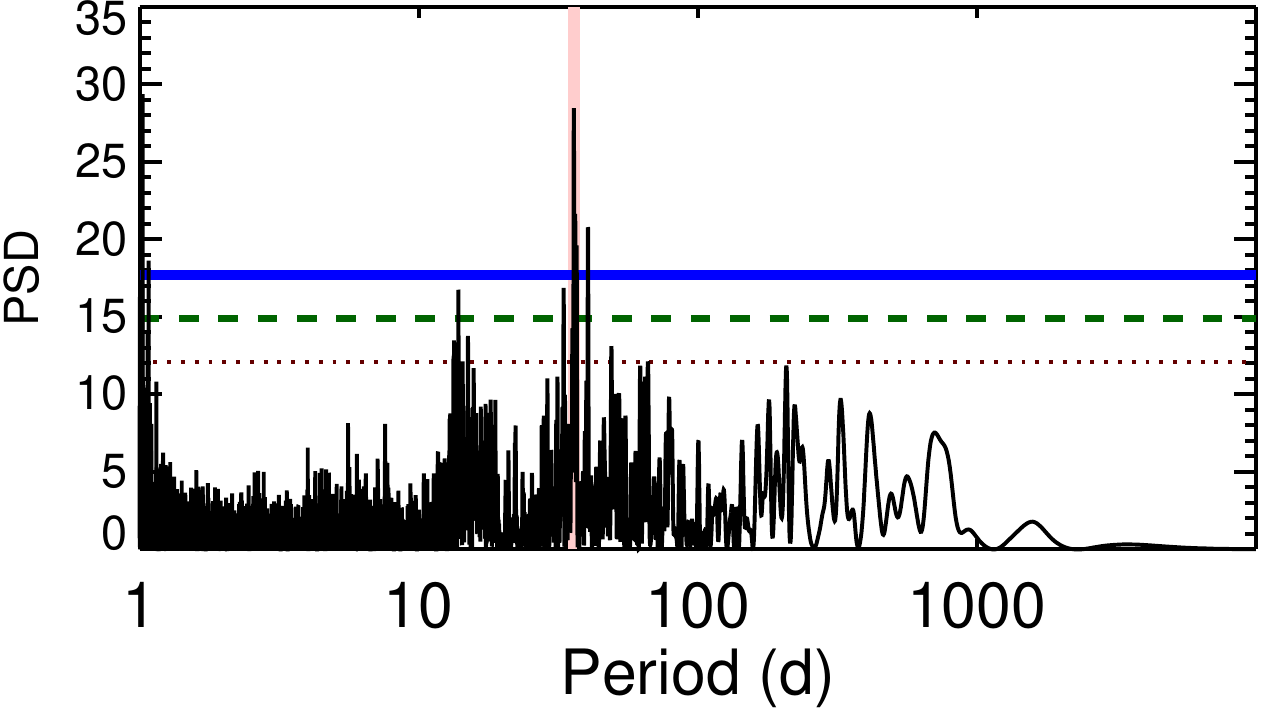}
        \caption{GLS periodogram of the variations of the FWHM time-series. The shaded region shows the detected periodicity of 36 d.}
        \label{fwhm_gls}
\end{figure}

\begin{figure}
\includegraphics[width=9.cm]{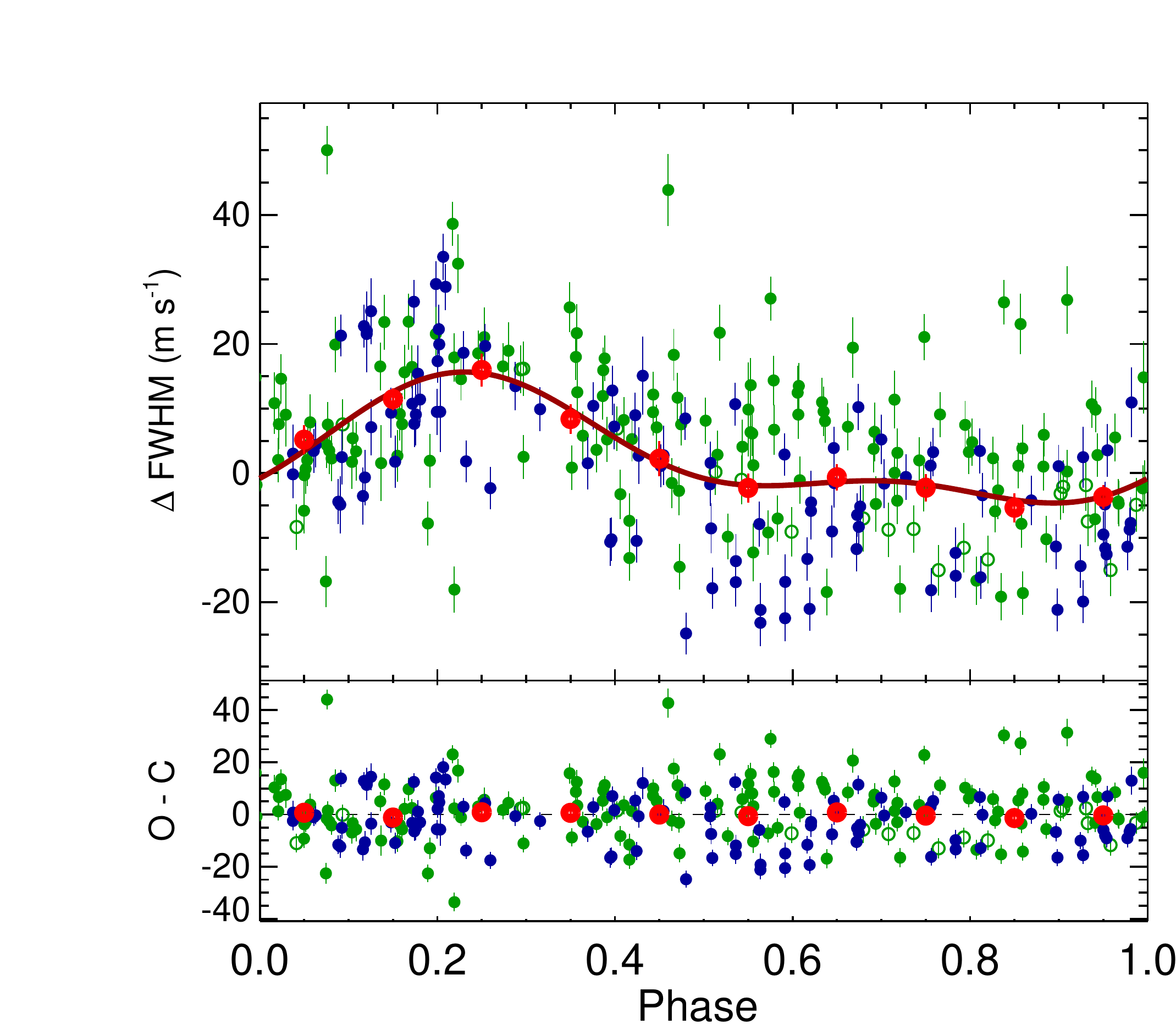}
\caption{Phase-folded curve of the FWHM time-series using the 36-day signal. See Fig.~\ref{rv_phase} for reference.}
\label{fwhm_phase}
\end{figure}

The periodogram FWHM time-series show a single significant signal at the period of 36 days (see Fig. ~\ref{fwhm_gls}). Using a double-sinusoidal fit, we measure a period of 36.00 $\pm$ 0.03 days and a semi-amplitude of 6.6 $\pm$ 1.4 m s$^{-1}$ (see Fig.~\ref{fwhm_phase}). The phase-folded fit is highly non-sinusoidal, very similar to the 35.7-day signal found in the RV dataset. The signal is consistent with the expected rotation period of a main-sequence K-dwarf of this activity level and with previous measurements that used a subset of the same data \citep{Masca2015}. No more significant signals are found after subtracting the 36-day signal.

\subsection{Analysis of the bisector span  variability}

\begin{figure}
        \includegraphics[width=9.0cm]{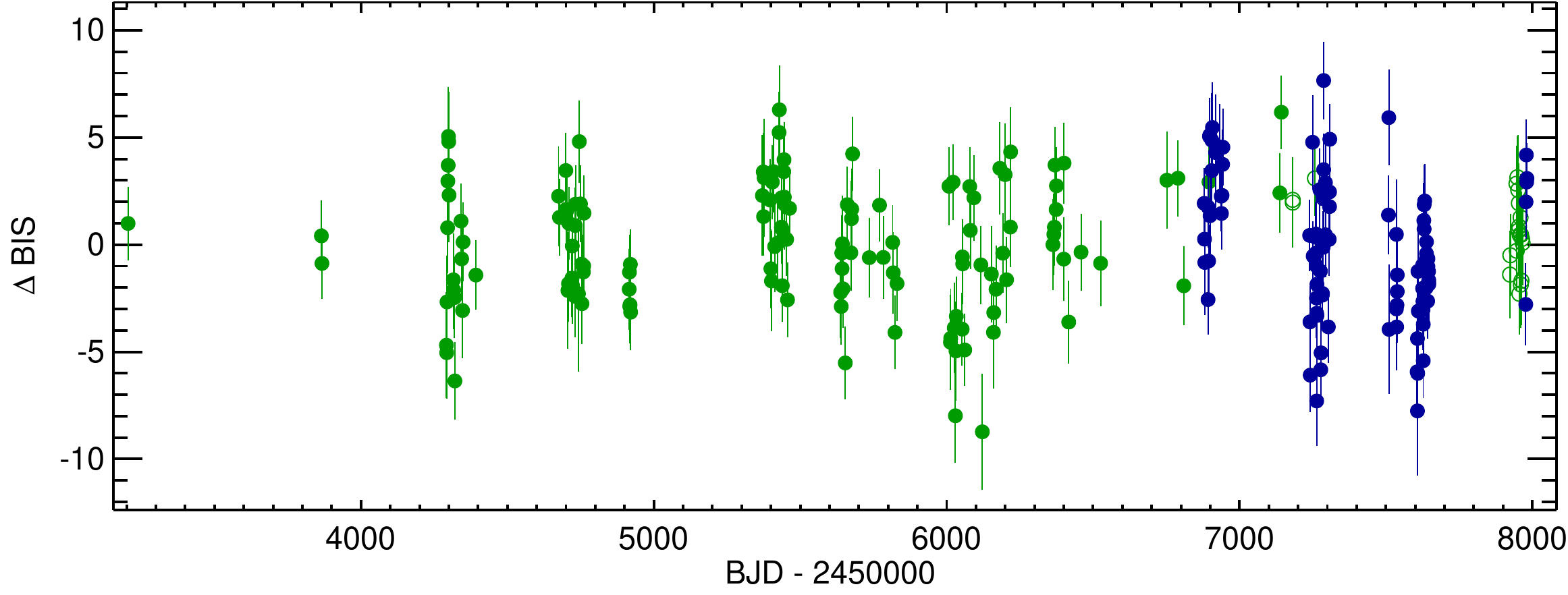}
        \caption{Time-series for variations of the bisector span of the CCF of the star HD 176986, obtained by combining HARPS and HARPS-N data. See Fig.~\ref{RV_timeseries} for reference.}
        \label{bis_timeseries}
\end{figure}

\begin{figure*}
        \includegraphics[width=18.0cm]{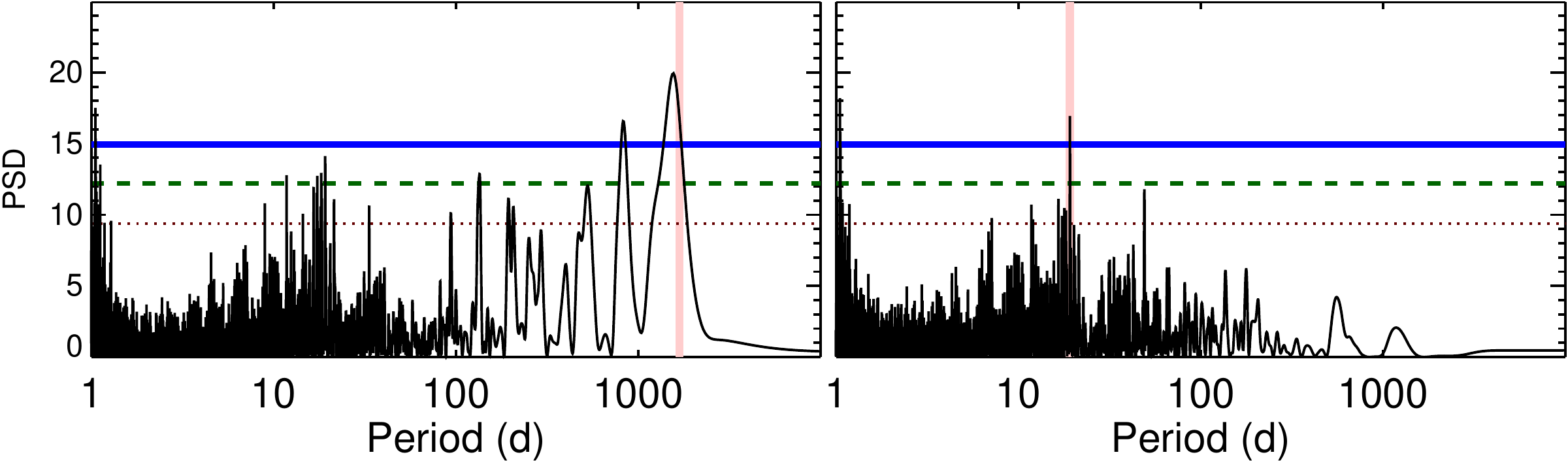}
        \caption{GLS periodogram of the variations of the bisector span time-series. The shaded region shows the detected periodicities at 1800 days (left panel) and 19 days (right panel).}
        \label{bis_gls}
\end{figure*}

\begin{figure*}
\begin{minipage}{0.5\textwidth}
        \centering
        \includegraphics[width=9.cm]{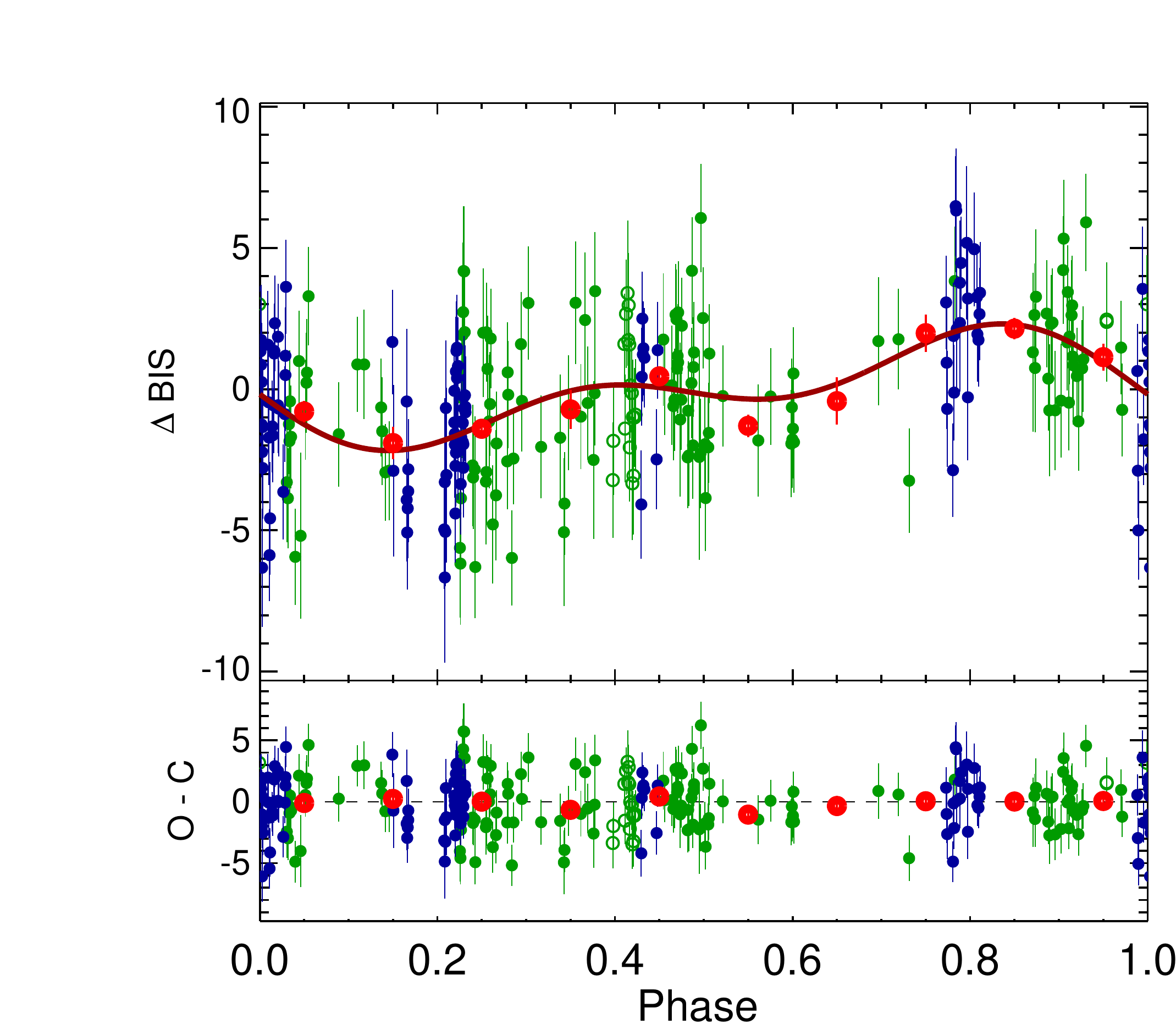}
\end{minipage}%
\begin{minipage}{0.5\textwidth}
        \centering
        \includegraphics[width=9.cm]{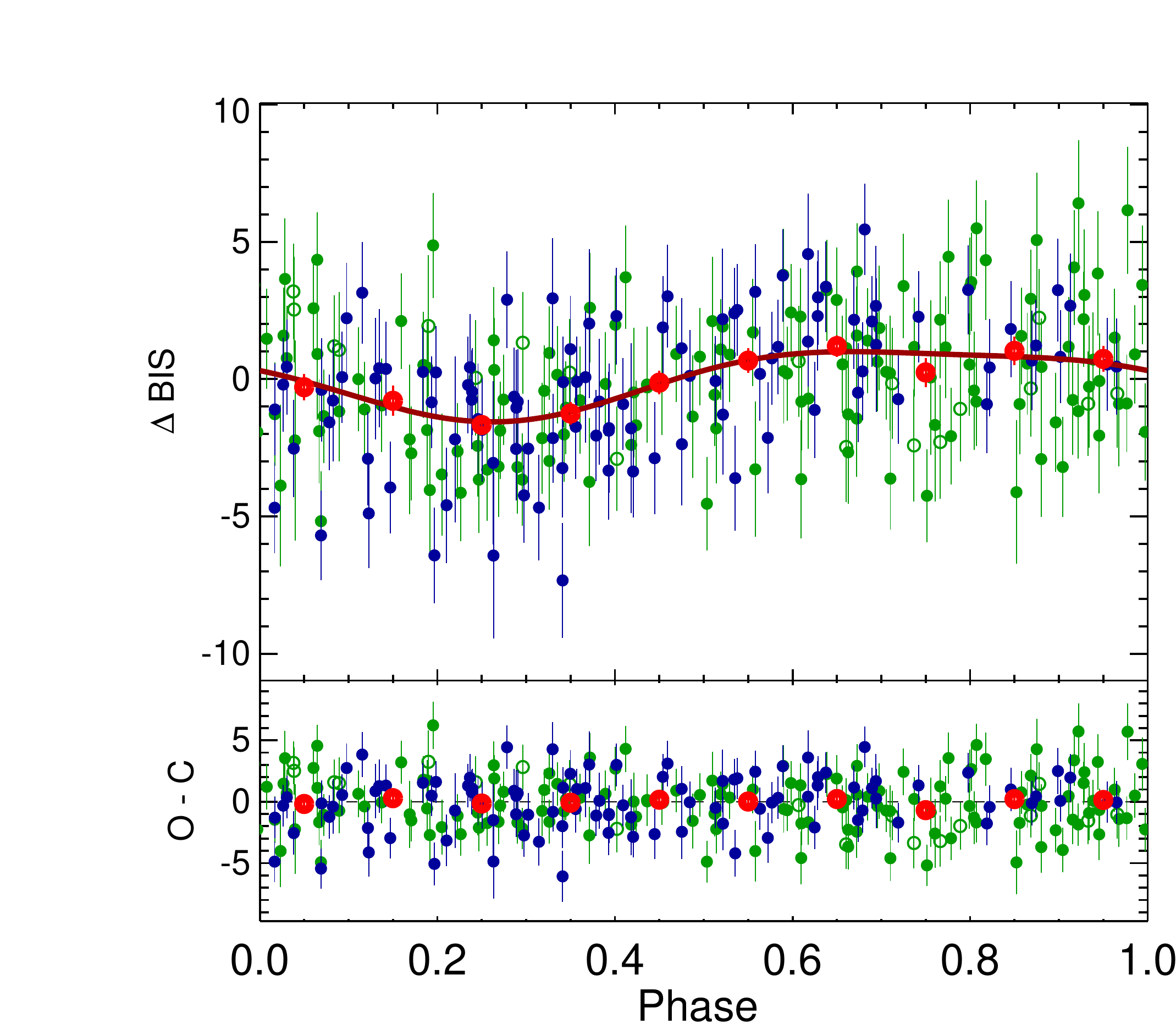}
\end{minipage}%

\caption{Phase-folded curve of the bisector span time-series using  the 1670-day signal (left panel) and the 19.2-day signal (right panel). See Fig.~\ref{rv_phase} for reference.}
\label{bis_phase}
\end{figure*}

For the bisector span, we have a mean measurement of 14.82 m s$^{-1}$ with an RMS of the measurements of 2.8 m s$^{-1}$ and a typical uncertainty of 1.8 m s$^{-1}$, using \citet{Boisse2009} as a reference (see Fig.~\ref{bis_timeseries}). The periodogram shows two significant signals at 1670 $\pm$ 29 days and 19.20 $\pm$ 0.01 days (Fig.~\ref{bis_gls}) with semi-amplitudes of 1.41 $\pm$ 0.39 m s$^{-1}$ and 1.22 $\pm$ 0.30 m s$^{-1}$. Figure ~\ref{bis_phase} shows the best fits to both signals using a double-sinusoidal model. The signal is approximately one day apart from the first harmonic of the rotation period of the star. We did not find any more significant signals in the residuals of the bisector span data.  The scarcity of points at phase 0.6-0.7 in the long-period signal leaves some doubts about its true nature. It is possible that it is mostly a window effect.

\subsection{Analysis of the Ca II H$\&$K variability}

\begin{figure}
        \includegraphics[width=9.0cm]{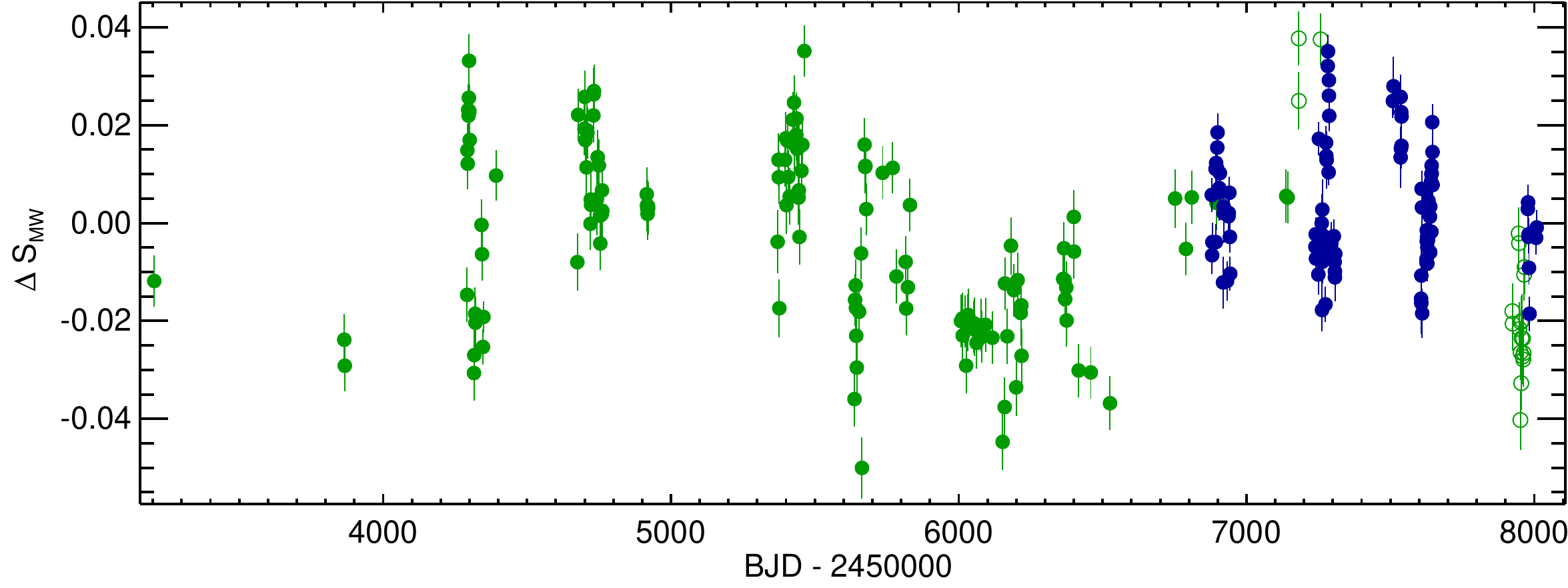}
        \caption{S$_{MW}$ index time series for the star HD 176986,
obtained by combining HARPS and HARPS-N data. See Fig.~\ref{RV_timeseries} for reference.}
        \label{smw_timeseries}
\end{figure}

\begin{figure*}
        \includegraphics[width=18.0cm]{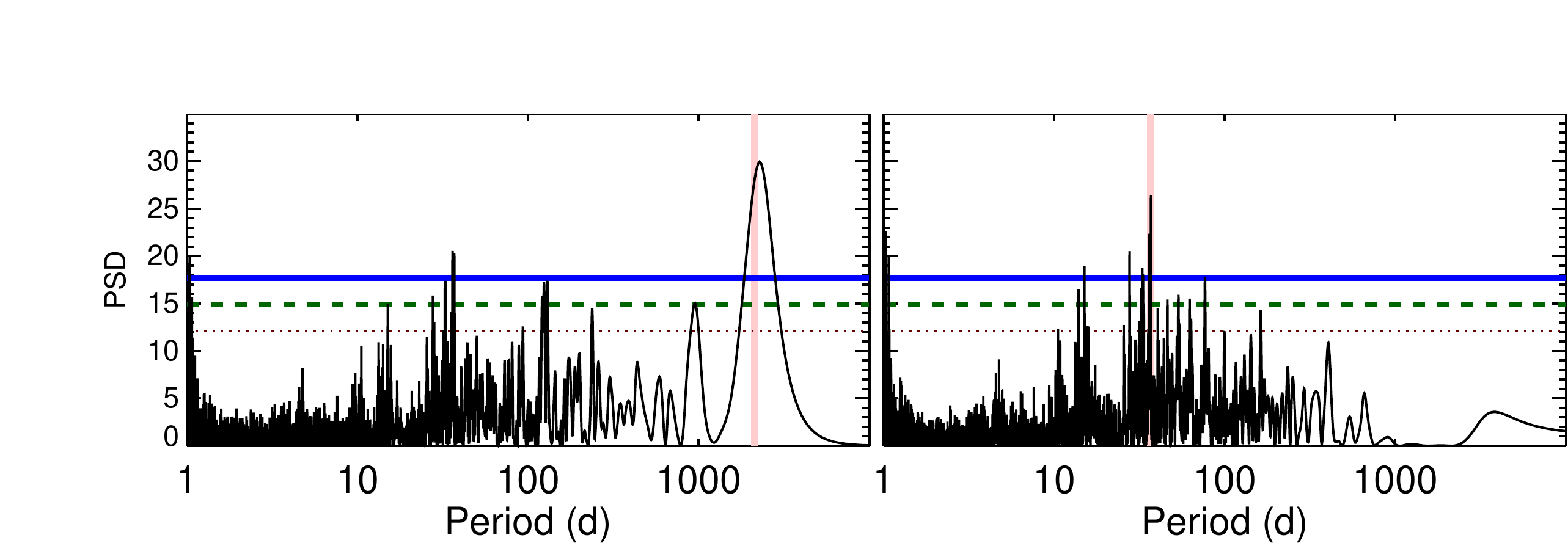}
        \caption{Periodogram for the S$_{MW}$ index time series for the star HD 176986. The shaded region highlights the detected periodicities of 6 yr (left) and 36 days (right).}
        \label{smw_gls}
\end{figure*}

We measured a mean S$_{MW}$ value of 0.2916 with an RMS of the measurements of 0.0165 and a typical uncertainty of 0.0053 per exposure (see Fig.~\ref{smw_timeseries}). A long-period signal is clearly visible in the time-series, with a length consistent with a solar-like magnetic cycle. The periodogram shows two dominant signals in the data (Fig.~\ref{smw_gls}) at 2100 days and 36 days. Some other apparently significant signals are visible, but all the long-period signals disappeared when we fitted the long-term signal, and the short-period signal disappeared when we fitted the 36 d signal. 

\begin{figure*}
\begin{minipage}{0.5\textwidth}
        \centering
        \includegraphics[width=9.cm]{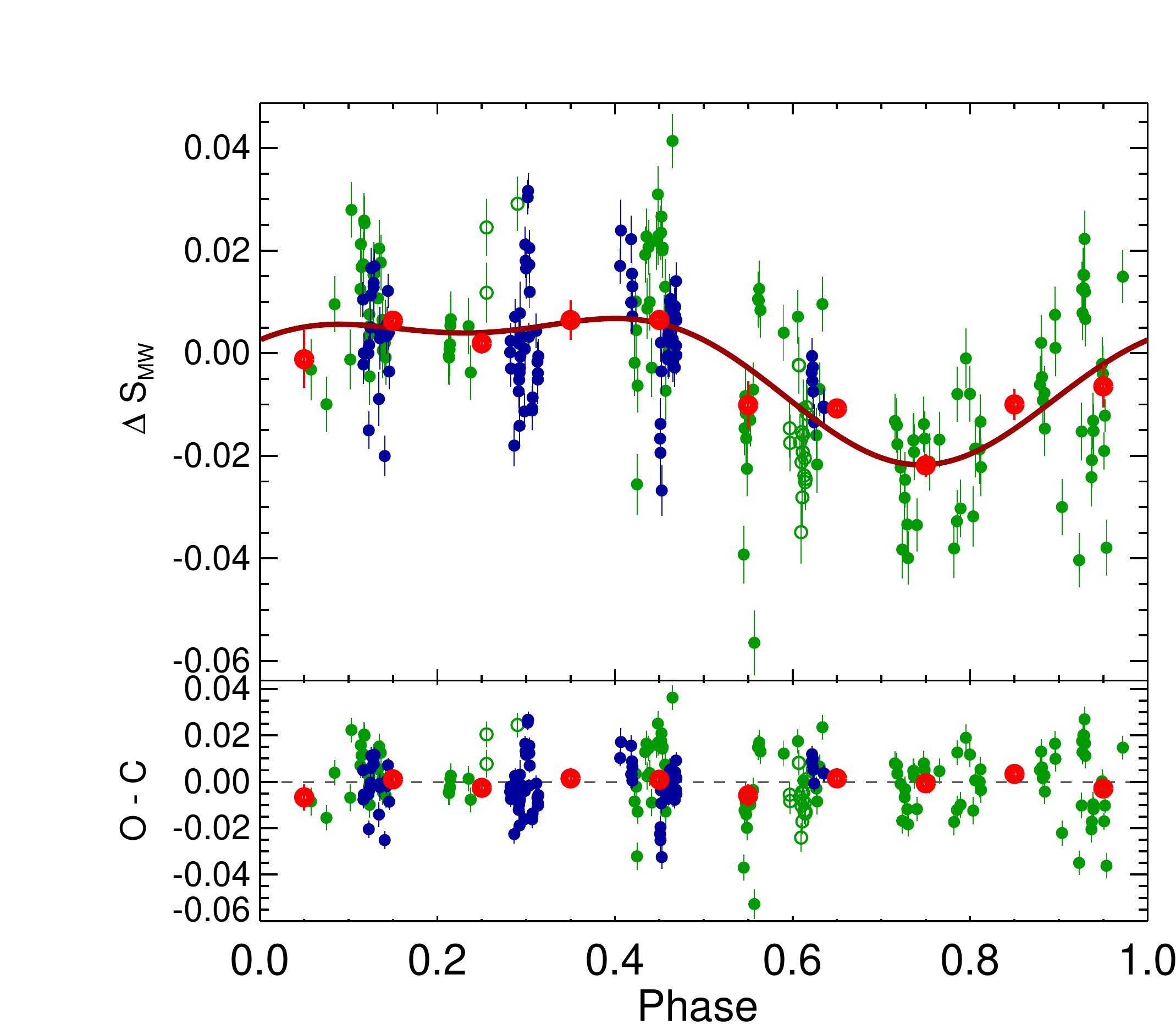}
\end{minipage}%
\begin{minipage}{0.5\textwidth}
        \centering
        \includegraphics[width=9.cm]{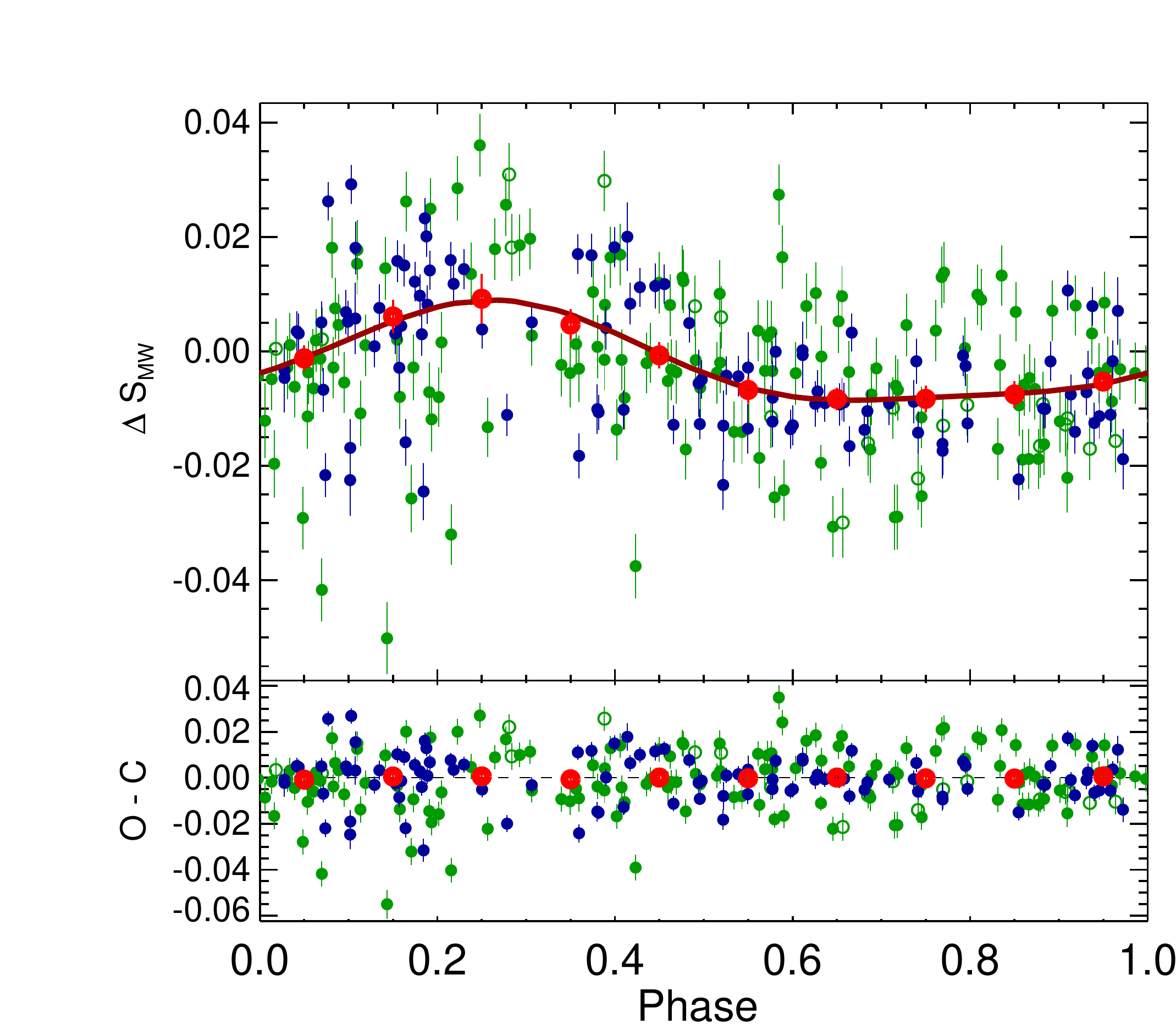}
\end{minipage}%
\caption{Phase-folded curve of the S$_{MW}$ index time series using the 6.0 yr signal (left panel) and the 36.02 d signal (right panel). See Fig.~\ref{rv_phase} for reference.}
\label{smw_phase}
\end{figure*}

The combined fit of both signals, using double sinusoidals at periods P and P/2, returns a period of 5.67 $\pm$ 0.11 yr with a semi-amplitude of 0.0153 $\pm$ 0.0018 for the long-period signal and  a period of 36.02 $\pm$ 0.02 days with a semi-amplitude of 0.0068 $\pm$ 0.0016 for the short-period signal. These measurements are consistent with the solar-like cycle and the rotation period of the star, as is also supported by the FWHM and BIS analysis. Figure~\ref{smw_phase} shows the individual fits of the two signals. They are clearly not sinusoidal, as it is often the case for activity-induced signals. 

\subsection{Analysis of the H$_{\alpha}$ variability}

\begin{figure}
        \includegraphics[width=9.0cm]{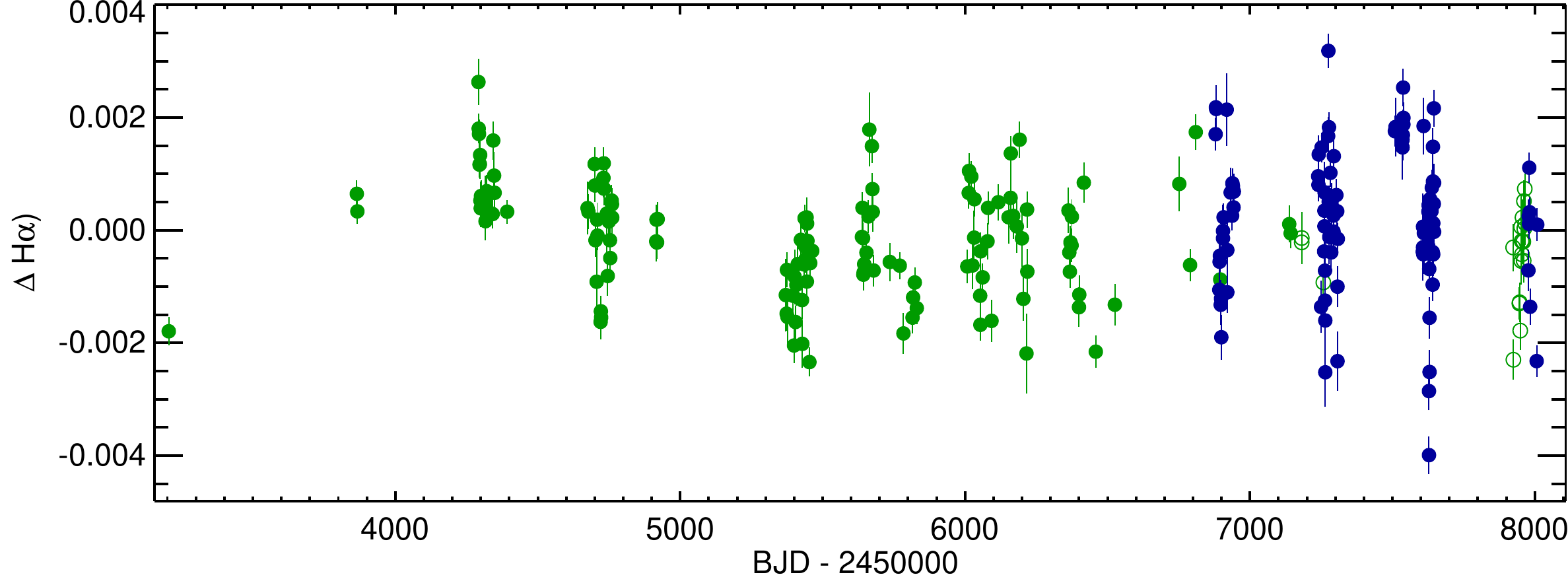}
        \caption{H$_{\alpha}$ index time-series for the star HD 176986, obtained by combining HARPS and HARPS-N data. See Fig.~\ref{RV_timeseries} for reference.}
        \label{ha_timeseries}
\end{figure}

\begin{figure*}
        \includegraphics[width=18.0cm]{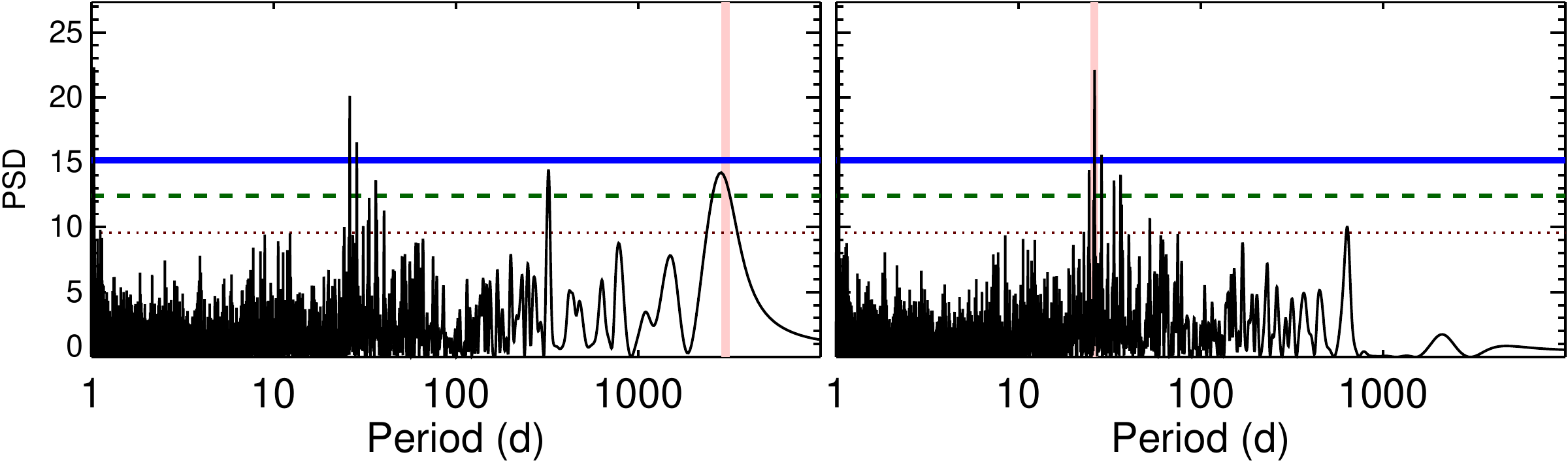}
        \caption{Periodogram for the H$\alpha$ index time-series for the star HD 176986. The shaded region highlights the detected periodicities of 2400 days (left panel) and 26 days (right panel).}
        \label{ha_gls}
\end{figure*}

For the H$\alpha$ index, we measured a mean value of 0.228356 with an RMS of the measurements of 0.00108 with a typical uncertainty of the measurements of 0.00032 (see Fig. ~\ref{ha_timeseries}). The periodogram shows two dominant signals at about 2400 and 26 days (Fig.~\ref{ha_gls}). The 26-day signal is more significant than the signal at 2400 days, but inadequate sampling of long-period signals often creates ghost signals in the periodograms when not properly modelled. These ghost signals can sometimes appear
to be more significant than the real signals. These ghost signals usually vanish when the long-period signal is fitted and subtracted. We opted to subtract the 2400-day signal before the 26-day signal in order to remove the possible ghost signals that are associated with the 2400-day signal.  

\begin{figure*}
\begin{minipage}{0.5\textwidth}
        \centering
        \includegraphics[width=9.cm]{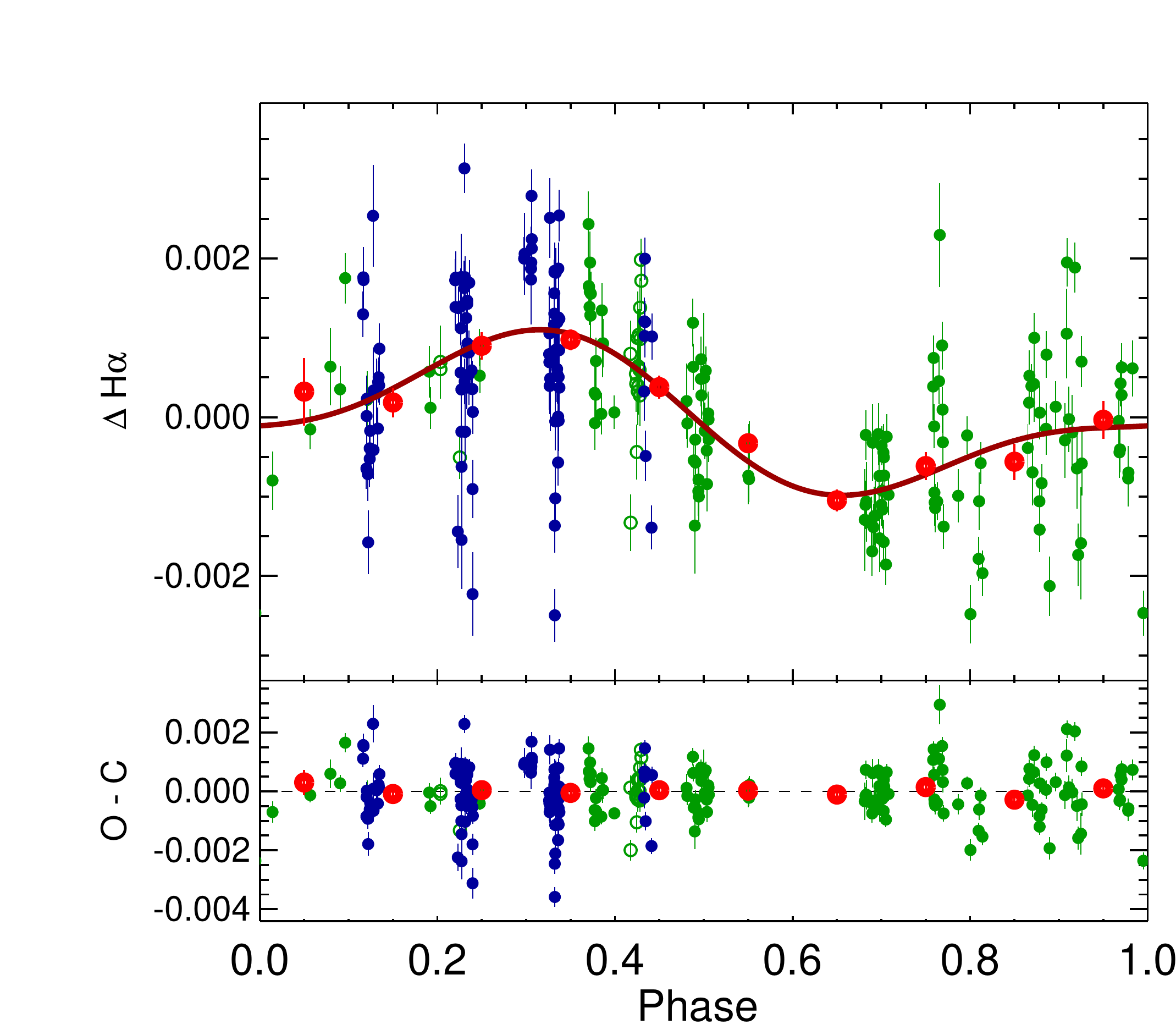}
\end{minipage}%
\begin{minipage}{0.5\textwidth}
        \centering
        \includegraphics[width=9.cm]{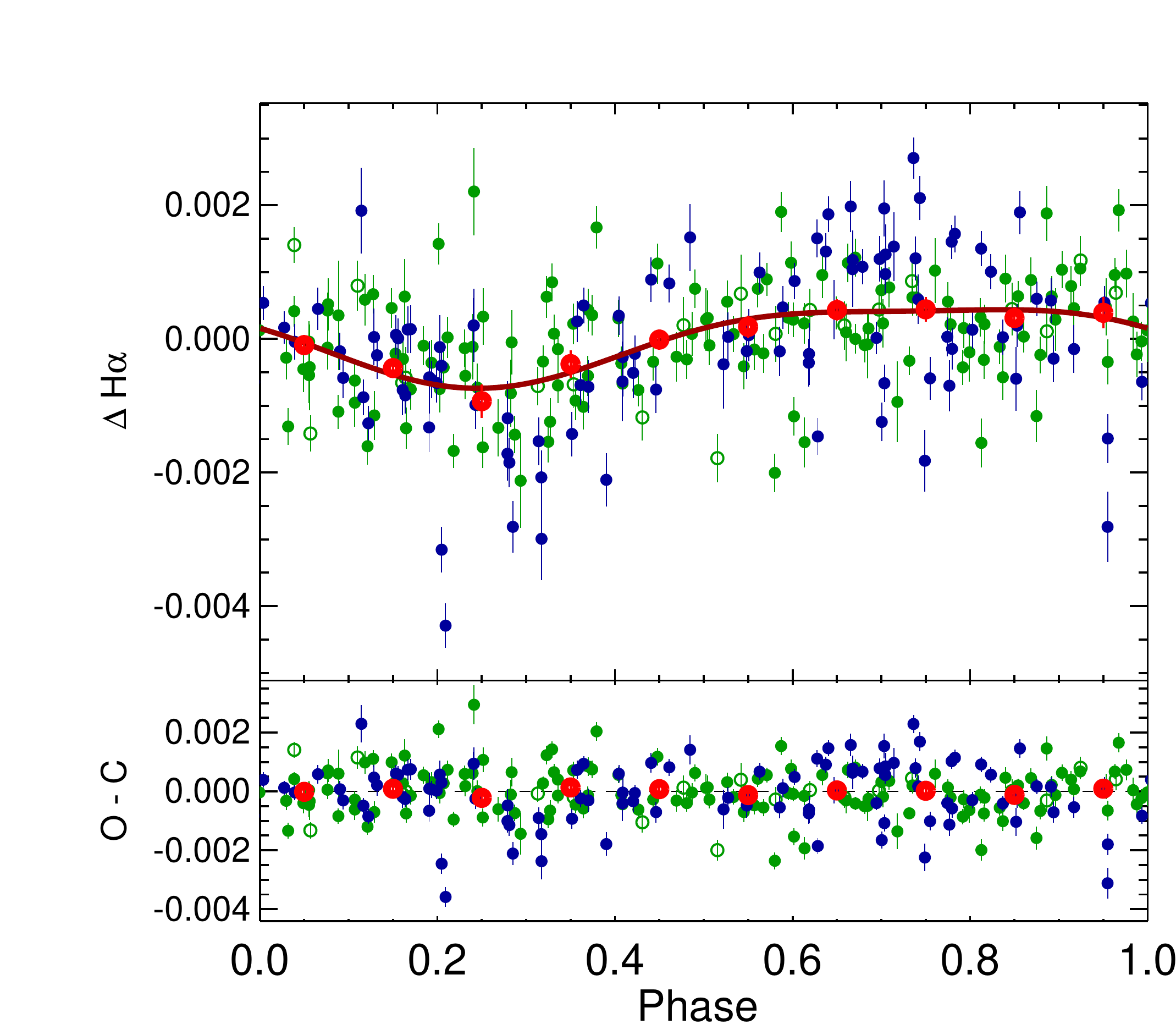}
\end{minipage}%
        
\caption{Phase-folded curve of the H$\alpha$ index time-series using the 2400 d signal (left panel) and the 26 d signal (right panel). See Fig.~\ref{rv_phase} for reference.}
\label{ha_phase}
\end{figure*}

The fit of the signals using double sinusoidals at periods P and P/2 returns periods of 8.21 $\pm$ 0.24 years and 26.20 $\pm$ 0.01 days with a semi-amplitude of 0.00022 $\pm$ 0.00013 and 0.00067 $\pm$ 0.00011. Figure~\ref{ha_phase} shows the phase-folded fits to the data. 

\subsection{Analysis of the photometric variability}

\begin{figure}
        \includegraphics[width=9.0cm]{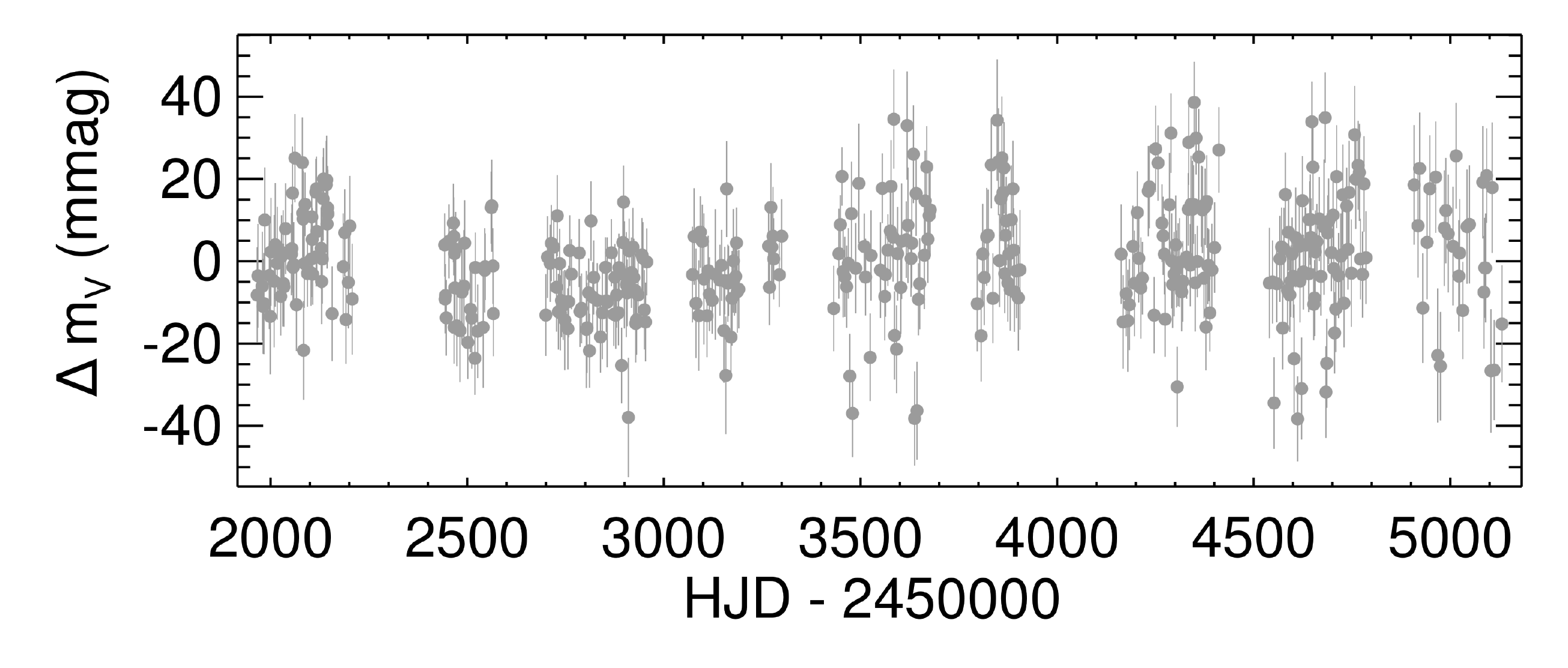}
        \caption{ASAS m$_{V}$ light curve for the star HD 176986. }
        \label{mv_timeseries}
\end{figure}

\begin{figure}
        \includegraphics[width=9.0cm]{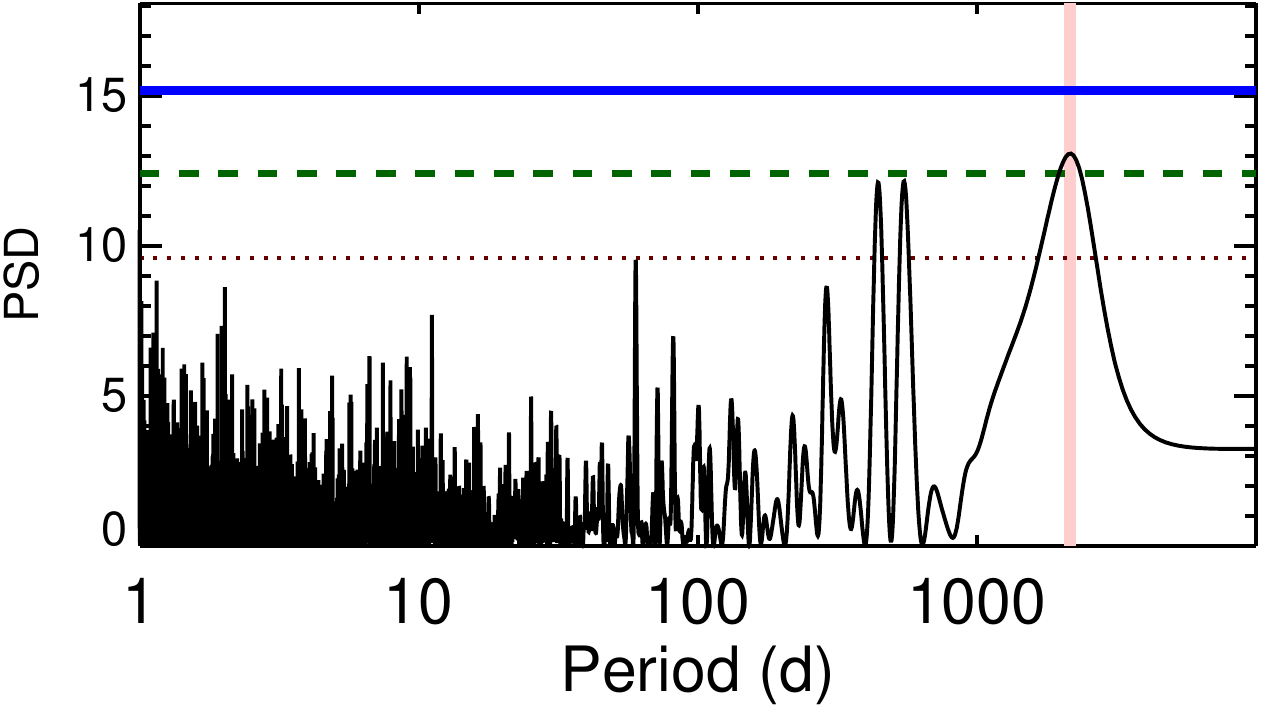}
        \includegraphics[width=9.0cm]{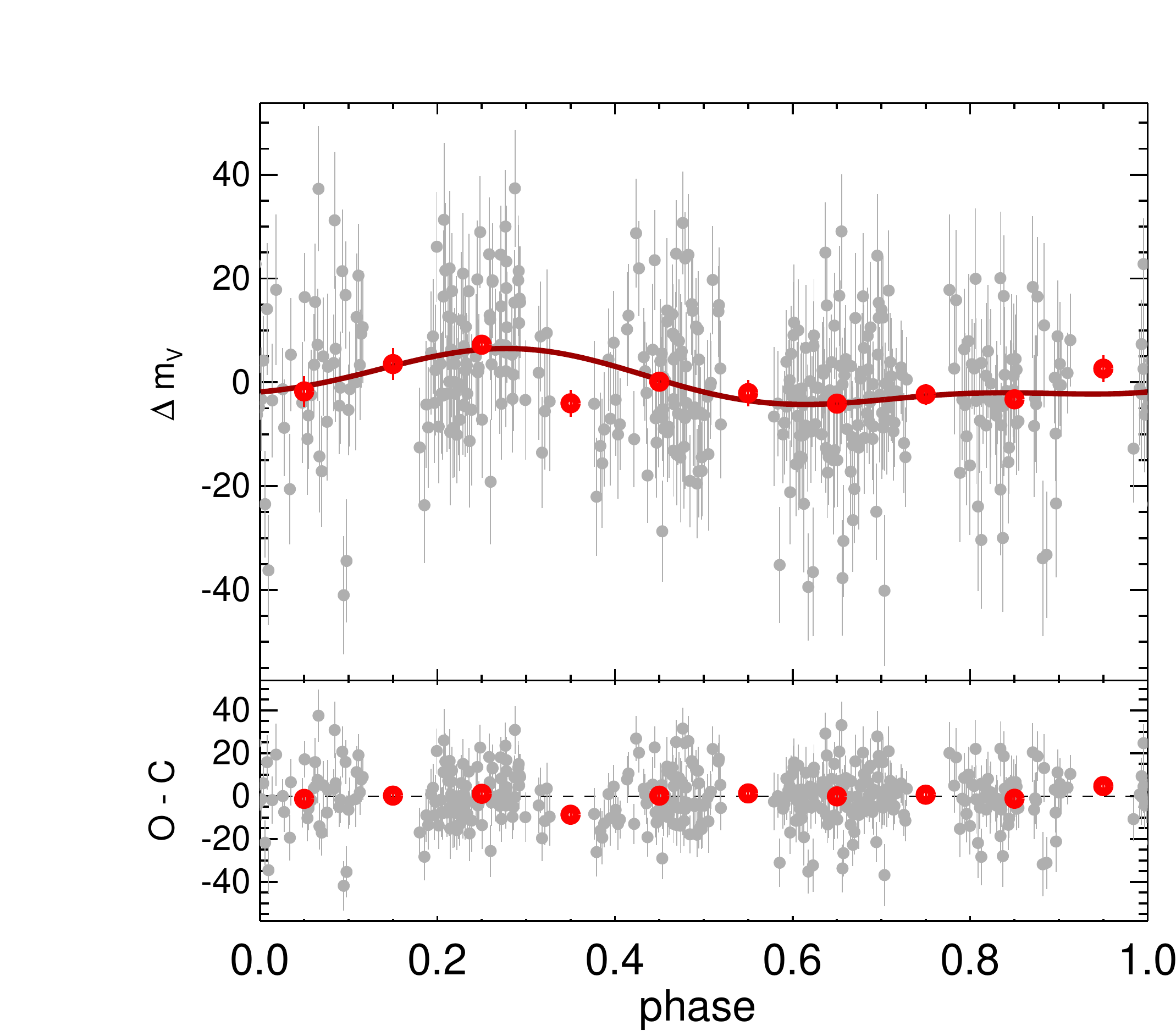}
        \caption{GLS periodogram of the ASAS light curve (top panel). The shaded region shows the periods estimated by the least-squares double-sinusoidal fit. The bottom panel shows the phase-folded curve of the data using the period highlighted in the top panel. Grey dots show the individual measurements after subtracting a long-period linear trend. See Fig.~\ref{rv_phase} for reference. }
        \label{mv_fit}
\end{figure}

The ASAS m$_{V}$ light curve consists of 411 nightly exposures over a baseline of 8.7 years, measuring an average m$_{V}$ of 8.46 with an RMS of 13.4 mmag and a typical uncertainty of the measurements of 10.3 mmag. Figure~\ref{mv_timeseries} shows the available light curve. We detected a significant long-period signal at 5.09 $\pm$ 0.26 yr with a semi-amplitude of 5.6 $\pm$ 1.3 mmag. Figure ~\ref{mv_fit} shows the periodogram and the phase-folded curve.  No more significant signals arise after subtracting the  long-period signal. 

\section{Analysis and interpretation of the detected periodicities} 
\label{sec_orig}

The previous analysis unveils several periodic signals ranging from 6 days to 8 years. Some of them are exclusive to a single dataset, while others are common to different data series. We can summarize the detected signals as a 6.5 d signal detected in the RV data; a 16.8 d signal detected in the RV data; a 19 d signal detected  in the bisector span data; a 26 d signal detected in the  H$_{\alpha}$ index data; a 36 d signal detected simultaneously in the RV, FWHM, and S$_{MW}$ index data; and a group of signals ranging from 4 to 8 yr detected in the bisector, S$_{MW}$ index , H$_{\alpha}$ index data and in the m$_{V}$ light curve. Table~\ref{signals} shows the periods and uncertainties of all the detected signals. 

\begin {table*}
\begin{center}
\caption {Summary of the detected signals in the different datasets of HD 176986. Errors are the results of the least-squares minimization fit. \label{tab:signals}}
    \begin{tabular}{ l  l  l l l l l l l l l l l  } \hline
Dataset  & 6.5 d  & 16.8 d  & 19 d & 26 d & 36 d &    4-8 yr \\ \hline
RV &  6.49 $\pm$ 0.01 &   16.82 $\pm$ 0.01 & & & 35.74 $\pm$ 0.02    \\
FWHM & & & & & 36.00 $\pm$ 0.03   \\
BIS & & &  19.20 $\pm$ 0.01 & & & 4.6 $\pm$ 0.1\\
S$_{MW}$&  && & & 36.02 $\pm$ 0.02   & 5.7 $\pm$ 0.1 \\
H$_{\alpha}$& &&& 26.20 $\pm$ 0.01  && 8.2 $\pm$ 0.2\\
m$_{V}$ &   &&&&&5.1 $\pm$ 0.3 \\
 \hline \\
\label{signals}
\end{tabular}  
\end{center}
\end{table*}

Signals that are simultaneously detected in RV and in one or several of the activity indicators are in most cases induced by stellar activity. Although there is no physical reason why a planet cannot orbit a star  with a period equal to the rotation or the magnetic cycle of the star, separating the two signals in order to prove the Keplerian origin of the RV signal is often very difficult. Based on this assumption, we can separate the RV signals into two different groups. One  group is probably associated with stellar activity (35.8 d), and the other group might have a Keplerian origin (6.5 d and 16.8 d). Some remaining signals are only present in the activity proxies (19 and 26 d, and three long-period signals). 

Given the spectral type of the star (K2.5), its mean level of stellar activity (log$_{10}$ (R$_{HK}^{'})$ = -- 4.90) and the previous measurements of \citet{Masca2015,Masca2017b}, which
were  performed over a much shorter baseline,  we expect a rotation period of 37 $\pm$ 6 days \citep{Masca2016}. We can safely say that the different signals at around 36 days are all related to the rotation of the star. The amplitude of the RV signal at this period is consistent with the measurement of the activity-induced signal measured in \citet{Masca2017b},  even though the measured periods are different (it has to be noted that the period was calculated using a subset of the data used in this work). The 40 d period measured in \citet{Masca2017b} seems to be the 1 yr alias of the real rotation signal. 

The 19-day signal present in the bisector span data is very close to half of the rotation period. The combination of the geometrical distribution of active regions and the inclination of the rotation axis of the star can easily create a pattern in which the rotation signal and its first harmonic are super-imposed \citep{Boisse2011,Dumusque2011, Dumusque2014, Masca2017b}. The highly non-sinusoidal shape of the signals identified at the rotation period (see Figs.~\ref{rv_phase}, \ref{fwhm_phase}, and \ref{smw_phase}) suggests a very inhomogeneous distribution of activity regions that might lead to the detection of signals at the harmonics of the rotation period. 

The different long-period signals present both in the bisector, S$_{MW}$ index, H$\alpha$ index time-series, and in the m$_{V}$ light curve are compatible with a solar-like cycle. The length of that cycle is still not well defined, considering the scatter of the different measurements, but the S$_{MW}$ index measurement is probably the most accurate measurement, as it has the largest amplitude. A longer baseline is needed to completely establish the cycle length, but based on our measurements, we assume that it is close to five years. We did not find a similar signal in the RV data, which supports the conclusions of \citet{Santos2010}.

The origin of the 26 d signal in the H$_{\alpha}$ index might be indicating activity regions that lie closer to the equator than those producing the 36 d signal. The Sun's differential rotation causes the rotation period to change from 25 days at the equator to 36 days at the poles. We might be seeing a similar behaviour, although this would imply a very extreme scenario. The H$_{\alpha}$ measurement showing the evolution of equatorial active regions, while every other activity proxy showing the evolution of polar active regions. This seems a rather unlikely scenario. The variations in H$_{\alpha}$ index for K-type stars are very small and sometimes harder to interpret than in the case of M-dwarfs or the variations in S$_{MW}$ index. Given the low amplitude measured of the variations in H$_{\alpha}$ index, we cannot rule out that the 26 d signal is not a real signal, but a combination of the 19 d and 36 d signals.  

The signals at 6.49 d have no counterpart in the time-series of activity proxies, which favours the interpretation of a Keplerian origin. The 16.82 d signal does not have a direct counterpart in any of the activity proxy time-series, but it lies close to the first harmonic of the rotation signal, which means that
we need to be careful when we interpret this signal.

As a test of the origin of the detected signals, we studied the Spearman correlation coefficient between the different time-series and the different individual signals. To measure the correlations between individual signals, we created individual datasets by subtracting all the other contributions, and we isolated the studied signal. The significance of the correlation was evaluated by performing 10~000 bootstrap simulations \citep{Endl2001}. No correlation was found between any of the RV datasets and the bisector or H$\alpha$ datasets. There is evidence for a weak correlation between the raw RV data and the FWHM and S$_{MW}$ index data. The correlation became non-significant when we tested the datasets that included only the 6.49-day or 16.82-day signals, or both. On the other hand, the correlation increased when we
tested the dataset that included the 35.74-day RV signal. To further test this correlation, we fitted and subtracted a linear relationship between the FWHM and the RV data. The resulting dataset shows the 6.49- and 16.82-day signals intact, while the amplitude of the 35.74-day signal decreases significantly. The same test with the S$_{MW}$ index data produced the same effect.  This favours the interpretation that the 35.74-day RV signal shares an origin with the 36-day activity signals. We did not find a correlation between any of the datasets that included only the 6.49- and/or 16.82-day signals  and any of the activity index time-series. This favours a Keplerian interpretation for the two signals.

\begin{figure*}
\begin{minipage}{0.5\textwidth}
        \centering
        \includegraphics[width=9.cm]{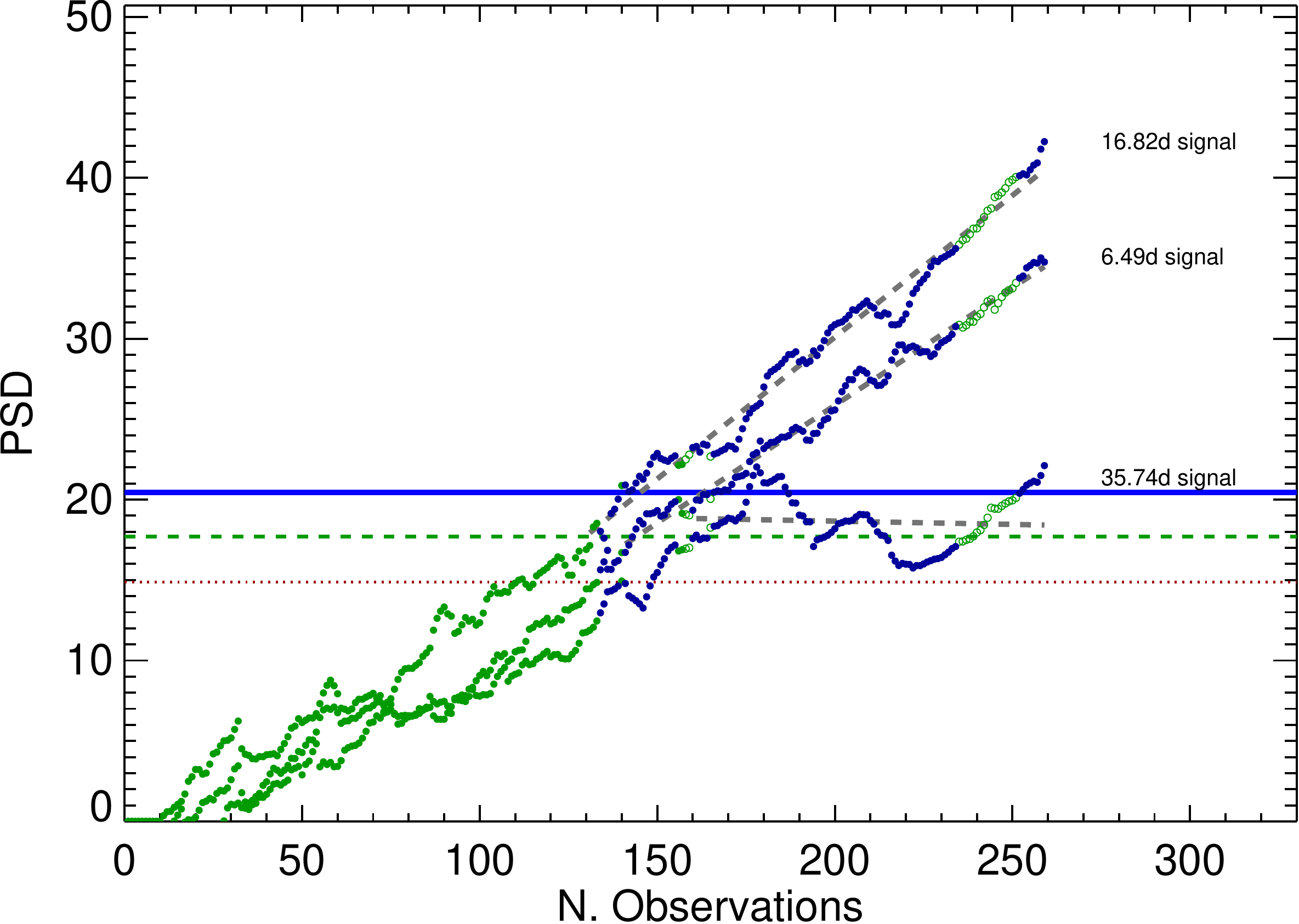}
\end{minipage}%
\begin{minipage}{0.5\textwidth}
        \centering
        \includegraphics[width=9.cm]{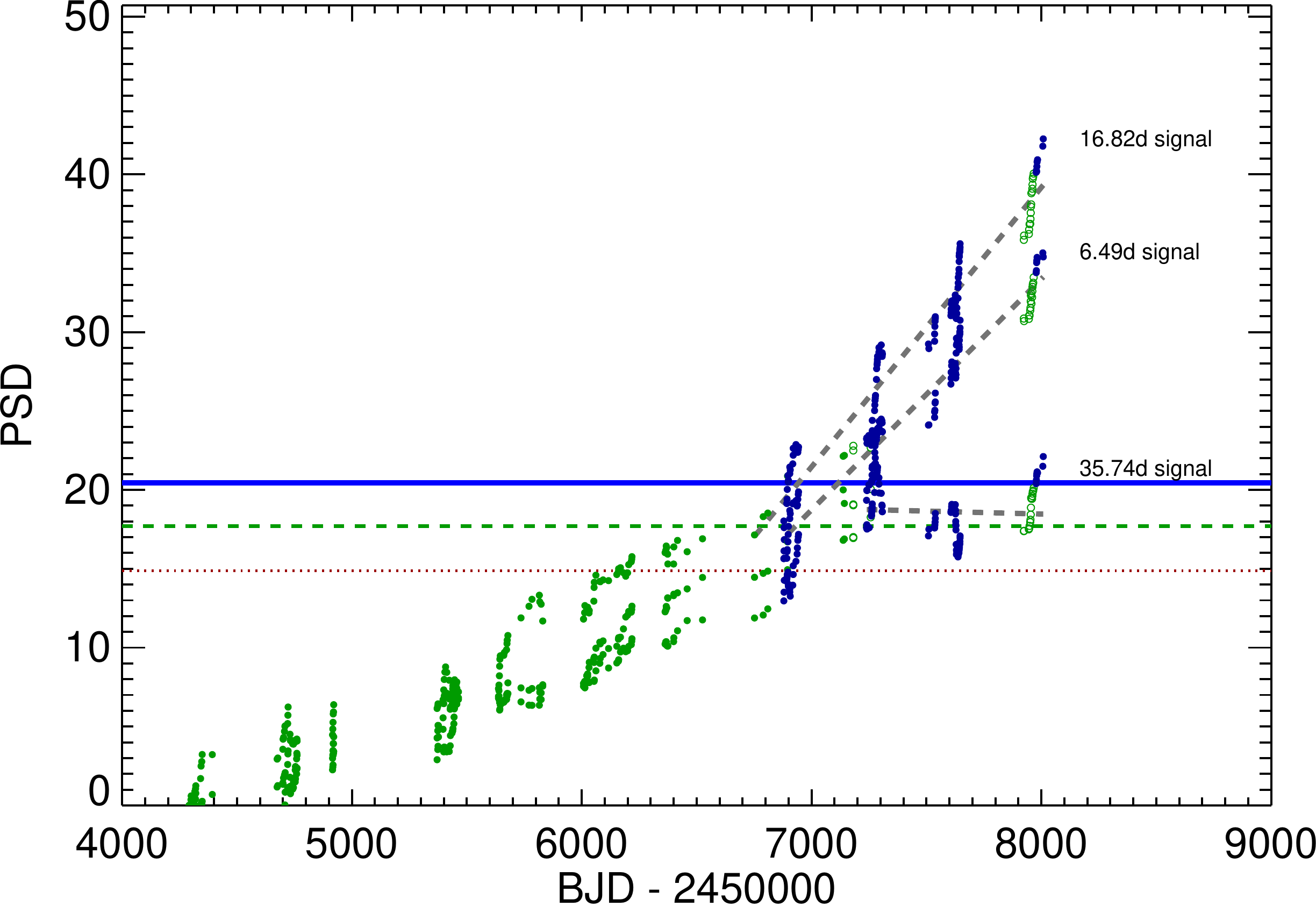}
\end{minipage}%

\caption{Evolution of the significance of the detected signals as a function of the number of measurements (left panel) and the date of the observations (right panel). Green filled dots, green empty dots, and blue filled dots show the HARPS pre-upgrade, HARPS post-upgrade, and HARPS-N measurements, respectively. The horizontal red dotted line, green dashed line, and blue solid line show the 10\%, 1\%, and 0.1\% FAP thresholds. The grey dashed lines show a linear fit to the evolution of the PSD after the signals cross the 1\% FAP threshold.}
\label{evol}
\end{figure*}

\begin{figure*}
\begin{minipage}{0.5\textwidth}
        \centering
        \includegraphics[width=9.cm]{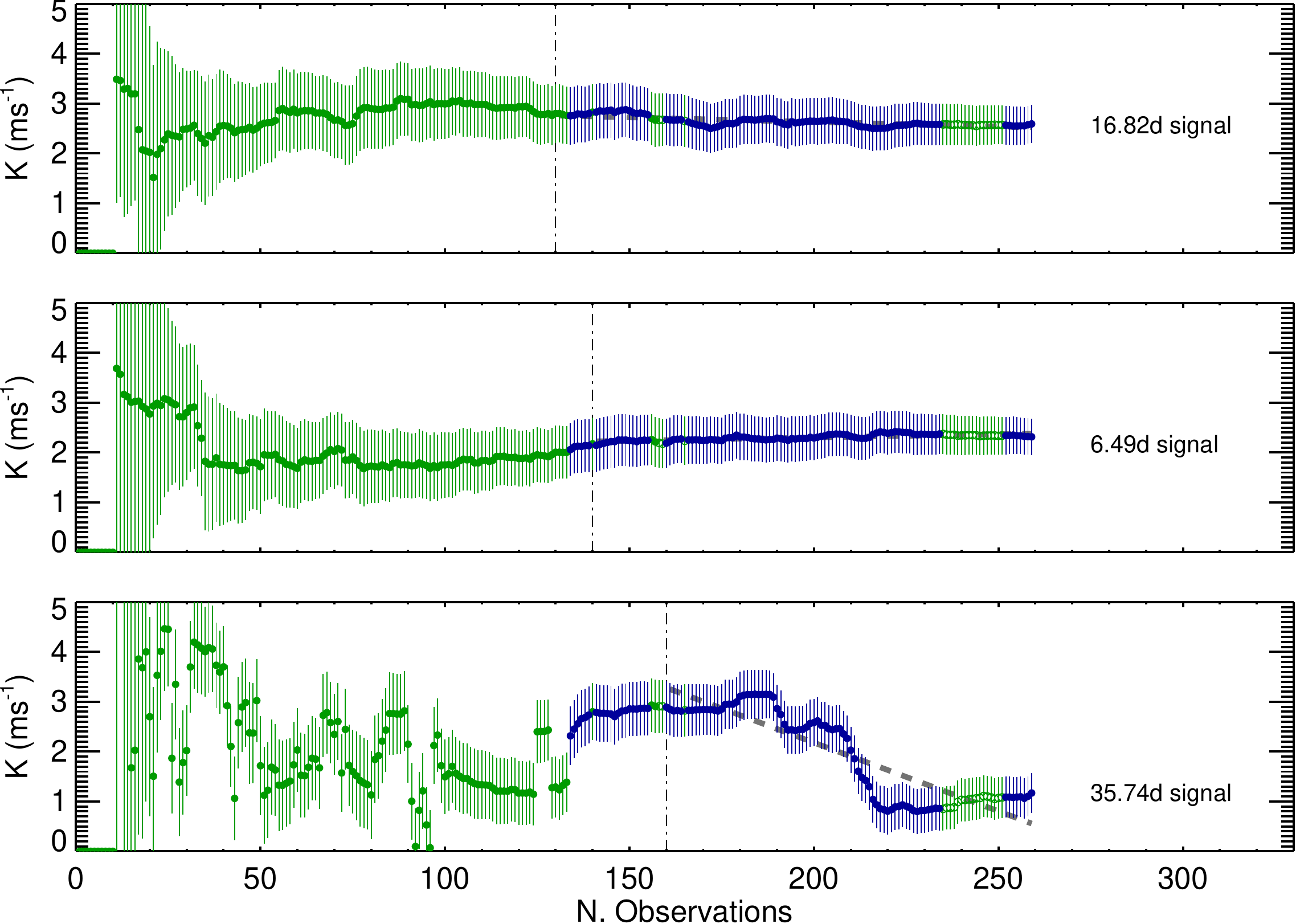}
\end{minipage}%
\begin{minipage}{0.5\textwidth}
        \centering
        \includegraphics[width=9.cm]{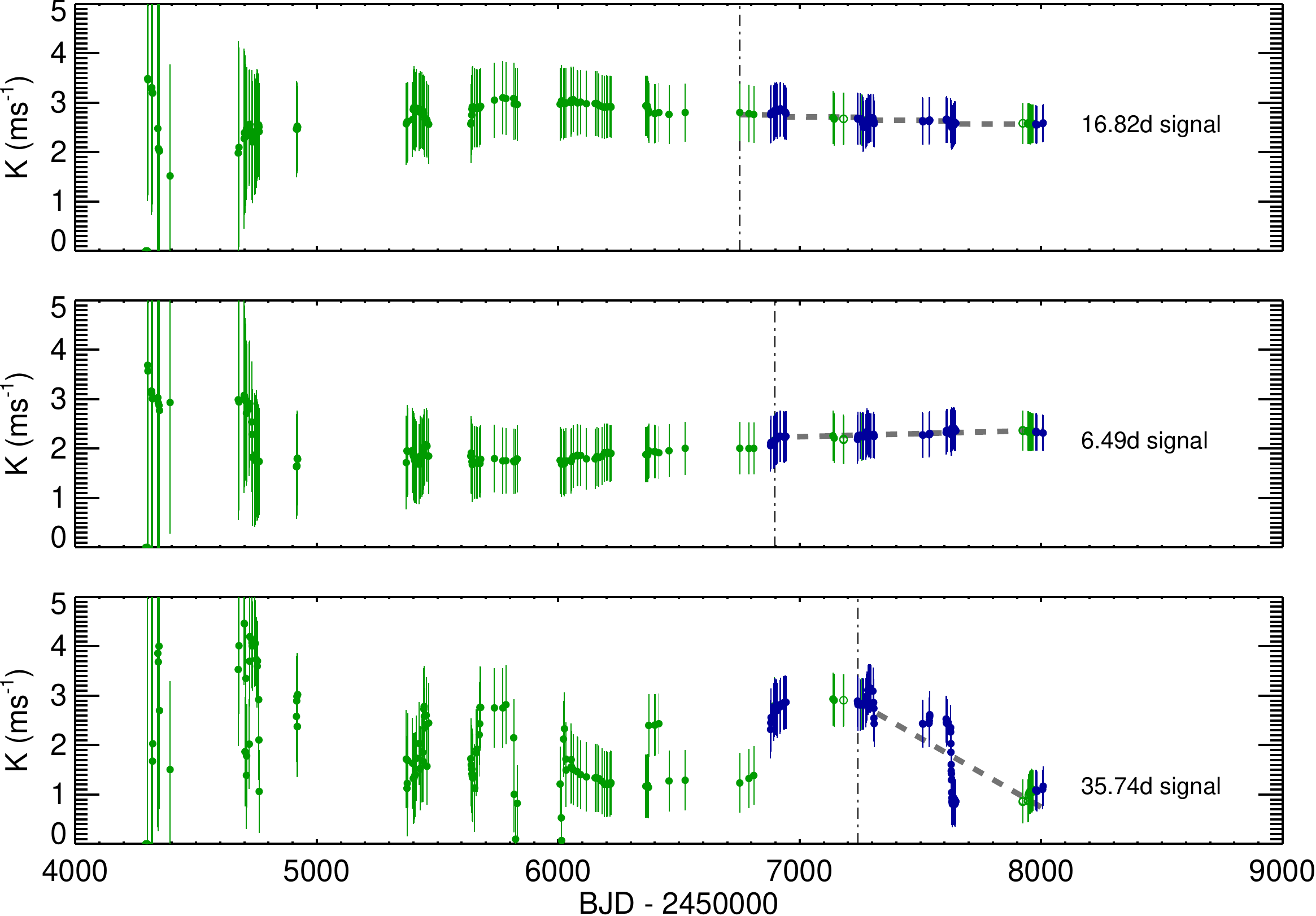}
\end{minipage}%

\caption{Evolution of the measured semi-amplitude of every detected signal as a function of the number of measurements (left panel) and the date of the observations (right panel), using a double-sinusoidal model as in Section~\ref{sect_rv}. Green filled dots, green empty dots, and blue filled dots show the HARPS pre-upgrade, HARPS post-upgrade, and HARPS-N measurements, respectively. The vertical black dashed line shows the moment at which each signal crosses the 1\% FAP threshold. The grey dashed lines show a linear fit to the evolution of the PSD after the signals cross the 1\% FAP threshold.}
\label{evol_k}
\end{figure*}

Keplerian signals are deterministic and consistent in time. When an individual signal is measured, the significance of the detection is expected to increase steadily with the number of observations
and the measured period is expected to be stable over time. However, in the case of an activity-related signal, this is not necessarily the case. As the stellar surface is not static and the configuration of active regions may change with time, changes in the modulation phase and in the detected period are expected \citep{Affer2016, Masca2017, Masca2017c, Mortier2017}. Even the disappearance of the signal at certain seasons is possible.  To study the evolution of the two signals, we performed a simultaneous fit of the detected periodicities in each of the time-series, and then used the derived parameters to subtract the contribution of one of them, which left the other "isolated". We then performed the stacked periodograms using a very narrow frequency window around each of the signals. Figure~\ref{evol} shows the evolution of the significance of the detection of all the signals after subtracting all the other detected components, against the number of observations and the date of observation. When the signal-to-noise ratio is high enough (after the signal crosses the 1\% FAP threshold), the evolution of the 6.5d and 16.8 d singals is steady and continuous. The slope of the 16.8 d signal is slightly steeper, which is to be expected given its larger amplitude.  The behaviour of the 35 d signal, on the other hand, is very erratic. It almost disappears after crossing the 0.1\% FAP threshold and recovers later. The result of this test favours the interpretation that the 6.49 d and 16.82 d signals are Keplerian signals and that the 36 d signal is an activity signal. Following the same reasoning as for the previous test, we expect the measured semi-amplitude of planetary signals to converge with the number of observations. Before they become significant, the 6.49 d and 16.82 d signals already show amplitudes that are compatible with the final measurement (see Fig.~\ref{evol_k}). The amplitude of these signals shows very small variability even before the signals cross the 1\% FAP threshold, and slopes that are compatible with flat lines. On the other hand, the 35.74-day signal shows a very steep negative slope and a variability greater than 77\% in its final amplitude. The signal amplitude increases significantly after 130 measurements (2014) and then decreases again after 200 measurements (2016).   By studying the evolution against the date of observations, we can also see the importance of the high-cadence observation campaigns. The slope of the significance is greatly increased after the start of the  HARPS-N nightly cadence campaigns, which means that the time needed for the detection of short-period planets decreases significantly, similar to what happened with the discovery of Proxima b \citep{AngladaEscude2016}.

\begin{figure*}
        \includegraphics[width=18cm]{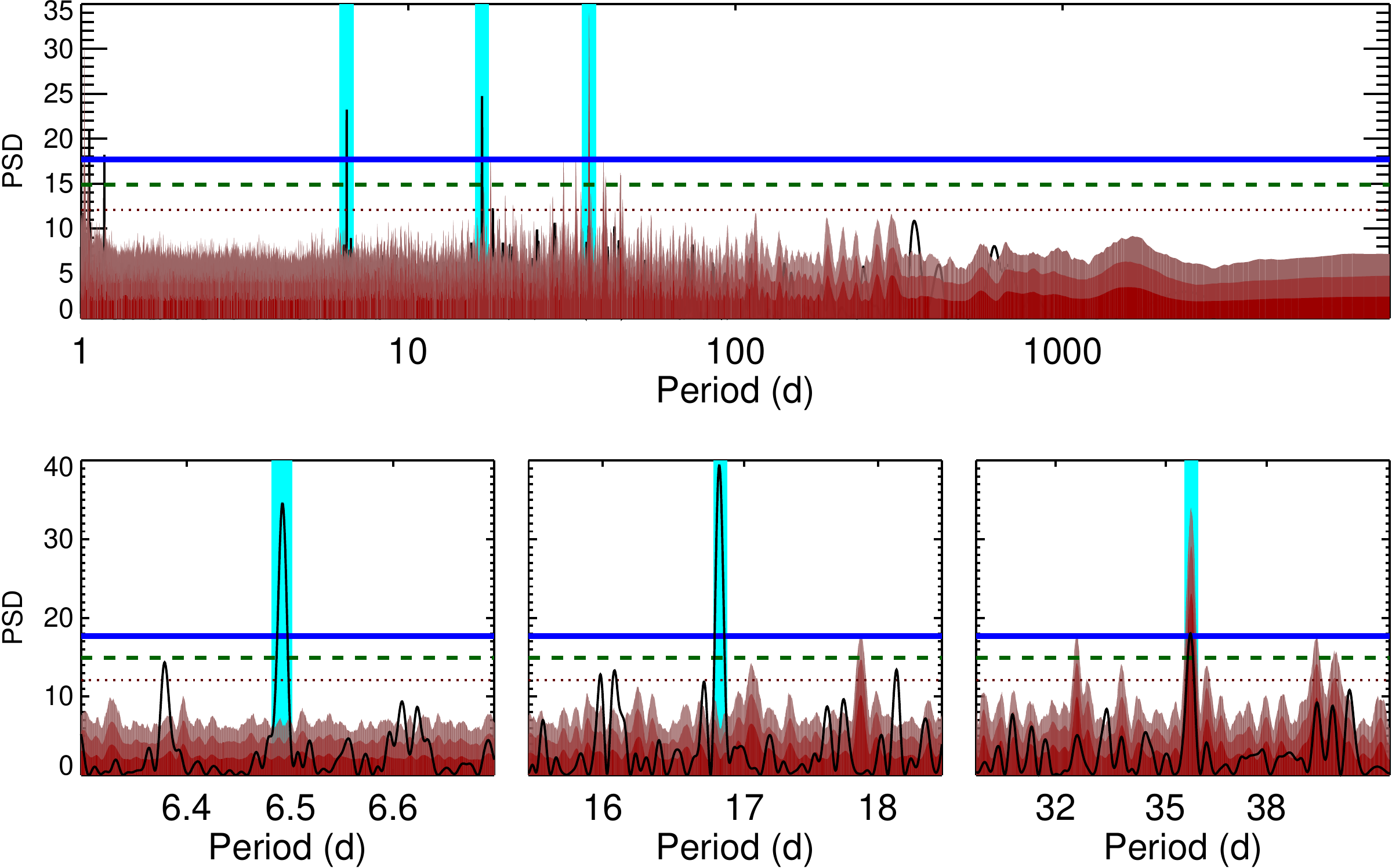}
\caption{Periodogram of the synthetic data created by injecting the 35.7 d signal, along with a low-amplitude 6 yr signal and a white-noise model. The black line shows the real RV periodogram. The dark red, red, and light red regions show the PSD level that
is reached at every individual frequency at least in 10\%, 1\%, and  0.1\% of the simulations. The vertical light blue lines show the position of the detected RV signals. The bottom panels are zoom-ins around the regions of the three detected RV signals. The black line in this case shows the RV periodogram after subtracting the contributions of all the other signals. }
\label{noise}
\end{figure*}

Finally, an analysis of the spectral window ruled out that the peaks in the periodogram are artefacts of the time sampling alone. Following \citet{Rajpaul2016}, we checked that no features appeared at 6.5, 16.8, or 35.7 days even after masking the oversaturated regions of the power spectrum. We then tried to re-create the planet candidate signals by injecting the detected signals corresponding to P$_{\rm Rot}$ and P$_{\rm Rot}$/2 along with a low-amplitude signal (< 1 m s$^{-1}$) at the period of the magnetic cycle (6 yr). We performed 10000 simulations with randomized periods and amplitudes (compatible with the measured uncertainties) and randomized phase shifts along with a white-noise model with an RMS of 1.9 m s$^{-1}$. Figure~\ref{noise} shows the results of these simulations; the periods attributed to the planetary candidates are highlighted.  We were never able to generate the signals at 6.5 or 16.8. It seems very unlikely that any of the candidate planetary signals are artefacts of the sampling or ghost signals created by the combination of activity signals. This test also shows that for future studies of this star, it is important to be very careful with periodicities that are detected between 25 and 45 days, as the combination of the different activity signals detected might generate several artefacts in that range.  It is also important to note that it seems possible to generate a 26-day peak in the periodogram by combining the 35.7 d signal and a signal at half that period with the appropriate phase difference and amplitude, which might explain the 26-day signal in the H$_{\alpha}$ index time-series that does not appear in any other time-series.

After these tests, we conclude that the signals at 6.5 days and 16.8 days are of Keplerian origin, while the signal at 35.8 days is
of stellar origin.

\section{The planetary system of HD 176986}

The analysis of the RV time-series and of the activity indicators leads us to conclude that the best explanation of the observed data is that two planets orbit the star HD 176986. The RV series also shows two different activity signals that are related to the stellar rotation and its first harmonic. 

In order to calculate the final parameters and  quantify the uncertainties of the orbital parameters of the planets, we performed a Bayesian analysis using the MCMC code built in the DACE platform, which is described in  \citet{Diaz2014, Diaz2016}. We fitted an RV model composed of three Keplerians;  we decided to model the activity signals as Keplerians. The MCMC analysis was performed with uniform priors for all variables, with the exception of the stellar mass, for which  a Gaussian prior was used. To take the uncertainties due to instrumental systematics and stellar signals not estimated during the analysis process into account, we included in the MCMC analysis a white-noise jitter parameter that was added quadratically to the individual RV error bars. 

\begin{figure*}
\includegraphics[width=18.cm]{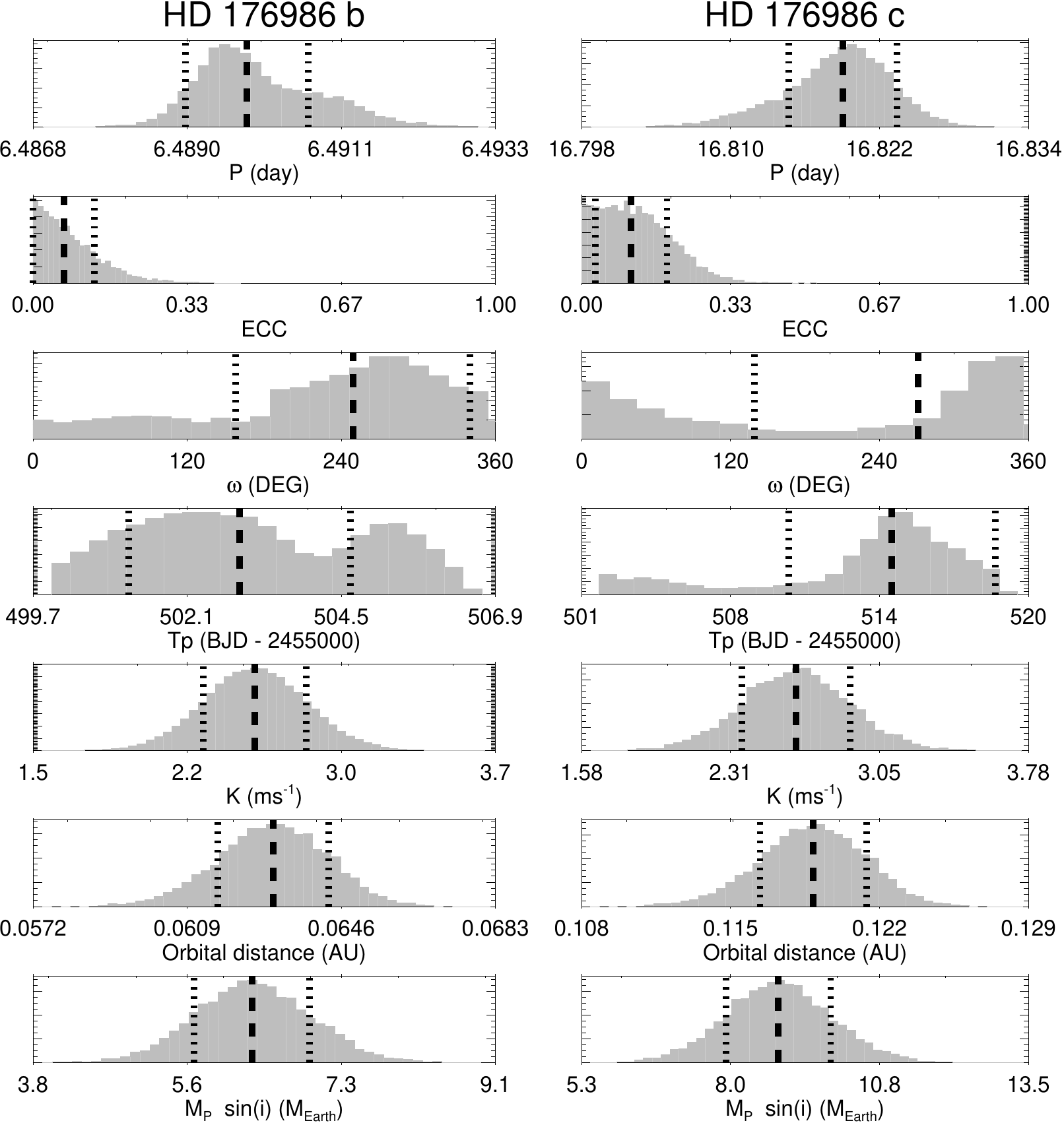}

\caption{Posterior distribution of the model parameters of the planetary companions of the K dwarf star HD 176986. The vertical dashed line shows the median value of the distribution, and the dotted lines represent the 1$\sigma$ values.}
\label{planets_mcmc}
\end{figure*}

Figures~\ref{planets_mcmc} shows the posterior distribution of the parameters of the planetary system, and in Table~\ref{MCMC_par} we show the final parameters and uncertainties obtained with the Bayesian MCMC analysis. It is important to note that the obtained amplitude for the stellar-induced RV signal is very different from the amplitude measured in previous sections. This is a side effect of using a Keplerian to model the signal instead of the double-sinusoidal model. %

\begin{table*}
\begin{center}
\caption {Parameters of the HD 176986 planetary system\label{MCMC_par}}
\begin{tabular}{l | c c | c c }
\hline
\textbf{Parameter} & \textbf{HD 176986 b} & \textbf{Priors (uniform)} & \textbf{HD 176986 c} & \textbf{Priors (uniform)}\\
\hline
$P_{\rm planet}$ [d]             & 6.48980 $\pm$ 0.00086 &  4 -- 9 &  16.8191 $\pm$ 0.0044 & 14 -- 19   \\
$e$                              & 0.066 $\pm$ 0.066 & 0 -- 1 & 0.111 $\pm$ 0.080 & 0 -- 1\\
$\omega$ [deg]                   & 249 $\pm$ 91 & 0 -- 360 & 270 $\pm$ 130 & 0 -- 360\\
$T_{P}$ [BJD -- 2455000]         & 502.9 $\pm$ 1.7 & & 514.4 $\pm$ 4.3 & \\
$K_{\rm planet}$ [ms$^{-1}$]     & 2.56 $\pm$ 0.24 & 1 -- 5 & 2.63 $\pm$ 0.27  & 1 -- 5\\
$a$ [AU]                         & 0.06296 $\pm$ 0.00013 & & 0.11878 $\pm$ 0.00025 &  \\
$m_p \sin i$ [M$_{\rm Earth}$]   & 5.74 $\pm$ 0.66 & &  9.18 $\pm$ 0.97  & \\ \\
\hline
\textbf{Parameter} & \textbf{Activity (Rot)} & \textbf{Priors (uniform)} \\ 
\hline
$P$ [d]             & 35.727 $\pm$ 0.013  & 30 -- 40\\
$K$ [ms$^{-1}$]     &  2.90 $\pm$ 0.62    &  0 -- 6 \\ \\
\hline
$RV ~noise$ [ms$^{-1}$]  &  1.98 $\pm$ 0.12\\
\hline
\end{tabular}  
\end{center}
\end{table*}

\subsection{Gaussian processes analysis}

Gaussian processes (GP) are becoming increasingly common in the analysis of stellar activity in RV time-series. Here we treated the stellar activity term as correlated quasi-periodic noise, using the approach based on GPs (see \citet{Rasmussen2006} and \citet{Roberts2012}). The stellar noise is described by a covariance functional form, and the parameters have some correspondence to the physical phenomena to be modelled. The GP framework can be used to characterize the activity signal without requiring a detailed knowledge of the distribution of active regions on the stellar surface, their lifetimes, or their temperature contrast. 

To model the short-term activity, we used the publicly available GEORGE Python library for GP regression \citep{Ambikasaran2014}, and adopted the quasi-periodic kernel described by the covariance matrix

\begin{equation}
\begin{split}
 k(t,t') = h^2 \cdot \exp \left  (- {{(t - t')^2}\over{2 \lambda^2}}  - {{\sin^2 (\pi (t - t')/\theta) }\over{ 2 \omega^2}} \right) +\\
 + (\sigma^2_{RV}(t) + \sigma^2_{j}) \cdot \delta_{t,t'}
\end{split}
,\end{equation}

where $t$ and $t'$ indicate two different epochs. This kernel has been widely used with great success in recent exoplanet publications \citep{Haywood2014, Faria2016, Affer2016}, as its functional form encompasses some known properties of stellar activity. It
is composed of a periodic term coupled to an exponential decay term, and the kernel describes a recurrent signal linked to the stellar rotation and takes the finite lifetime of the active regions and evolution of their size into account. The parameters $h$ in the covariance function represent the amplitude of the correlations, $\theta$ is the rotation period of the star, $\omega$ is the amplitude scale of the periodic component (linked to the evolution in size of active regions), and $\lambda$ is the correlation decay timescale (related to the lifetime of the active regions). The equation also includes a term of uncorrelated noise, added quadratically to the diagonal of the covariance matrix. $\sigma^2_{RV}(t)$ is the RV uncertainty at time t, and $\sigma^2_{j}$ is the additional noise fit in the model to take other noise sources into account
that are not included in the internal errors and in the stellar activity framework. 

For the parameters of the covariance kernel, we assigned log-uniform priors in sensible ranges. For $\theta$, the parameter that can be interpreted as the stellar rotation period, we have several previous measurements that place it at 36 days. Table~\ref{GP_par} shows the parameters obtained through the GP analysis and the priors used for each parameter. The GP parameters for the two Keplerian signals are consistent with those obtained with the MCMC analysis. The GP MCMC converged towards the previously measured periods. For the case of the activity signal, the measurement of the rotation period coincides with our previous measurements, and the decay timescale also seems to be consistent with what is expected for a main-sequence star \citep{Giles2017}.

\begin{table*}
\begin{center}
\caption {Parameters of the HD 176986 planetary system given by the Gaussian processes analysis\label{GP_par}}
\begin{tabular}{l | c c | c c }
\hline
\textbf{Parameter} & \textbf{HD 176986 b} & \textbf{Priors (uniform)} & \textbf{HD 176986 c} & \textbf{Priors (uniform)}\\
\hline
$P_{\rm planet}$ [d]             & 6.4897 $\pm$ 0.0065 &  4 -- 9 &  16.822 $\pm$ 0.014 & 14 -- 19   \\
$K_{\rm planet}$ [ms$^{-1}$]     & 2.27 $\pm$ 0.32 & 1 -- 5 & 2.66 $\pm$ 0.31  & 1 -- 5\\ \\

\hline
\textbf{Parameter} & \textbf{Activity (Rot)} & \textbf{Priors (uniform)} \\ 
\hline
$\theta$ [d]             & 35.60 $\pm$ 0.43  & 33 -- 38\\
$h$ [ms$^{-1}$]     &  3.1 $\pm$ 1.4 & 0 -- 5 \\
$\lambda$ [d]             & 49.0 $\pm$ 8.0  & 10 -- 100 \\
$\omega$             & 0.42 $\pm$ 0.40  & 0 -- 1\\
\\
\hline
$RV ~noise$ [ms$^{-1}$]  &  1.14 $\pm$ 0.89\\
\hline
\end{tabular}  
\end{center}
\end{table*}

\subsection{Dynamical analysis}

The two HD 176986 planets are very close to each other. Planet-planet interactions might occur when planets oribt so closely to each other, especially somewhat massive planets, which might lead to unstable configurations. A necessary step when analysing a system like this is to test for its long-term stability. We tested the stability of the whole system using the DACE built-in implementation of the GENGA N-body simulations code \citep{GrimmStadel2014}. The global analysis shows the system to be stable over scales of giga years, and shows the semi-major axis and eccentricities of the planets to be very stable along the whole simulation, and the peak-to-peak variation of the eccentricities is smaller than 10$^{-5}$ and 10$^{-4}$ for planets b and c, respectively.  

To test the regions of the parameter space where the system might be stable, we then built stability maps for each of the two planets. For each planet, the system was integrated on a regular 2D mesh of initial conditions, with varying semi-major axis and eccentricity, while the other parameters were kept at their estimated values. For each initial condition, the system was integrated over 10$^{4}$ yr and a stability criterion was derived with the frequency analysis of the mean longitude \citep{Laskar1990, Laskar1993}. In all cases, the 1 $\sigma$ confidence interval is fully compatibly with regions where the system would be stable for timescales comparable with or much longer than the age of the system.

\section{Discussion}

We detected a system of two planets with semi-amplitudes of 2.56 m s$^{-1}$ and 2.63 m s$^{-1}$  . Based on the stellar mass of 0.789 M$_{\odot}$, this is converted into $m_{p} \sin i$ of 5.74 M$_{\oplus}$  and 9.18 M$_{\oplus}$, and the planets orbit with periods of 6.49 d and 16.82 d around HD 176986, a K2.5-type star with a mean rotation period of about 36 days and a 6 yr activity cycle. The two planets would be at orbital distances of 0.063 AU  and 0.119 AU. Following \citet{Queloz1998, Weise2010}, we estimated the $v$ sin $i$ to be 1.10 $\pm$ 0.20 km s$^{-1}$, which means a $P$ sin $i$ of 36 $\pm$ 4 days. This means that the rotation axis of the star is compatible with an edge-on configuration. If the orbital and rotation axis were aligned, this would mean that our estimation for the minimum masses would be very close to the true masses of the planets, and would make the star a very interesting candidate for a transit follow-up. 

The two detected planets may be super-Earths, and HD 176986 c has a greater chance of being a mini-Neptune. The two planets are too close to their host star to be considered habitable. For the two planets HD 176986 b and c, we measured an effective flux of 83.5 $\pm$ 6.9 and 23.5 $\pm$ 4.5 times the flux received by Earth, rendering all of them most likely inhabitable, independently of the planetary conditions. Table~\ref{tab:hab_limits} shows the limits of the habitable zone of HD 176986 for a variety of scenarios.

\begin{table}
\begin{center}
\caption{Limits of the habitable zone for HD 176986 \label{tab:hab_limits}}
\begin{tabular}[center]{l c c c}
\hline
Outer limit & Distance [AU] & Period [d] & Ref. \\ \hline
Early Mars & 1.097 $\pm$ 0.067 & 474 $\pm$ 53 & 1\\ 
Max. greenhouse & 1.054 $\pm$ 0.066 & 446 $\pm$ 51 & 1\\ \\ \hline
Inner limit & Distance [AU] & Period [d] \\ \hline
Runaway greenhouse & 0.588 $\pm$ 0.027 & 186 $\pm$ 21 & 1 \\
Recent Venus & 0.445 $\pm$ 0.030 & 122 $\pm$ 13 & 1\\
Dry world & 0.341 $\pm$ 0.021 & 82.1 $\pm$ 9.2 & 2\\
Dry world (high alb.) & 0.225 $\pm$ 0.013 & 43.9 $\pm$ 4.8 & 2 \\
\hline
\end{tabular}
\end{center}
\textbf{References:} 1 - \citet{Kopparapu2013}, 2 - \citet{Zsom2013}
\end{table}

HD 176986 is a quiet early K-dwarf, with a rotation period of 36 days, which is a typical rotation period for a quiet star of its kind \citep{Masca2016}. Its rotation-induced RV signal has a semi-amplitude of 1.2 ms$^{-1}$ (double sinusoidal), 2.9 ms$^{-1}$ (Keplerian), or 3.1 ms$^{-1}$ (GP), depending on the model used; this signal does not remain stable with time. The double-sinusoidal model gives the smallest amplitude, and the GP analysis gives the largest amplitude. We find a second activity signal at 19 days that is probably related to the first harmonic of the rotation period and that only appears in the bisector span time-series.  The star exhibits a 6 yr magnetic cycle for which we did not detect any measurable induced RV signal, supporting the conclusions of \citet{Santos2010}.

Given the rms of the residuals of the LS-fit or the MCMC, there is still room for the detection of more planets in this system. These hypothetical planets, if present, should be smaller than 6 M$_{\rm Earth}$ for periods shorter than 180 days. For the periods corresponding to the conservative habitable zone, planets of up to 10 M$_{\rm Earth}$ could have been undetected. The region between 20-50 days would be difficult to explore. The rotation signal and the artefacts it creates strongly contaminate the periodogram from $\sim$ 20 to at least $\sim$ 50 days. Between $\sim$ 50 to $\sim$ 100 days, a possible planet should be smaller than 4 M$_{\rm Earth}$ to not appear even as a candidate signal in our data. The GP analysis, on the other hand, leaves a smaller room for additional planets. There could still be very low-mass planets in the system, but the probability of finding them seems rather low. 

\section{Conclusions}

We have analysed 250 high-resolution spectra taken with HARPS and HARPS-N along with photometric observations in V band to study the planetary companions around the K-dwarf star HD 176986 and its stellar activity. We detected the presence of three significant RV signals. 

We conclude that two signals at periods of 6.49 d and 16.82 d are caused by two planets with minimum masses of 5.74 M$_{\oplus}$ and 9.18 M$_{\oplus}$ in orbits with semi-major axes of 0.063 AU and 0.119 AU. Our estimation of the $v$ sin $i$ makes the system compatible with an edge-on configuration, which means that the true masses of the planets might be very close to the minimum masses; the system is a very promising candidate for a transit follow-up.

The remaining signal is the rotation-induced RV signal at 36 days. The measured rotation period and the amplitude of the rotation-induced signal match the expected results of \citet{Masca2015, Masca2016, Masca2017b}. We also found evidence for a  magnetic cycle of $\sim$6 yr, which needs future observations to be better constrained.

We performed a dynamical analysis of the system and found it to be stable over long periods of time, comparable to the age of the system.

\section*{Acknowledgements}

 A.S.M acknowledges financial support from the  Swiss National Science Foundation (SNSF). This work has been partially financed by the Spanish Ministry project MINECO AYA2014-56359-P. J.I.G.H. acknowledges financial support from the Spanish MINECO under the 2013 Ram\'on y Cajal program MINECO RYC-2013-14875. PF and NCS acknowledge support by Funda\c{c}\~ao para a Ci\^encia e a Tecnologia (FCT) through Investigador FCT contracts of reference IF/01037/2013/CP1191/CT0001 and IF/00169/2012/CP0150/CT0002, respectively, and POPH/FSE (EC) by FEDER funding through the program ``Programa Operacional de Factores de Competitividade - COMPETE''. PF further acknowledges support from Funda\c{c}\~ao para a Ci\^encia e a Tecnologia (FCT) in the form of an exploratory project of reference IF/01037/2013/CP1191/CT0001.  This work is based on data obtained via the  HARPS public database at the European Southern Observatory (ESO). This research has made extensive use of the SIMBAD database, operated at CDS, Strasbourg, France, and NASA's Astrophysics Data System. This publication makes use of The Data \& Analysis Center for Exoplanets (DACE), which is a facility based at the University of Geneva (CH) dedicated to extrasolar planets data visualisation, exchange and analysis. DACE is a platform of the Swiss National Centre of Competence in Research (NCCR) PlanetS, federating the Swiss expertise in Exoplanet research. The DACE platform is available at https://dace.unige.ch.

\bibliography{biblio}

\onecolumn
\begin{appendix}
\section{Full Dataset} \label{append_a}

\begin{longtable}{@{\extracolsep{\fill}}llccccccccccc@{}}
\caption {Full available dataset. RVs are given in the barycentric reference frame after subtracting the secular acceleration. RV uncertainties include photon noise, calibration, and telescope-related uncertainties. The mean values for the different quantities are RV = 37.74 km s$^{-1}$, FWHM = 6.13 km s$^{-1}$, Bis Span = 14.82 m s$^{-1}$, S$_{MW}$ index = 0.2916, and  H$_{\alpha}$ = 0.22836. \label{tab_a1}}\\
\endfirsthead
\multicolumn{9}{c}{\tablename\ \thetable\ -- \textit{Continued from previous page}} \\
\hline
\endhead
\hline \multicolumn{9}{r}{\textit{Continued on next page}} \\
\endfoot
\hline
\endlastfoot
\hline

BJD - 2450000 & $\Delta$ V$_{r}$  & $\Delta$ FWHM  &  $\Delta$ BIS Span  & $\Delta$ S$_{MW}$ index & $\Delta$ H$_{\alpha}$ index & Dataset\\ 
 (d) & (ms$^{-1}$) & (ms$^{-1}$) & (ms$^{-1}$)  \\ \hline
3204.7248 & -4.23 $\pm$ 0.94 & 17.6 $\pm$ 3.4 & 2.1 $\pm$ 1.7 & -0.0112 $\pm$ 0.0053 & -0.00174 $\pm$ 0.00025 & HARPS pre-Upgrade \\
3864.8198 & -0.24 $\pm$ 0.92 & 13.6 $\pm$ 3.3 & 1.3 $\pm$ 1.7 & -0.0244 $\pm$ 0.0052 & 0.00068 $\pm$ 0.00024 & HARPS pre-Upgrade \\
3866.8045 & -1.26 $\pm$ 0.91 & 11.7 $\pm$ 3.3 & -0.0 $\pm$ 1.6 & -0.0298 $\pm$ 0.0052 & 0.00037 $\pm$ 0.00022 & HARPS pre-Upgrade \\
4291.7601 & 4.31 $\pm$ 1.35 & 2.7 $\pm$ 4.9 & -4.0 $\pm$ 2.5 & -0.0155 $\pm$ 0.0056 & 0.00265 $\pm$ 0.00041 & HARPS pre-Upgrade \\
4292.7528 & 5.96 $\pm$ 1.21 & 1.6 $\pm$ 4.3 & -4.4 $\pm$ 2.2 & 0.0140 $\pm$ 0.0053 & 0.00183 $\pm$ 0.00026 & HARPS pre-Upgrade \\
4293.7692 & 4.70 $\pm$ 1.21 & 13.6 $\pm$ 4.3 & -2.0 $\pm$ 2.2 & 0.0113 $\pm$ 0.0053 & 0.00173 $\pm$ 0.00026 & HARPS pre-Upgrade \\
4295.7580 & -1.97 $\pm$ 1.20 & 17.1 $\pm$ 4.3 & 1.5 $\pm$ 2.1 & 0.0223 $\pm$ 0.0052 & 0.00119 $\pm$ 0.00023 & HARPS pre-Upgrade \\
4296.7440 & -4.14 $\pm$ 1.22 & 17.1 $\pm$ 4.4 & 3.6 $\pm$ 2.2 & 0.0211 $\pm$ 0.0053 & 0.00119 $\pm$ 0.00027 & HARPS pre-Upgrade \\
4297.8454 & -1.13 $\pm$ 1.35 & 15.2 $\pm$ 4.9 & 4.4 $\pm$ 2.5 & 0.0247 $\pm$ 0.0056 & 0.00136 $\pm$ 0.00039 & HARPS pre-Upgrade \\
4298.7485 & 0.10 $\pm$ 1.27 & 26.1 $\pm$ 4.6 & 5.7 $\pm$ 2.3 & 0.0323 $\pm$ 0.0055 & 0.00055 $\pm$ 0.00033 & HARPS pre-Upgrade \\
4299.8099 & 1.11 $\pm$ 1.28 & 14.7 $\pm$ 4.6 & 5.5 $\pm$ 2.3 & 0.0220 $\pm$ 0.0055 & 0.00041 $\pm$ 0.00034 & HARPS pre-Upgrade \\
4300.7898 & -1.66 $\pm$ 1.23 & 12.6 $\pm$ 4.4 & 3.0 $\pm$ 2.2 & 0.0161 $\pm$ 0.0053 & 0.00063 $\pm$ 0.00028 & HARPS pre-Upgrade \\
4315.6880 & -5.73 $\pm$ 1.14 & -11.3 $\pm$ 4.3 & -1.0 $\pm$ 2.2 & -0.0315 $\pm$ 0.0057 & 0.00056 $\pm$ 0.00044 & HARPS pre-Upgrade \\
4316.6734 & -5.04 $\pm$ 0.99 & -24.4 $\pm$ 3.7 & -1.4 $\pm$ 1.8 & -0.0278 $\pm$ 0.0054 & 0.00018 $\pm$ 0.00033 & HARPS pre-Upgrade \\
4319.7678 & 0.18 $\pm$ 1.02 & -23.2 $\pm$ 3.8 & -1.8 $\pm$ 1.9 & -0.0194 $\pm$ 0.0054 & 0.00041 $\pm$ 0.00034 & HARPS pre-Upgrade \\
4320.7665 & 0.34 $\pm$ 0.99 & -25.7 $\pm$ 3.6 & -5.7 $\pm$ 1.8 & -0.0212 $\pm$ 0.0053 & 0.00071 $\pm$ 0.00029 & HARPS pre-Upgrade \\
4341.7002 & -4.68 $\pm$ 0.96 & -19.9 $\pm$ 3.5 & 1.8 $\pm$ 1.7 & -0.0013 $\pm$ 0.0053 & 0.00031 $\pm$ 0.00026 & HARPS pre-Upgrade \\
4343.7126 & -3.86 $\pm$ 1.02 & -9.5 $\pm$ 3.8 & 0.0 $\pm$ 1.9 & -0.0072 $\pm$ 0.0054 & 0.00161 $\pm$ 0.00034 & HARPS pre-Upgrade \\
4346.7273 & -5.24 $\pm$ 1.24 & -5.5 $\pm$ 4.7 & -0.8 $\pm$ 2.4 & -0.0262 $\pm$ 0.0037 & 0.00099 $\pm$ 0.00043 & HARPS pre-Upgrade \\
4348.6110 & -7.35 $\pm$ 1.00 & -7.9 $\pm$ 3.7 & 0.8 $\pm$ 1.8 & -0.0200 $\pm$ 0.0032 & 0.00068 $\pm$ 0.00031 & HARPS pre-Upgrade \\
4392.5347 & -4.02 $\pm$ 0.90 & -13.0 $\pm$ 3.2 & -0.8 $\pm$ 1.6 & 0.0088 $\pm$ 0.0052 & 0.00035 $\pm$ 0.00021 & HARPS pre-Upgrade \\
4674.8100 & 6.09 $\pm$ 1.23 & 10.5 $\pm$ 4.7 & 2.8 $\pm$ 2.3 & -0.0087 $\pm$ 0.0058 & 0.00040 $\pm$ 0.00047 & HARPS pre-Upgrade \\
4677.7051 & 2.78 $\pm$ 0.98 & 12.1 $\pm$ 3.6 & 1.8 $\pm$ 1.8 & 0.0214 $\pm$ 0.0054 & 0.00034 $\pm$ 0.00031 & HARPS pre-Upgrade \\
4699.5821 & -7.10 $\pm$ 0.96 & 9.0 $\pm$ 3.5 & 4.0 $\pm$ 1.8 & 0.0186 $\pm$ 0.0054 & 0.00118 $\pm$ 0.00030 & HARPS pre-Upgrade \\
4700.6907 & -5.13 $\pm$ 0.96 & 6.9 $\pm$ 3.5 & 2.2 $\pm$ 1.8 & 0.0251 $\pm$ 0.0054 & 0.00080 $\pm$ 0.00029 & HARPS pre-Upgrade \\
4702.7098 & -2.79 $\pm$ 0.97 & 0.4 $\pm$ 3.6 & 2.1 $\pm$ 1.8 & 0.0163 $\pm$ 0.0054 & -0.00017 $\pm$ 0.00030 & HARPS pre-Upgrade \\
4706.6641 & -1.44 $\pm$ 1.42 & -2.7 $\pm$ 5.5 & -1.6 $\pm$ 2.7 & 0.0107 $\pm$ 0.0064 & -0.00090 $\pm$ 0.00060 & HARPS pre-Upgrade \\
4708.6214 & 1.50 $\pm$ 0.97 & 4.4 $\pm$ 3.5 & -1.3 $\pm$ 1.8 & 0.0183 $\pm$ 0.0054 & 0.00020 $\pm$ 0.00030 & HARPS pre-Upgrade \\
4709.6845 & 3.14 $\pm$ 0.93 & -1.0 $\pm$ 3.4 & 1.5 $\pm$ 1.7 & 0.0177 $\pm$ 0.0053 & -0.00009 $\pm$ 0.00026 & HARPS pre-Upgrade \\
4720.6335 & 0.81 $\pm$ 0.98 & -16.3 $\pm$ 3.6 & -1.0 $\pm$ 1.8 & -0.0008 $\pm$ 0.0054 & -0.00162 $\pm$ 0.00031 & HARPS pre-Upgrade \\
4721.5778 & 3.14 $\pm$ 0.95 & -13.5 $\pm$ 3.5 & 0.5 $\pm$ 1.7 & 0.0041 $\pm$ 0.0053 & -0.00143 $\pm$ 0.00028 & HARPS pre-Upgrade \\
4722.5681 & 3.25 $\pm$ 0.97 & -11.5 $\pm$ 3.6 & -1.4 $\pm$ 1.8 & 0.0031 $\pm$ 0.0054 & -0.00154 $\pm$ 0.00031 & HARPS pre-Upgrade \\
4730.6539 & 4.62 $\pm$ 1.02 & 8.8 $\pm$ 3.8 & -1.9 $\pm$ 1.9 & 0.0213 $\pm$ 0.0055 & 0.00094 $\pm$ 0.00034 & HARPS pre-Upgrade \\
4731.6270 & 1.76 $\pm$ 0.96 & 9.4 $\pm$ 3.5 & 1.4 $\pm$ 1.8 & 0.0256 $\pm$ 0.0053 & 0.00120 $\pm$ 0.00028 & HARPS pre-Upgrade \\
4732.6304 & 2.51 $\pm$ 0.98 & 7.4 $\pm$ 3.6 & 2.4 $\pm$ 1.8 & 0.0263 $\pm$ 0.0054 & 0.00075 $\pm$ 0.00030 & HARPS pre-Upgrade \\
4742.5817 & 3.70 $\pm$ 1.86 & 0.6 $\pm$ 7.3 & -1.8 $\pm$ 3.6 & 0.0042 $\pm$ 0.0074 & 0.00030 $\pm$ 0.00082 & HARPS pre-Upgrade \\
4745.5683 & 0.84 $\pm$ 1.03 & 1.8 $\pm$ 3.8 & 5.3 $\pm$ 1.9 & 0.0128 $\pm$ 0.0055 & -0.00080 $\pm$ 0.00035 & HARPS pre-Upgrade \\
4749.5102 & -4.56 $\pm$ 0.98 & -7.3 $\pm$ 3.6 & 2.4 $\pm$ 1.8 & 0.0111 $\pm$ 0.0054 & 0.00016 $\pm$ 0.00030 & HARPS pre-Upgrade \\
4753.5315 & -0.95 $\pm$ 0.97 & -8.1 $\pm$ 3.5 & -0.4 $\pm$ 1.8 & 0.0009 $\pm$ 0.0054 & -0.00017 $\pm$ 0.00030 & HARPS pre-Upgrade \\
4754.5442 & 0.42 $\pm$ 1.01 & -8.2 $\pm$ 3.7 & -2.2 $\pm$ 1.9 & -0.0048 $\pm$ 0.0055 & -0.00049 $\pm$ 0.00034 & HARPS pre-Upgrade \\
4759.5214 & 5.31 $\pm$ 0.95 & -7.2 $\pm$ 3.5 & -0.8 $\pm$ 1.7 & 0.0011 $\pm$ 0.0053 & 0.00053 $\pm$ 0.00027 & HARPS pre-Upgrade \\
4760.5679 & 4.53 $\pm$ 0.97 & -15.1 $\pm$ 3.6 & -0.5 $\pm$ 1.8 & 0.0060 $\pm$ 0.0053 & 0.00047 $\pm$ 0.00029 & HARPS pre-Upgrade \\
4761.5327 & 1.51 $\pm$ 0.96 & -1.8 $\pm$ 3.5 & 2.0 $\pm$ 1.7 & 0.0019 $\pm$ 0.0053 & 0.00023 $\pm$ 0.00029 & HARPS pre-Upgrade \\
4915.8957 & -3.58 $\pm$ 1.02 & -3.7 $\pm$ 3.8 & -1.6 $\pm$ 1.9 & 0.0054 $\pm$ 0.0055 & -0.00020 $\pm$ 0.00036 & HARPS pre-Upgrade \\
4916.8756 & -3.54 $\pm$ 0.95 & -4.3 $\pm$ 3.5 & -0.8 $\pm$ 1.7 & 0.0032 $\pm$ 0.0053 & 0.00019 $\pm$ 0.00027 & HARPS pre-Upgrade \\
4917.8881 & -5.76 $\pm$ 0.94 & -4.2 $\pm$ 3.4 & -2.4 $\pm$ 1.7 & 0.0026 $\pm$ 0.0053 & -0.00021 $\pm$ 0.00027 & HARPS pre-Upgrade \\
4918.8893 & -8.66 $\pm$ 0.96 & -2.4 $\pm$ 3.5 & -2.4 $\pm$ 1.8 & 0.0014 $\pm$ 0.0053 & -0.00021 $\pm$ 0.00029 & HARPS pre-Upgrade \\
4919.8816 & -5.94 $\pm$ 0.91 & -1.9 $\pm$ 3.3 & -0.5 $\pm$ 1.6 & 0.0031 $\pm$ 0.0052 & 0.00020 $\pm$ 0.00023 & HARPS pre-Upgrade \\
4920.8737 & -5.59 $\pm$ 0.97 & -1.4 $\pm$ 3.6 & -2.7 $\pm$ 1.8 & 0.0026 $\pm$ 0.0054 & 0.00020 $\pm$ 0.00030 & HARPS pre-Upgrade \\
5370.6961 & -1.71 $\pm$ 1.46 & 8.9 $\pm$ 5.6 & 2.6 $\pm$ 2.8 & -0.0034 $\pm$ 0.0065 & -0.00116 $\pm$ 0.00064 & HARPS pre-Upgrade \\
5373.6783 & -4.89 $\pm$ 0.99 & -2.4 $\pm$ 3.6 & 3.7 $\pm$ 1.8 & 0.0134 $\pm$ 0.0054 & -0.00149 $\pm$ 0.00030 & HARPS pre-Upgrade \\
5374.7670 & -3.27 $\pm$ 1.26 & -2.5 $\pm$ 4.5 & 1.6 $\pm$ 2.3 & 0.0098 $\pm$ 0.0054 & -0.00072 $\pm$ 0.00031 & HARPS pre-Upgrade \\
5376.6336 & 2.50 $\pm$ 1.26 & 1.7 $\pm$ 4.8 & 3.4 $\pm$ 2.4 & -0.0169 $\pm$ 0.0059 & -0.00155 $\pm$ 0.00050 & HARPS pre-Upgrade \\
5396.7172 & 6.01 $\pm$ 1.02 & -2.5 $\pm$ 3.8 & 2.3 $\pm$ 1.9 & 0.0134 $\pm$ 0.0055 & -0.00118 $\pm$ 0.00034 & HARPS pre-Upgrade \\
5399.7489 & 6.64 $\pm$ 1.00 & -0.8 $\pm$ 3.7 & -0.9 $\pm$ 1.8 & 0.0179 $\pm$ 0.0054 & -0.00206 $\pm$ 0.00031 & HARPS pre-Upgrade \\
5401.7121 & 7.92 $\pm$ 1.23 & 17.5 $\pm$ 4.7 & -1.4 $\pm$ 2.3 & 0.0042 $\pm$ 0.0059 & -0.00084 $\pm$ 0.00049 & HARPS pre-Upgrade \\
5404.7505 & 1.19 $\pm$ 0.95 & 4.3 $\pm$ 3.5 & 3.2 $\pm$ 1.7 & 0.0174 $\pm$ 0.0054 & -0.00164 $\pm$ 0.00029 & HARPS pre-Upgrade \\
5407.7305 & 1.13 $\pm$ 1.03 & 9.1 $\pm$ 3.8 & 3.7 $\pm$ 1.9 & 0.0100 $\pm$ 0.0056 & -0.00098 $\pm$ 0.00037 & HARPS pre-Upgrade \\
5412.7368 & 2.70 $\pm$ 1.12 & 10.2 $\pm$ 4.2 & 0.2 $\pm$ 2.1 & 0.0060 $\pm$ 0.0057 & -0.00061 $\pm$ 0.00043 & HARPS pre-Upgrade \\
5423.6545 & 0.40 $\pm$ 1.08 & 13.1 $\pm$ 4.0 & 0.3 $\pm$ 2.0 & 0.0217 $\pm$ 0.0056 & -0.00018 $\pm$ 0.00039 & HARPS pre-Upgrade \\
5427.7046 & 7.42 $\pm$ 1.03 & 9.1 $\pm$ 3.8 & 5.5 $\pm$ 1.9 & 0.0252 $\pm$ 0.0055 & -0.00125 $\pm$ 0.00035 & HARPS pre-Upgrade \\
5428.6779 & 6.85 $\pm$ 1.11 & 7.3 $\pm$ 4.2 & 6.5 $\pm$ 2.1 & 0.0167 $\pm$ 0.0058 & -0.00203 $\pm$ 0.00043 & HARPS pre-Upgrade \\
5435.6213 & 2.16 $\pm$ 0.94 & -1.8 $\pm$ 3.4 & 1.1 $\pm$ 1.7 & 0.0186 $\pm$ 0.0053 & -0.00062 $\pm$ 0.00027 & HARPS pre-Upgrade \\
5436.6024 & 1.30 $\pm$ 0.92 & -2.8 $\pm$ 3.3 & 2.4 $\pm$ 1.7 & 0.0220 $\pm$ 0.0053 & 0.00020 $\pm$ 0.00025 & HARPS pre-Upgrade \\
5438.6584 & -1.30 $\pm$ 0.93 & 0.9 $\pm$ 3.4 & -1.7 $\pm$ 1.7 & 0.0158 $\pm$ 0.0053 & -0.00030 $\pm$ 0.00025 & HARPS pre-Upgrade \\
5439.6028 & -1.81 $\pm$ 0.93 & -4.8 $\pm$ 3.4 & 0.9 $\pm$ 1.7 & 0.0167 $\pm$ 0.0052 & -0.00053 $\pm$ 0.00024 & HARPS pre-Upgrade \\
5443.6464 & -2.75 $\pm$ 1.05 & 2.6 $\pm$ 3.9 & 3.6 $\pm$ 2.0 & 0.0059 $\pm$ 0.0055 & 0.00021 $\pm$ 0.00036 & HARPS pre-Upgrade \\
5444.6754 & -1.89 $\pm$ 0.96 & -5.3 $\pm$ 3.5 & 4.2 $\pm$ 1.8 & 0.0073 $\pm$ 0.0053 & -0.00093 $\pm$ 0.00028 & HARPS pre-Upgrade \\
5445.5846 & -2.88 $\pm$ 0.97 & -0.6 $\pm$ 3.6 & 2.5 $\pm$ 1.8 & 0.0162 $\pm$ 0.0054 & 0.00011 $\pm$ 0.00030 & HARPS pre-Upgrade \\
5446.6111 & 1.21 $\pm$ 1.12 & -3.2 $\pm$ 4.2 & 2.1 $\pm$ 2.1 & -0.0022 $\pm$ 0.0057 & -0.00020 $\pm$ 0.00043 & HARPS pre-Upgrade \\
5453.5663 & 1.18 $\pm$ 0.94 & -2.3 $\pm$ 3.4 & 0.5 $\pm$ 1.7 & 0.0114 $\pm$ 0.0053 & -0.00236 $\pm$ 0.00026 & HARPS pre-Upgrade \\
5456.5276 & -2.98 $\pm$ 0.96 & -1.2 $\pm$ 3.5 & -2.3 $\pm$ 1.8 & 0.0167 $\pm$ 0.0053 & -0.00060 $\pm$ 0.00029 & HARPS pre-Upgrade \\
5463.5851 & 11.60 $\pm$ 0.94 & 22.3 $\pm$ 3.4 & 1.9 $\pm$ 1.7 & 0.0358 $\pm$ 0.0053 & -0.00038 $\pm$ 0.00026 & HARPS pre-Upgrade \\
5637.8838 & 0.46 $\pm$ 1.13 & -10.0 $\pm$ 4.3 & -2.0 $\pm$ 2.1 & -0.0347 $\pm$ 0.0056 & -0.00014 $\pm$ 0.00042 & HARPS pre-Upgrade \\
5639.9205 & -5.10 $\pm$ 0.97 & -17.1 $\pm$ 3.5 & -2.7 $\pm$ 1.8 & -0.0145 $\pm$ 0.0053 & 0.00038 $\pm$ 0.00028 & HARPS pre-Upgrade \\
5641.8770 & -6.25 $\pm$ 0.96 & -12.4 $\pm$ 3.5 & -0.2 $\pm$ 1.8 & -0.0115 $\pm$ 0.0053 & -0.00016 $\pm$ 0.00028 & HARPS pre-Upgrade \\
5642.9046 & -4.75 $\pm$ 1.24 & -14.8 $\pm$ 4.5 & -0.9 $\pm$ 2.2 & -0.0161 $\pm$ 0.0053 & -0.00081 $\pm$ 0.00028 & HARPS pre-Upgrade \\
5643.8968 & -1.34 $\pm$ 0.98 & -9.6 $\pm$ 3.6 & 0.2 $\pm$ 1.8 & -0.0218 $\pm$ 0.0053 & -0.00077 $\pm$ 0.00030 & HARPS pre-Upgrade \\
5645.8895 & -3.30 $\pm$ 0.99 & -20.9 $\pm$ 3.6 & -1.9 $\pm$ 1.8 & -0.0283 $\pm$ 0.0053 & -0.00062 $\pm$ 0.00031 & HARPS pre-Upgrade \\
5653.8408 & -0.50 $\pm$ 0.93 & -21.0 $\pm$ 3.4 & -5.3 $\pm$ 1.7 & -0.0168 $\pm$ 0.0052 & -0.00042 $\pm$ 0.00024 & HARPS pre-Upgrade \\
5660.8446 & 0.32 $\pm$ 0.97 & 2.0 $\pm$ 3.6 & 2.0 $\pm$ 1.8 & -0.0049 $\pm$ 0.0053 & 0.00022 $\pm$ 0.00029 & HARPS pre-Upgrade \\
5663.8222 & 0.55 $\pm$ 1.52 & -0.8 $\pm$ 5.9 & -4.8 $\pm$ 2.9 & -0.0487 $\pm$ 0.0063 & 0.00176 $\pm$ 0.00066 & HARPS pre-Upgrade \\
5672.8883 & -4.35 $\pm$ 0.97 & 15.5 $\pm$ 3.5 & -0.2 $\pm$ 1.8 & 0.0174 $\pm$ 0.0054 & 0.00147 $\pm$ 0.00030 & HARPS pre-Upgrade \\
5674.8567 & -2.26 $\pm$ 0.97 & 9.9 $\pm$ 3.5 & 1.4 $\pm$ 1.8 & 0.0128 $\pm$ 0.0054 & 0.00071 $\pm$ 0.00029 & HARPS pre-Upgrade \\
5675.8459 & -0.19 $\pm$ 1.03 & 9.4 $\pm$ 3.8 & 1.8 $\pm$ 1.9 & 0.0130 $\pm$ 0.0055 & 0.00030 $\pm$ 0.00035 & HARPS pre-Upgrade \\
5678.8719 & -4.96 $\pm$ 0.95 & 3.9 $\pm$ 3.5 & 4.4 $\pm$ 1.7 & 0.0042 $\pm$ 0.0053 & -0.00074 $\pm$ 0.00028 & HARPS pre-Upgrade \\
5735.7979 & 6.64 $\pm$ 1.00 & 14.7 $\pm$ 3.7 & -0.5 $\pm$ 1.8 & 0.0118 $\pm$ 0.0055 & -0.00059 $\pm$ 0.00034 & HARPS pre-Upgrade \\
5770.6859 & 3.00 $\pm$ 0.92 & 3.8 $\pm$ 3.4 & 2.0 $\pm$ 1.7 & 0.0129 $\pm$ 0.0052 & -0.00065 $\pm$ 0.00024 & HARPS pre-Upgrade \\
5783.6255 & 1.47 $\pm$ 1.04 & -3.0 $\pm$ 3.9 & -0.5 $\pm$ 1.9 & -0.0092 $\pm$ 0.0055 & -0.00186 $\pm$ 0.00037 & HARPS pre-Upgrade \\
5815.5781 & -2.29 $\pm$ 0.95 & -0.5 $\pm$ 3.5 & 0.2 $\pm$ 1.7 & -0.0061 $\pm$ 0.0053 & -0.00158 $\pm$ 0.00028 & HARPS pre-Upgrade \\
5817.5465 & -0.32 $\pm$ 1.03 & -4.6 $\pm$ 3.8 & -1.2 $\pm$ 1.9 & -0.0156 $\pm$ 0.0055 & -0.00122 $\pm$ 0.00036 & HARPS pre-Upgrade \\
5823.5407 & -1.32 $\pm$ 0.94 & -10.5 $\pm$ 3.4 & -4.0 $\pm$ 1.7 & -0.0113 $\pm$ 0.0053 & -0.00095 $\pm$ 0.00026 & HARPS pre-Upgrade \\
5830.5069 & 8.64 $\pm$ 0.96 & 7.8 $\pm$ 3.5 & -1.7 $\pm$ 1.7 & 0.0056 $\pm$ 0.0053 & -0.00141 $\pm$ 0.00029 & HARPS pre-Upgrade \\
6007.8750 & 1.26 $\pm$ 1.26 & 4.8 $\pm$ 4.5 & 2.8 $\pm$ 2.3 & -0.0175 $\pm$ 0.0053 & -0.00068 $\pm$ 0.00030 & HARPS pre-Upgrade \\
6012.8884 & -0.01 $\pm$ 1.24 & -4.3 $\pm$ 4.4 & -4.5 $\pm$ 2.2 & -0.0169 $\pm$ 0.0053 & 0.00063 $\pm$ 0.00028 & HARPS pre-Upgrade \\
6013.8710 & -0.99 $\pm$ 1.01 & 2.2 $\pm$ 3.7 & -4.3 $\pm$ 1.9 & -0.0204 $\pm$ 0.0054 & 0.00102 $\pm$ 0.00032 & HARPS pre-Upgrade \\
6021.8655 & -1.96 $\pm$ 0.96 & 0.6 $\pm$ 3.5 & 2.9 $\pm$ 1.8 & -0.0174 $\pm$ 0.0053 & 0.00092 $\pm$ 0.00028 & HARPS pre-Upgrade \\
6025.8504 & -3.53 $\pm$ 1.12 & 0.2 $\pm$ 4.2 & -3.9 $\pm$ 2.1 & -0.0265 $\pm$ 0.0056 & -0.00066 $\pm$ 0.00042 & HARPS pre-Upgrade \\
6031.8293 & 3.85 $\pm$ 1.22 & 10.8 $\pm$ 4.6 & -4.9 $\pm$ 2.3 & -0.0177 $\pm$ 0.0058 & -0.00016 $\pm$ 0.00048 & HARPS pre-Upgrade \\
6032.8637 & 1.74 $\pm$ 1.00 & 10.2 $\pm$ 3.7 & -3.3 $\pm$ 1.8 & -0.0161 $\pm$ 0.0054 & 0.00052 $\pm$ 0.00031 & HARPS pre-Upgrade \\
6052.9229 & 6.63 $\pm$ 0.93 & 0.9 $\pm$ 3.4 & -3.9 $\pm$ 1.7 & -0.0184 $\pm$ 0.0053 & -0.00120 $\pm$ 0.00026 & HARPS pre-Upgrade \\
6053.7833 & 7.09 $\pm$ 0.97 & -6.6 $\pm$ 3.5 & -0.6 $\pm$ 1.8 & -0.0177 $\pm$ 0.0053 & -0.00171 $\pm$ 0.00028 & HARPS pre-Upgrade \\
6054.8282 & 1.18 $\pm$ 1.02 & -3.7 $\pm$ 3.8 & -0.9 $\pm$ 1.9 & -0.0190 $\pm$ 0.0054 & -0.00041 $\pm$ 0.00034 & HARPS pre-Upgrade \\
6061.7772 & -8.95 $\pm$ 0.93 & -9.8 $\pm$ 3.4 & -4.9 $\pm$ 1.7 & -0.0217 $\pm$ 0.0052 & -0.00087 $\pm$ 0.00024 & HARPS pre-Upgrade \\
6078.8089 & 3.57 $\pm$ 1.00 & 5.1 $\pm$ 3.7 & 2.7 $\pm$ 1.8 & -0.0180 $\pm$ 0.0054 & -0.00023 $\pm$ 0.00033 & HARPS pre-Upgrade \\
6080.7016 & -1.35 $\pm$ 0.98 & -2.1 $\pm$ 3.6 & 0.7 $\pm$ 1.8 & -0.0204 $\pm$ 0.0053 & 0.00036 $\pm$ 0.00029 & HARPS pre-Upgrade \\
6092.8891 & -3.44 $\pm$ 1.07 & -0.2 $\pm$ 4.0 & 2.2 $\pm$ 2.0 & -0.0179 $\pm$ 0.0055 & -0.00164 $\pm$ 0.00037 & HARPS pre-Upgrade \\
6116.8123 & -2.56 $\pm$ 0.99 & -6.7 $\pm$ 3.7 & -1.0 $\pm$ 1.8 & -0.0204 $\pm$ 0.0054 & 0.00046 $\pm$ 0.00032 & HARPS pre-Upgrade \\
6152.7067 & 1.54 $\pm$ 1.19 & 8.5 $\pm$ 4.5 & -1.4 $\pm$ 2.2 & -0.0415 $\pm$ 0.0058 & 0.00019 $\pm$ 0.00047 & HARPS pre-Upgrade \\
6159.7076 & -1.04 $\pm$ 1.36 & 23.9 $\pm$ 5.2 & -4.1 $\pm$ 2.6 & -0.0344 $\pm$ 0.0061 & 0.00054 $\pm$ 0.00057 & HARPS pre-Upgrade \\
6160.7084 & -0.09 $\pm$ 0.99 & 7.7 $\pm$ 3.6 & -3.2 $\pm$ 1.8 & -0.0091 $\pm$ 0.0054 & 0.00133 $\pm$ 0.00031 & HARPS pre-Upgrade \\
6168.6507 & -1.54 $\pm$ 1.09 & 6.1 $\pm$ 4.1 & -2.1 $\pm$ 2.0 & -0.0199 $\pm$ 0.0056 & 0.00022 $\pm$ 0.00041 & HARPS pre-Upgrade \\
6181.6258 & 4.53 $\pm$ 1.15 & 18.5 $\pm$ 4.3 & 3.5 $\pm$ 2.2 & -0.0013 $\pm$ 0.0059 & 0.00003 $\pm$ 0.00047 & HARPS pre-Upgrade \\
6191.5771 & -2.19 $\pm$ 1.00 & 4.1 $\pm$ 3.7 & -0.4 $\pm$ 1.9 & -0.0103 $\pm$ 0.0054 & 0.00157 $\pm$ 0.00032 & HARPS pre-Upgrade \\
6199.5747 & -2.62 $\pm$ 1.26 & 7.4 $\pm$ 4.8 & 3.2 $\pm$ 2.4 & -0.0302 $\pm$ 0.0059 & -0.00018 $\pm$ 0.00051 & HARPS pre-Upgrade \\
6204.5614 & -1.80 $\pm$ 1.08 & -0.8 $\pm$ 4.0 & -1.7 $\pm$ 2.0 & -0.0082 $\pm$ 0.0056 & -0.00126 $\pm$ 0.00039 & HARPS pre-Upgrade \\
6215.5426 & 0.66 $\pm$ 1.45 & 40.2 $\pm$ 5.6 & -1.5 $\pm$ 2.8 & -0.0149 $\pm$ 0.0067 & -0.00223 $\pm$ 0.00071 & HARPS pre-Upgrade \\
6217.5388 & -1.45 $\pm$ 1.00 & -0.8 $\pm$ 3.7 & 0.8 $\pm$ 1.9 & -0.0133 $\pm$ 0.0054 & 0.00033 $\pm$ 0.00032 & HARPS pre-Upgrade \\
6218.5506 & -0.85 $\pm$ 1.11 & 0.4 $\pm$ 4.2 & 4.3 $\pm$ 2.1 & -0.0236 $\pm$ 0.0056 & -0.00077 $\pm$ 0.00041 & HARPS pre-Upgrade \\
6362.9080 & -1.26 $\pm$ 1.13 & 7.7 $\pm$ 4.2 & -0.1 $\pm$ 2.1 & -0.0072 $\pm$ 0.0056 & 0.00031 $\pm$ 0.00041 & HARPS pre-Upgrade \\
6365.8644 & -1.45 $\pm$ 1.03 & 3.6 $\pm$ 3.8 & 0.4 $\pm$ 1.9 & -0.0010 $\pm$ 0.0054 & -0.00044 $\pm$ 0.00032 & HARPS pre-Upgrade \\
6367.8908 & -3.24 $\pm$ 0.98 & -2.2 $\pm$ 3.6 & 0.7 $\pm$ 1.8 & -0.0075 $\pm$ 0.0053 & -0.00078 $\pm$ 0.00029 & HARPS pre-Upgrade \\
6369.9193 & 0.39 $\pm$ 0.98 & -8.3 $\pm$ 3.6 & 3.6 $\pm$ 1.8 & -0.0113 $\pm$ 0.0053 & -0.00026 $\pm$ 0.00029 & HARPS pre-Upgrade \\
6373.8851 & -1.65 $\pm$ 0.99 & -13.9 $\pm$ 3.7 & 1.5 $\pm$ 1.8 & -0.0089 $\pm$ 0.0054 & -0.00031 $\pm$ 0.00030 & HARPS pre-Upgrade \\
6374.8950 & -0.80 $\pm$ 0.99 & -16.3 $\pm$ 3.6 & 2.6 $\pm$ 1.8 & -0.0157 $\pm$ 0.0054 & 0.00020 $\pm$ 0.00030 & HARPS pre-Upgrade \\
6399.8608 & -2.38 $\pm$ 1.05 & 0.2 $\pm$ 3.9 & -0.8 $\pm$ 2.0 & 0.0056 $\pm$ 0.0055 & -0.00141 $\pm$ 0.00035 & HARPS pre-Upgrade \\
6400.8430 & -0.72 $\pm$ 1.02 & 2.5 $\pm$ 3.8 & 3.7 $\pm$ 1.9 & -0.0015 $\pm$ 0.0055 & -0.00119 $\pm$ 0.00034 & HARPS pre-Upgrade \\
6416.8694 & -4.63 $\pm$ 1.04 & -6.2 $\pm$ 3.9 & -3.7 $\pm$ 1.9 & -0.0257 $\pm$ 0.0055 & 0.00080 $\pm$ 0.00035 & HARPS pre-Upgrade \\
6458.9077 & -5.36 $\pm$ 0.98 & -25.6 $\pm$ 3.6 & -0.5 $\pm$ 1.8 & -0.0259 $\pm$ 0.0053 & -0.00220 $\pm$ 0.00029 & HARPS pre-Upgrade \\
6525.7069 & 4.62 $\pm$ 1.07 & -25.6 $\pm$ 4.0 & -1.0 $\pm$ 2.0 & -0.0318 $\pm$ 0.0055 & -0.00137 $\pm$ 0.00037 & HARPS pre-Upgrade \\
6751.9109 & -0.19 $\pm$ 1.20 & 8.7 $\pm$ 4.5 & 2.8 $\pm$ 2.3 & 0.0113 $\pm$ 0.0060 & 0.00077 $\pm$ 0.00049 & HARPS pre-Upgrade \\
6789.8224 & -1.54 $\pm$ 0.97 & -5.4 $\pm$ 3.6 & 2.8 $\pm$ 1.8 & 0.0012 $\pm$ 0.0054 & -0.00068 $\pm$ 0.00029 & HARPS pre-Upgrade \\
6809.7469 & -1.49 $\pm$ 1.00 & -8.5 $\pm$ 3.7 & -2.2 $\pm$ 1.8 & 0.0119 $\pm$ 0.0054 & 0.00168 $\pm$ 0.00032 & HARPS pre-Upgrade \\
6879.4661 & -4.21 $\pm$ 0.91 & 6.0 $\pm$ 3.3 & 1.9 $\pm$ 1.7 & 0.0017 $\pm$ 0.0034 & 0.00174 $\pm$ 0.00028 & HARPS-N \\
6880.4586 & -5.96 $\pm$ 1.30 & 7.4 $\pm$ 4.7 & 0.2 $\pm$ 2.3 & -0.0106 $\pm$ 0.0038 & 0.00221 $\pm$ 0.00039 & HARPS-N \\
6881.4452 & -6.93 $\pm$ 1.09 & 8.4 $\pm$ 4.1 & -0.9 $\pm$ 2.0 & -0.0079 $\pm$ 0.0039 & 0.00218 $\pm$ 0.00042 & HARPS-N \\
6892.4272 & 0.49 $\pm$ 0.91 & 2.4 $\pm$ 3.3 & -2.6 $\pm$ 1.6 & 0.0071 $\pm$ 0.0033 & -0.00103 $\pm$ 0.00026 & HARPS-N \\
6893.4413 & 0.20 $\pm$ 1.02 & 18.2 $\pm$ 3.8 & 1.7 $\pm$ 1.9 & -0.0079 $\pm$ 0.0037 & -0.00053 $\pm$ 0.00037 & HARPS-N \\
6894.4448 & 4.82 $\pm$ 0.93 & 14.6 $\pm$ 3.4 & -0.8 $\pm$ 1.7 & 0.0083 $\pm$ 0.0035 & -0.00042 $\pm$ 0.00029 & HARPS-N \\
6895.6440 & 9.32 $\pm$ 1.04 & 10.3 $\pm$ 3.9 & 2.6 $\pm$ 1.9 & 0.0113 $\pm$ 0.0056 & -0.00094 $\pm$ 0.00036 & HARPS pre-Upgrade \\
6897.4528 & 1.93 $\pm$ 0.97 & 11.9 $\pm$ 3.5 & 5.0 $\pm$ 1.8 & 0.0072 $\pm$ 0.0038 & -0.00129 $\pm$ 0.00036 & HARPS-N \\
6898.4386 & -3.92 $\pm$ 1.23 & 7.3 $\pm$ 4.4 & 5.1 $\pm$ 2.2 & 0.0115 $\pm$ 0.0036 & -0.00118 $\pm$ 0.00033 & HARPS-N \\
6899.4502 & -3.21 $\pm$ 1.27 & 7.4 $\pm$ 4.6 & 1.3 $\pm$ 2.3 & 0.0146 $\pm$ 0.0039 & -0.00187 $\pm$ 0.00040 & HARPS-N \\
6905.4280 & -5.46 $\pm$ 1.23 & -1.3 $\pm$ 4.4 & 4.8 $\pm$ 2.2 & 0.0032 $\pm$ 0.0043 & -0.00011 $\pm$ 0.00036 & HARPS-N \\
6905.4339 & -5.82 $\pm$ 1.33 & -0.0 $\pm$ 4.9 & 3.4 $\pm$ 2.4 & 0.0024 $\pm$ 0.0058 & 0.00002 $\pm$ 0.00046 & HARPS-N \\
6907.4247 & -4.35 $\pm$ 0.90 & -0.7 $\pm$ 3.3 & 5.4 $\pm$ 1.6 & 0.0063 $\pm$ 0.0033 & 0.00026 $\pm$ 0.00025 & HARPS-N \\
6918.4294 & -10.16 $\pm$ 1.41 & 15.3 $\pm$ 5.4 & 4.2 $\pm$ 2.7 & -0.0160 $\pm$ 0.0053 & 0.00217 $\pm$ 0.00064 & HARPS-N \\
6920.4275 & -7.56 $\pm$ 1.00 & 4.1 $\pm$ 3.7 & 3.4 $\pm$ 1.8 & -0.0004 $\pm$ 0.0042 & -0.00107 $\pm$ 0.00037 & HARPS-N \\
6920.4331 & -6.74 $\pm$ 1.18 & 7.3 $\pm$ 4.5 & -0.1 $\pm$ 2.2 & -0.0018 $\pm$ 0.0061 & -0.00032 $\pm$ 0.00054 & HARPS-N \\
6932.3832 & 0.59 $\pm$ 1.08 & 5.7 $\pm$ 4.0 & 4.5 $\pm$ 2.0 & -0.0156 $\pm$ 0.0040 & 0.00070 $\pm$ 0.00044 & HARPS-N \\
6937.3617 & 0.15 $\pm$ 0.91 & 5.6 $\pm$ 3.3 & 2.2 $\pm$ 1.6 & -0.0023 $\pm$ 0.0034 & 0.00029 $\pm$ 0.00026 & HARPS-N \\
6938.3613 & -2.01 $\pm$ 0.92 & 14.7 $\pm$ 3.3 & 1.4 $\pm$ 1.7 & -0.0017 $\pm$ 0.0033 & 0.00086 $\pm$ 0.00026 & HARPS-N \\
6940.3596 & 3.62 $\pm$ 0.91 & 6.9 $\pm$ 3.3 & 2.2 $\pm$ 1.6 & 0.0025 $\pm$ 0.0033 & 0.00081 $\pm$ 0.00025 & HARPS-N \\
6942.3581 & 3.50 $\pm$ 0.91 & 7.9 $\pm$ 3.3 & 3.7 $\pm$ 1.6 & -0.0065 $\pm$ 0.0033 & 0.00044 $\pm$ 0.00024 & HARPS-N \\
6943.3543 & 2.08 $\pm$ 0.98 & 14.2 $\pm$ 3.6 & 4.5 $\pm$ 1.8 & -0.0140 $\pm$ 0.0035 & 0.00072 $\pm$ 0.00032 & HARPS-N \\
7137.8406 & 0.58 $\pm$ 1.01 & 31.6 $\pm$ 3.7 & 2.0 $\pm$ 1.9 & 0.0143 $\pm$ 0.0055 & 0.00004 $\pm$ 0.00033 & HARPS pre-Upgrade \\
7142.9354 & -0.41 $\pm$ 0.94 & 20.2 $\pm$ 3.4 & 5.8 $\pm$ 1.7 & 0.0140 $\pm$ 0.0053 & -0.00012 $\pm$ 0.00026 & HARPS pre-Upgrade \\
7181.7016 & 5.32 $\pm$ 1.00 & 10.9 $\pm$ 3.7 & 1.9 $\pm$ 1.8 & 0.0284 $\pm$ 0.0055 & -0.00019 $\pm$ 0.00035 & HARPS post-Upgrade \\
7181.8042 & 5.23 $\pm$ 1.12 & 11.0 $\pm$ 4.2 & 1.8 $\pm$ 2.1 & 0.0156 $\pm$ 0.0058 & -0.00012 $\pm$ 0.00046 & HARPS post-Upgrade \\
7239.4698 & -4.05 $\pm$ 0.92 & -20.3 $\pm$ 3.3 & 0.3 $\pm$ 1.7 & -0.0065 $\pm$ 0.0034 & 0.00098 $\pm$ 0.00027 & HARPS-N \\
7240.5095 & -1.96 $\pm$ 0.92 & -19.0 $\pm$ 3.3 & -3.8 $\pm$ 1.7 & -0.0040 $\pm$ 0.0035 & 0.00083 $\pm$ 0.00029 & HARPS-N \\
7241.4639 & -0.81 $\pm$ 0.95 & -11.7 $\pm$ 3.5 & -6.3 $\pm$ 1.7 & -0.0089 $\pm$ 0.0036 & 0.00137 $\pm$ 0.00033 & HARPS-N \\
7249.5427 & 1.59 $\pm$ 1.17 & 16.3 $\pm$ 4.4 & 4.6 $\pm$ 2.2 & -0.0121 $\pm$ 0.0041 & -0.00134 $\pm$ 0.00046 & HARPS-N \\
7251.4932 & -0.40 $\pm$ 0.90 & 2.7 $\pm$ 3.2 & -0.7 $\pm$ 1.6 & 0.0156 $\pm$ 0.0034 & 0.00149 $\pm$ 0.00026 & HARPS-N \\
7257.6128 & 1.82 $\pm$ 0.95 & 1.2 $\pm$ 3.5 & 2.9 $\pm$ 1.7 & 0.0287 $\pm$ 0.0053 & -0.00091 $\pm$ 0.00027 & HARPS post-Upgrade \\
7260.4141 & 0.76 $\pm$ 0.90 & -24.0 $\pm$ 3.2 & 0.2 $\pm$ 1.6 & -0.0088 $\pm$ 0.0034 & -0.00036 $\pm$ 0.00027 & HARPS-N \\
7261.4313 & -0.10 $\pm$ 1.04 & -7.8 $\pm$ 3.9 & 0.3 $\pm$ 1.9 & -0.0016 $\pm$ 0.0059 & 0.00037 $\pm$ 0.00048 & HARPS-N \\
7261.4781 & 0.67 $\pm$ 0.88 & -17.0 $\pm$ 3.2 & -0.6 $\pm$ 1.6 & -0.0087 $\pm$ 0.0026 & 0.00009 $\pm$ 0.00022 & HARPS-N \\
7262.4296 & -0.67 $\pm$ 1.01 & -16.1 $\pm$ 3.7 & -2.7 $\pm$ 1.9 & -0.0193 $\pm$ 0.0044 & 0.00069 $\pm$ 0.00041 & HARPS-N \\
7262.4353 & 0.27 $\pm$ 1.12 & -12.8 $\pm$ 4.2 & -1.2 $\pm$ 2.1 & -0.0089 $\pm$ 0.0061 & 0.00069 $\pm$ 0.00055 & HARPS-N \\
7263.4296 & -1.36 $\pm$ 0.98 & -22.4 $\pm$ 3.6 & -3.4 $\pm$ 1.8 & -0.0094 $\pm$ 0.0044 & -0.00123 $\pm$ 0.00040 & HARPS-N \\
7263.4354 & -2.08 $\pm$ 1.11 & -20.4 $\pm$ 4.2 & -7.5 $\pm$ 2.1 & 0.0012 $\pm$ 0.0062 & -0.00070 $\pm$ 0.00055 & HARPS-N \\
7264.4290 & 4.76 $\pm$ 0.98 & -21.7 $\pm$ 3.6 & -3.5 $\pm$ 1.8 & -0.0082 $\pm$ 0.0044 & -0.00158 $\pm$ 0.00041 & HARPS-N \\
7264.4346 & 5.39 $\pm$ 1.14 & -16.1 $\pm$ 4.3 & -2.0 $\pm$ 2.1 & -0.0040 $\pm$ 0.0064 & -0.00250 $\pm$ 0.00062 & HARPS-N \\
7274.4131 & -7.57 $\pm$ 1.03 & -3.5 $\pm$ 3.8 & 2.4 $\pm$ 1.9 & -0.0181 $\pm$ 0.0036 & 0.00169 $\pm$ 0.00034 & HARPS-N \\
7275.4132 & -13.91 $\pm$ 0.96 & -10.6 $\pm$ 3.5 & 0.2 $\pm$ 1.8 & -0.0057 $\pm$ 0.0035 & 0.00320 $\pm$ 0.00031 & HARPS-N \\
7276.4118 & -9.28 $\pm$ 0.92 & -13.7 $\pm$ 3.3 & -1.4 $\pm$ 1.7 & 0.0150 $\pm$ 0.0035 & 0.00053 $\pm$ 0.00029 & HARPS-N \\
7277.4100 & -5.37 $\pm$ 0.90 & -10.9 $\pm$ 3.3 & -6.0 $\pm$ 1.6 & 0.0123 $\pm$ 0.0034 & 0.00185 $\pm$ 0.00026 & HARPS-N \\
7278.4428 & 2.63 $\pm$ 1.07 & -7.0 $\pm$ 4.0 & -5.2 $\pm$ 2.0 & 0.0115 $\pm$ 0.0059 & -0.00010 $\pm$ 0.00048 & HARPS-N \\
7282.4250 & 4.65 $\pm$ 0.90 & 22.0 $\pm$ 3.3 & -2.5 $\pm$ 1.6 & 0.0307 $\pm$ 0.0034 & 0.00104 $\pm$ 0.00025 & HARPS-N \\
7283.3603 & 2.24 $\pm$ 0.92 & 23.5 $\pm$ 3.3 & -2.5 $\pm$ 1.7 & 0.0337 $\pm$ 0.0034 & 0.00046 $\pm$ 0.00027 & HARPS-N \\
7285.3833 & 4.61 $\pm$ 0.92 & 27.2 $\pm$ 3.4 & -0.3 $\pm$ 1.7 & 0.0090 $\pm$ 0.0027 & -0.00037 $\pm$ 0.00027 & HARPS-N \\
7286.3458 & 1.98 $\pm$ 0.91 & 18.0 $\pm$ 3.3 & 1.9 $\pm$ 1.7 & 0.0278 $\pm$ 0.0035 & 0.00056 $\pm$ 0.00029 & HARPS-N \\
7286.4111 & 2.79 $\pm$ 0.92 & 20.6 $\pm$ 3.3 & 2.1 $\pm$ 1.7 & 0.0247 $\pm$ 0.0035 & 0.00051 $\pm$ 0.00030 & HARPS-N \\
7287.3937 & -1.55 $\pm$ 0.93 & 19.3 $\pm$ 3.4 & 3.3 $\pm$ 1.7 & 0.0205 $\pm$ 0.0031 & -0.00011 $\pm$ 0.00028 & HARPS-N \\
7293.3446 & -3.63 $\pm$ 1.01 & -10.0 $\pm$ 3.7 & 2.7 $\pm$ 1.9 & -0.0054 $\pm$ 0.0043 & -0.00000 $\pm$ 0.00038 & HARPS-N \\
7293.3965 & -2.60 $\pm$ 0.93 & -9.6 $\pm$ 3.4 & 0.3 $\pm$ 1.7 & -0.0059 $\pm$ 0.0031 & 0.00028 $\pm$ 0.00028 & HARPS-N \\
7294.4132 & -4.44 $\pm$ 0.93 & -9.9 $\pm$ 3.4 & -0.0 $\pm$ 1.7 & -0.0054 $\pm$ 0.0034 & 0.00133 $\pm$ 0.00029 & HARPS-N \\
7303.3575 & 1.20 $\pm$ 0.92 & -5.9 $\pm$ 3.4 & -4.0 $\pm$ 1.7 & -0.0039 $\pm$ 0.0034 & 0.00064 $\pm$ 0.00028 & HARPS-N \\
7306.3577 & 3.13 $\pm$ 0.93 & -17.5 $\pm$ 3.4 & 0.0 $\pm$ 1.7 & -0.0092 $\pm$ 0.0037 & 0.00036 $\pm$ 0.00036 & HARPS-N \\
7307.3487 & -2.72 $\pm$ 1.07 & -11.8 $\pm$ 4.0 & 1.6 $\pm$ 2.0 & -0.0111 $\pm$ 0.0061 & -0.00231 $\pm$ 0.00053 & HARPS-N \\
7307.3543 & -0.99 $\pm$ 0.94 & -15.3 $\pm$ 3.4 & 2.3 $\pm$ 1.7 & -0.0123 $\pm$ 0.0043 & -0.00098 $\pm$ 0.00037 & HARPS-N \\
7308.3648 & -4.99 $\pm$ 0.92 & -15.5 $\pm$ 3.3 & 4.7 $\pm$ 1.7 & -0.0075 $\pm$ 0.0034 & -0.00013 $\pm$ 0.00028 & HARPS-N \\
7508.6888 & -9.17 $\pm$ 1.00 & 10.9 $\pm$ 3.7 & 1.1 $\pm$ 1.8 & 0.0252 $\pm$ 0.0034 & 0.00177 $\pm$ 0.00027 & HARPS-N \\
7510.6951 & 0.68 $\pm$ 1.56 & 15.5 $\pm$ 6.1 & -4.2 $\pm$ 3.0 & 0.0283 $\pm$ 0.0060 & 0.00184 $\pm$ 0.00052 & HARPS-N \\
7535.6823 & 1.15 $\pm$ 1.34 & 25.6 $\pm$ 5.1 & 0.2 $\pm$ 2.6 & 0.0139 $\pm$ 0.0062 & 0.00148 $\pm$ 0.00057 & HARPS-N \\
7535.6879 & 0.98 $\pm$ 1.16 & 7.6 $\pm$ 4.4 & -3.3 $\pm$ 2.2 & 0.0263 $\pm$ 0.0046 & 0.00162 $\pm$ 0.00044 & HARPS-N \\
7536.6580 & 2.85 $\pm$ 1.08 & 2.3 $\pm$ 4.0 & -4.1 $\pm$ 2.0 & 0.0158 $\pm$ 0.0042 & 0.00170 $\pm$ 0.00045 & HARPS-N \\
7537.6699 & 7.08 $\pm$ 0.96 & 11.9 $\pm$ 3.5 & -3.1 $\pm$ 1.8 & 0.0232 $\pm$ 0.0037 & 0.00254 $\pm$ 0.00033 & HARPS-N \\
7538.6079 & 4.55 $\pm$ 0.98 & 34.0 $\pm$ 3.6 & -2.5 $\pm$ 1.8 & 0.0163 $\pm$ 0.0033 & 0.00189 $\pm$ 0.00026 & HARPS-N \\
7538.7047 & 6.64 $\pm$ 0.97 & 29.3 $\pm$ 3.6 & -1.7 $\pm$ 1.8 & 0.0223 $\pm$ 0.0034 & 0.00201 $\pm$ 0.00027 & HARPS-N \\
7606.5080 & -4.78 $\pm$ 1.12 & 3.2 $\pm$ 4.2 & -6.2 $\pm$ 2.1 & -0.0144 $\pm$ 0.0038 & -0.00037 $\pm$ 0.00040 & HARPS-N \\
7607.5155 & -3.12 $\pm$ 1.53 & 22.2 $\pm$ 5.9 & -4.7 $\pm$ 3.0 & -0.0153 $\pm$ 0.0062 & -0.00029 $\pm$ 0.00060 & HARPS-N \\
7607.5209 & -6.50 $\pm$ 1.56 & 22.7 $\pm$ 6.0 & -8.1 $\pm$ 3.0 & -0.0097 $\pm$ 0.0051 & 0.00007 $\pm$ 0.00059 & HARPS-N \\
7608.5056 & -2.44 $\pm$ 1.04 & 10.0 $\pm$ 3.8 & -6.3 $\pm$ 1.9 & 0.0081 $\pm$ 0.0037 & -0.00042 $\pm$ 0.00035 & HARPS-N \\
7609.5098 & 2.58 $\pm$ 1.26 & 9.8 $\pm$ 4.8 & -1.6 $\pm$ 2.4 & 0.0043 $\pm$ 0.0043 & 0.00186 $\pm$ 0.00050 & HARPS-N \\
7610.4957 & 1.25 $\pm$ 1.60 & 10.2 $\pm$ 6.2 & -3.4 $\pm$ 3.1 & -0.0174 $\pm$ 0.0050 & -0.00004 $\pm$ 0.00066 & HARPS-N \\
7625.3830 & -0.85 $\pm$ 0.92 & -12.6 $\pm$ 3.3 & -2.4 $\pm$ 1.7 & -0.0068 $\pm$ 0.0034 & 0.00013 $\pm$ 0.00028 & HARPS-N \\
7625.4766 & -1.14 $\pm$ 0.93 & -20.3 $\pm$ 3.4 & -1.3 $\pm$ 1.7 & -0.0062 $\pm$ 0.0035 & -0.00027 $\pm$ 0.00031 & HARPS-N \\
7626.3865 & -0.57 $\pm$ 1.08 & -8.3 $\pm$ 4.0 & -3.0 $\pm$ 2.0 & -0.0024 $\pm$ 0.0040 & 0.00034 $\pm$ 0.00042 & HARPS-N \\
7626.4766 & -1.54 $\pm$ 1.02 & -0.8 $\pm$ 3.8 & -3.4 $\pm$ 1.9 & -0.0003 $\pm$ 0.0037 & 0.00007 $\pm$ 0.00035 & HARPS-N \\
7627.3807 & -0.22 $\pm$ 0.95 & -11.0 $\pm$ 3.5 & -5.7 $\pm$ 1.7 & -0.0026 $\pm$ 0.0037 & 0.00045 $\pm$ 0.00035 & HARPS-N \\
7627.4712 & -1.70 $\pm$ 0.97 & -7.6 $\pm$ 3.6 & -4.0 $\pm$ 1.8 & -0.0016 $\pm$ 0.0038 & 0.00046 $\pm$ 0.00036 & HARPS-N \\
7628.3839 & -0.97 $\pm$ 1.04 & 6.0 $\pm$ 3.9 & -0.4 $\pm$ 1.9 & -0.0070 $\pm$ 0.0036 & -0.00285 $\pm$ 0.00034 & HARPS-N \\
7628.4993 & 0.92 $\pm$ 1.03 & -0.9 $\pm$ 3.8 & -3.2 $\pm$ 1.9 & -0.0038 $\pm$ 0.0036 & -0.00398 $\pm$ 0.00033 & HARPS-N \\
7629.3800 & 5.21 $\pm$ 0.96 & 0.2 $\pm$ 3.5 & 0.8 $\pm$ 1.8 & -0.0025 $\pm$ 0.0037 & -0.00068 $\pm$ 0.00036 & HARPS-N \\
7630.3821 & 2.17 $\pm$ 1.01 & 1.9 $\pm$ 3.7 & 0.4 $\pm$ 1.9 & -0.0022 $\pm$ 0.0038 & -0.00155 $\pm$ 0.00037 & HARPS-N \\
7630.4850 & 1.68 $\pm$ 1.02 & 4.0 $\pm$ 3.8 & 1.5 $\pm$ 1.9 & 0.0049 $\pm$ 0.0039 & -0.00251 $\pm$ 0.00039 & HARPS-N \\
7632.3816 & -2.23 $\pm$ 0.95 & 4.2 $\pm$ 3.5 & 1.7 $\pm$ 1.7 & 0.0058 $\pm$ 0.0036 & 0.00056 $\pm$ 0.00031 & HARPS-N \\
7632.4842 & -1.23 $\pm$ 0.94 & -2.7 $\pm$ 3.4 & -1.3 $\pm$ 1.7 & 0.0040 $\pm$ 0.0035 & -0.00039 $\pm$ 0.00030 & HARPS-N \\
7637.3894 & 3.13 $\pm$ 0.97 & -8.7 $\pm$ 3.6 & -2.3 $\pm$ 1.8 & -0.0007 $\pm$ 0.0040 & 0.00011 $\pm$ 0.00041 & HARPS-N \\
7637.4491 & 4.24 $\pm$ 0.97 & -4.1 $\pm$ 3.5 & -1.0 $\pm$ 1.8 & 0.0026 $\pm$ 0.0039 & 0.00035 $\pm$ 0.00039 & HARPS-N \\
7638.3624 & -0.60 $\pm$ 0.99 & -10.6 $\pm$ 3.6 & -0.7 $\pm$ 1.8 & -0.0047 $\pm$ 0.0038 & 0.00010 $\pm$ 0.00037 & HARPS-N \\
7638.4526 & 0.36 $\pm$ 0.92 & -8.0 $\pm$ 3.4 & -0.2 $\pm$ 1.7 & 0.0047 $\pm$ 0.0036 & 0.00076 $\pm$ 0.00032 & HARPS-N \\
7641.3727 & -1.51 $\pm$ 0.94 & 4.2 $\pm$ 3.4 & -1.6 $\pm$ 1.7 & 0.0098 $\pm$ 0.0035 & -0.00096 $\pm$ 0.00029 & HARPS-N \\
7641.4507 & -1.83 $\pm$ 0.95 & 5.2 $\pm$ 3.5 & -0.9 $\pm$ 1.7 & 0.0094 $\pm$ 0.0035 & -0.00038 $\pm$ 0.00028 & HARPS-N \\
7642.3706 & -2.42 $\pm$ 0.96 & -3.6 $\pm$ 3.5 & -3.0 $\pm$ 1.8 & 0.0113 $\pm$ 0.0036 & 0.00149 $\pm$ 0.00033 & HARPS-N \\
7642.4490 & -0.55 $\pm$ 1.24 & -4.1 $\pm$ 4.4 & -2.2 $\pm$ 2.2 & -0.0004 $\pm$ 0.0038 & 0.00087 $\pm$ 0.00037 & HARPS-N \\
7643.3693 & 2.02 $\pm$ 1.23 & -2.8 $\pm$ 4.4 & -1.4 $\pm$ 2.2 & 0.0131 $\pm$ 0.0039 & -0.00042 $\pm$ 0.00038 & HARPS-N \\
7643.4479 & 0.54 $\pm$ 1.21 & 0.1 $\pm$ 4.3 & -1.0 $\pm$ 2.2 & 0.0114 $\pm$ 0.0036 & 0.00013 $\pm$ 0.00031 & HARPS-N \\
7645.3705 & -1.38 $\pm$ 0.97 & 11.6 $\pm$ 3.5 & -1.6 $\pm$ 1.8 & 0.0093 $\pm$ 0.0036 & 0.00048 $\pm$ 0.00032 & HARPS-N \\
7645.4510 & -1.68 $\pm$ 0.98 & 8.4 $\pm$ 3.6 & -2.1 $\pm$ 1.8 & 0.0220 $\pm$ 0.0037 & 0.00217 $\pm$ 0.00033 & HARPS-N \\
7646.3662 & -1.92 $\pm$ 0.97 & 10.3 $\pm$ 3.6 & -2.0 $\pm$ 1.8 & 0.0159 $\pm$ 0.0037 & 0.00085 $\pm$ 0.00033 & HARPS-N \\
7646.4449 & -3.29 $\pm$ 1.03 & 23.1 $\pm$ 3.8 & -2.2 $\pm$ 1.9 & 0.0091 $\pm$ 0.0037 & -0.00002 $\pm$ 0.00035 & HARPS-N \\
7923.7976 & -0.94 $\pm$ 1.10 & -7.0 $\pm$ 4.1 & -2.3 $\pm$ 2.0 & -0.0215 $\pm$ 0.0057 & -0.00031 $\pm$ 0.00042 & HARPS post-Upgrade \\
7924.7978 & -1.36 $\pm$ 1.04 & -12.3 $\pm$ 3.8 & -0.9 $\pm$ 1.9 & -0.0241 $\pm$ 0.0055 & -0.00230 $\pm$ 0.00036 & HARPS post-Upgrade \\
7945.7043 & -1.07 $\pm$ 0.97 & -4.6 $\pm$ 3.6 & 2.4 $\pm$ 1.8 & -0.0055 $\pm$ 0.0053 & -0.00130 $\pm$ 0.00029 & HARPS post-Upgrade \\
7946.7629 & 0.42 $\pm$ 1.03 & -5.9 $\pm$ 3.8 & -0.7 $\pm$ 1.9 & -0.0074 $\pm$ 0.0055 & -0.00128 $\pm$ 0.00035 & HARPS post-Upgrade \\
7948.7909 & -2.69 $\pm$ 1.04 & -13.9 $\pm$ 3.8 & 2.7 $\pm$ 1.9 & -0.0250 $\pm$ 0.0055 & -0.00178 $\pm$ 0.00035 & HARPS post-Upgrade \\
7951.6973 & -4.36 $\pm$ 1.34 & -11.8 $\pm$ 5.1 & 2.1 $\pm$ 2.6 & -0.0435 $\pm$ 0.0061 & 0.00006 $\pm$ 0.00058 & HARPS post-Upgrade \\
7952.7219 & 0.09 $\pm$ 1.12 & -13.6 $\pm$ 4.2 & 0.2 $\pm$ 2.1 & -0.0297 $\pm$ 0.0057 & -0.00055 $\pm$ 0.00042 & HARPS post-Upgrade \\
7953.7369 & 2.11 $\pm$ 1.00 & -13.5 $\pm$ 3.7 & 1.5 $\pm$ 1.8 & -0.0236 $\pm$ 0.0054 & -0.00020 $\pm$ 0.00032 & HARPS post-Upgrade \\
7954.7522 & 1.70 $\pm$ 1.05 & -19.9 $\pm$ 3.9 & 0.4 $\pm$ 2.0 & -0.0360 $\pm$ 0.0055 & -0.00043 $\pm$ 0.00036 & HARPS post-Upgrade \\
7955.7725 & -1.01 $\pm$ 1.03 & -16.4 $\pm$ 3.8 & -2.7 $\pm$ 1.9 & -0.0268 $\pm$ 0.0054 & -0.00020 $\pm$ 0.00034 & HARPS post-Upgrade \\
7956.7489 & 1.21 $\pm$ 1.01 & -18.2 $\pm$ 3.7 & -0.0 $\pm$ 1.9 & -0.0232 $\pm$ 0.0054 & 0.00022 $\pm$ 0.00033 & HARPS post-Upgrade \\
7959.7006 & 0.68 $\pm$ 1.08 & -8.0 $\pm$ 4.0 & 0.8 $\pm$ 2.0 & -0.0305 $\pm$ 0.0056 & -0.00021 $\pm$ 0.00040 & HARPS post-Upgrade \\
7960.7290 & -0.54 $\pm$ 1.08 & -6.7 $\pm$ 4.0 & -2.3 $\pm$ 2.0 & -0.0268 $\pm$ 0.0056 & -0.00054 $\pm$ 0.00039 & HARPS post-Upgrade \\
7961.7240 & -1.60 $\pm$ 1.09 & -19.8 $\pm$ 4.1 & 0.0 $\pm$ 2.0 & -0.0311 $\pm$ 0.0055 & 0.00051 $\pm$ 0.00037 & HARPS post-Upgrade \\
7962.7643 & -3.45 $\pm$ 1.11 & -9.7 $\pm$ 4.2 & -2.1 $\pm$ 2.1 & -0.0298 $\pm$ 0.0055 & 0.00003 $\pm$ 0.00038 & HARPS post-Upgrade \\
7964.7226 & -2.96 $\pm$ 0.96 & -13.2 $\pm$ 3.5 & -0.2 $\pm$ 1.8 & -0.0137 $\pm$ 0.0053 & 0.00073 $\pm$ 0.00027 & HARPS post-Upgrade \\
7966.5916 & 2.09 $\pm$ 1.06 & 2.7 $\pm$ 3.9 & -0.4 $\pm$ 2.0 & -0.0122 $\pm$ 0.0056 & 0.00012 $\pm$ 0.00038 & HARPS post-Upgrade \\
7977.5291 & -1.29 $\pm$ 1.03 & 14.5 $\pm$ 3.8 & -3.2 $\pm$ 1.9 & 0.0071 $\pm$ 0.0038 & -0.00072 $\pm$ 0.00037 & HARPS-N \\
7978.4574 & -0.37 $\pm$ 0.95 & 10.7 $\pm$ 3.5 & 1.5 $\pm$ 1.7 & 0.0084 $\pm$ 0.0035 & 0.00024 $\pm$ 0.00030 & HARPS-N \\
7979.4902 & 2.99 $\pm$ 0.91 & 2.8 $\pm$ 3.3 & 3.7 $\pm$ 1.7 & 0.0014 $\pm$ 0.0034 & 0.00011 $\pm$ 0.00027 & HARPS-N \\
7980.4864 & 1.21 $\pm$ 0.91 & 10.2 $\pm$ 3.3 & 2.5 $\pm$ 1.7 & 0.0019 $\pm$ 0.0034 & 0.00111 $\pm$ 0.00027 & HARPS-N \\
7981.4933 & 0.34 $\pm$ 0.92 & -0.0 $\pm$ 3.3 & 2.6 $\pm$ 1.7 & -0.0049 $\pm$ 0.0034 & 0.00032 $\pm$ 0.00026 & HARPS-N \\
7983.4886 & -1.41 $\pm$ 0.96 & -6.2 $\pm$ 3.5 & 2.2 $\pm$ 1.8 & -0.0143 $\pm$ 0.0036 & -0.00136 $\pm$ 0.00032 & HARPS-N \\
8006.3821 & 8.36 $\pm$ 0.96 & 31.0 $\pm$ 3.5 & -2.2 $\pm$ 1.8 & 0.0014 $\pm$ 0.0034 & -0.00233 $\pm$ 0.00028 & HARPS-N \\
8008.3756 & 5.32 $\pm$ 0.94 & 21.4 $\pm$ 3.4 & 0.7 $\pm$ 1.7 & 0.0035 $\pm$ 0.0035 & 0.00010 $\pm$ 0.00029 & HARPS-N \\

 \hline

\end {longtable}
\end{appendix}
\label{lastpage}

\end{document}